\documentclass[11pt,a4paper]{article}
\pdfoutput = 1
\usepackage{jheppub}
\usepackage{ulem}
\usepackage{epsfig}
\usepackage{multirow}
\usepackage{slashed}
\usepackage{amsmath}
\usepackage{color}
\usepackage[dvipsnames]{xcolor}
\usepackage{subfigure}
\usepackage{textcomp}

\def\bea#1\eea{\begin{align}#1\end{align}}

\newcommand{\la}{\left\langle}
\newcommand{\ra}{\right\rangle}
\newcommand{\lc}{\left[}
\newcommand{\rc}{\right]}
\newcommand{\lp}{\left(}
\newcommand{\rp}{\right)}
\newcommand\abs[1]{\left|#1\right|}

\newcommand{\eP}{\texttt{ePump }}

\begin{document}

\preprint{MSUHEP-21-007}
	
\title{Determining the helicity structure of the nucleon at the Electron Ion Collider in China}

\date{\today}

\author[a,b,c]{Daniele Paolo Anderle,}
\emailAdd{dpa@m.scnu.edu.cn}
\affiliation[a]{Guangdong Provincial Key Laboratory of Nuclear Science,\\ Institute of Quantum Matter, South China Normal University, Guangzhou 510006, China}

\affiliation[b]{Guangdong-Hong Kong Joint Laboratory of Quantum Matter, Southern Nuclear Science Computing Center, South China Normal University, Guangzhou 510006, China}

\affiliation[c]{Department of Physics and Astronomy, University of California, Los Angeles, California 90095, USA}

\author[d]{Tie-Jiun Hou,}
\emailAdd{houtiejiun@mail.neu.edu.cn}
\affiliation[d]{Department of Physics, Northeastern University, Shenyang 110819, China}

\author[a,b]{Hongxi Xing,}
\emailAdd{hxing@m.scnu.edu.cn}

\author[e]{Mengshi Yan,}
\emailAdd{msyan@pku.edu.cn}
\affiliation[e]{Department of Physics and State Key Laboratory of Nuclear Physics and Technology,\\ Peking University, Beijing 100871, China}       

\author[f]{C.--P. Yuan}
\emailAdd{yuanch@msu.edu}
\affiliation[f]{Department of Physics and Astronomy, Michigan State University, East Lansing, MI 48824, US}       

\author[g,h]{and Yuxiang Zhao}
\emailAdd{yxzhao@impcas.ac.cn}
\affiliation[g]{Institute of Modern Physics, Chinese Academy of Sciences, Lanzhou 730000, China}
\affiliation[h]{University of Chinese Academy of Sciences, Beijing 100049, China}

\date{\today} 

\abstract{
Understanding how sea quarks behave inside a nucleon is one of the most important physics goals of the proposed Electron-Ion Collider
in China (EicC), which is designed to have a 3.5 GeV polarized electron beam (80\% polarization) colliding with a 20 GeV polarized proton beam (70\% polarization) at instantaneous luminosity of $2 \times 10^{33} {\rm cm}^{-2} {\rm s}^{-1}$. A specific topic at EicC is to understand the polarization of individual quarks inside a longitudinally polarized nucleon.
The potential of various future EicC data, including the inclusive and semi-inclusive deep inelastic scattering data from both doubly polarized electron-proton and electron-$^3{\rm He}$ collisions, to reduce the uncertainties of parton helicity distributions is explored at the next-to-leading order in QCD, using the Error PDF Updating Method Package ({\sc ePump}) which is based on the Hessian profiling method. 
We show that the semi-inclusive data are well able to provide good separation between flavour distributions, and to constrain their uncertainties in the $x>0.005$ region, especially when electron-$^3{\rm He}$ collisions, acting as effective electron-neutron
collisions, are taken into account.
To enable this study, we have generated a Hessian representation of the DSSV14 set of PDF replicas, named DSSV14H PDFs.}

\maketitle

\section{Introduction}
\label{sec_introduction}
Understanding the helicity structure of the nucleon in terms of quark and gluon degrees of freedom is of fundamental interest in modern hadronic physics. In the naive parton model, the proton spin is considered to be originated from its three valence quarks. This naive picture was first challenged by the pioneering measurements of polarized deep inelastic scattering (DIS) performed in the EMC experiment in the late 1980s ~\cite{Ashman:1987hv}, in which the spin carried by the three valence quarks has been shown to be much less than the expected $1/2$. Ever since then, tremendous experimental progress has been made in the past 32 years, including fixed target experiments of polarized lepton-proton and lepton-ion scatterings at SLAC, CERN, DESY, and JLAB~\cite{Aidala:2012mv}, as well as the seminal spin program of polarized proton-proton and proton-helium collisions at Relativistic Heavy Ion Collider (RHIC) ~\cite{Aschenauer:2015eha}. 

Identifying the gluon spin contribution and the quark flavour discriminated helicity distributions are key steps in order to precisely pin down the proton spin configurations.
The first evidence of the polarization of gluon inside the proton~\cite{deFlorian:2014yva} was shown in the RHIC spin program, particularly in the measurement of jet production in polarized proton-proton collisions~\cite{Adamczyk:2014ozi}, where, however, one still could not draw any reliable conclusions on the exact gluon contribution to the proton spin due to the limited kinematic coverage. Another interesting measurement at RHIC was the charged weak vector boson production in polarized proton-proton collisions~\cite{Aggarwal:2010vc,Adare:2010xa}, which was supposed to provide a sensitive channel to determine the flavour discriminated helicity distributions. However, again, due to the limited kinematic coverage, no conclusive statement could be made. As a result, the parton helicity distributions were left largely unconstrained.  

It has long been recognized that parton helicity distributions can be probed in polarized lepton-nucleon scatterings.
Recently, there have been several proposals to build a new generation of polarized Electron-Ion Collider (EIC) worldwide, such as an EIC at Brookhaven National Laboratory~\cite{EIC} and an EIC in China~\cite{Anderle:2021wcy}. The EIC machine in the US is designed to probe the parton helicity distribution in a relatively small $x$ region as compared to the ongoing experiments at JLab. The Electron-Ion Collider in China (EicC) is proposed to be constructed based on an upgraded heavy-ion accelerator, High-Intensity heavy-ion Accelerator Facility (HIAF) which is currently under construction in Guangdong Province of China, together with an additional electron ring. The EicC is designed to cover a center-of-mass energy range from 15 to 20 GeV with the luminosity of about $2 \times 10^{33} {\rm cm}^{-2} {\rm s}^{-1}$ in electron-proton collisions, which aims to bridge the kinematic coverage between EIC and JLab. In addition to a polarized electron beam with a polarization of $80\%$ and a polarized proton beam with a polarization of $70\%$, it also plans to offer polarized light-ion beams such as ${}^3{\rm He}$ with a polarization of $70\%$.

The design of the EicC offers unprecedented new opportunities to study the spin structure of the nucleon. In this paper, we present a quantitative study of the impact that future EicC inclusive and semi-inclusive deep inelastic scattering (SIDIS) measurements will have on the determination of various helicity distributions inside the nucleon as well as on the determination of their contributions to the proton spin. In our study, we take advantage of simulated DIS and SIDIS data from electron-proton and electron-${}^3{\rm He}$ collisions in order to disentangle the constraining power that each initial and final state combination has on the different flavour helicity distributions. During our discussion, it will become apparent that both proton target and effective neutron targets, such as ${}^3{\rm He}$, together with identified final states of well-known flavour content, such as pions and kaons, are needed to achieve a consistent reduction of uncertainties across different helicity distributions. 
We will also show that the high accuracy with which these processes are planned to be measured at the future EicC will allow for unprecedented precision in the extraction of quark helicity distributions in the region usually referred as the ``sea-quark'' region.

Apart from the intrinsic advancement of our knowledge of the proton spin content that comes with the improved overall precision of helicity distributions, reaching such precision in the sea sector also means that the EicC will help to clarify some intriguing problems and phenomena observed in the previous experiments, such as the asymmetry in the distribution of polarized light sea quarks, and the polarization of strange quarks inside a polarized nucleon, 
etc.~\cite{Bhalerao:2001rn,Peng:2003zm,Bourrely:2001du,Airapetian:2004zf,Airapetian:2008qf,Khorramian:2020gkr}. 

The paper is organized as follows. In the next section, we introduce the theoretical framework used to compute the Double-Spin-Asymmetries observable in DIS and SIDIS processes. In Sec.~\ref{sec_data} we present the methodology applied to generate EicC pseudo-data for the different processes considered. Our main discussions and findings are collected in Sec.~\ref{sec_fitting}. After introducing the {\tt ePump} tool~\cite{Schmidt:2018hvu,Hou:2019gfw} used to perform our analysis, we summarize in Sec.~\ref{sec_mc2hessian} the procedure used to convert replica sets of Parton Distribution Functions (PDFs) into equivalent hessian sets of them as this is a step needed in order to be able to use the chosen replica PDF set within the hessian framework utilized by the {\tt ePump} software. The details of the theoretical framework upon which the {\tt ePump} operates are summarized in Sec.~\ref{sec:hpm} and~\ref{sec:hmo}. We proceed by describing in Sec.~\ref{sec_tables} the specific choices that went into making the theory and data tables fed to {\tt ePump}. In Sec.~\ref{sec_discussion} we present our main results and discuss them. Discussion of the sensitivity of helicity distributions to specific data samples is presented in a more quantitative manner in Sec.~\ref{sec_optimize}. We conclude by summarizing our findings and remarks in Sec.~\ref{sec_summary}.

\section{Polarized Lepton-Nucleon scatterings to access helicity distributions}
\label{sec_DIS_SIDIS}
In the Deep-Inelastic Scattering (DIS) process, $e+p(n) \to e + X$, a nucleon such as a proton (neutron) is collided with an electron and gets destroyed into unobserved hadronic remnants $X$ while keeping track of the original electron bouncing off the nucleon. Making use of both longitudinally polarized electron and proton (or effective neutron) beams, a possible observable that can be measured during this process is the so-called Double-Spin-Asymmetry (DSA):
\begin{equation}
A_{LL}=\frac{d\sigma^{++} - d\sigma^{+-}}{d\sigma^{++} + d\sigma^{+-}} 
= \frac{1}{P_e P_p} \frac{N^{++} - N^{+-}}{N^{++} + N^{+-}},
\label{eq_asy}
\end{equation}
where the superscript ``$+$" and ``$-$" denotes the helicity of the two beams respectively, 
$P_e$ ($P_p$) means the polarization of electron (proton) beam, $N$ is the luminosity-normalized number of events in a specific
spin orientation state.

It is necessary to define the kinematic variables for discussions on the experimental observables. With
$k, k'$ denoting the four-momenta of the incoming and outgoing electron, $p$ denoting the four-momentum of the incoming nucleon, 
one can define the following kinematic variables: 
\begin{eqnarray}
Q^2 &=& -q^2 = -(k-k')^2, \\
x &=& \frac{Q^2}{2p \cdot q}, \\
y &=& \frac{q \cdot p}{k \cdot p}, \\
W_h &=& \sqrt{(p+q)^2}.
\end{eqnarray}

The measured DSA can be written down approximately as
\begin{equation}
    A_{LL} = D(y) A_1= \frac{y(y-2)}{y^2+2(1-y)(1+R)} A_1
\end{equation}
in a kinematic region where $x$ is small while the momentum transfer $Q^2$ is relatively high ~\cite{Aghasyan:2017vck,Aschenauer:2012ve}. The factor
$R$ is the cross section ratio between the absorption of a longitudinally polarized virtual photon and a transversely 
polarized virtual photon by a nucleon, $R=\frac{\sigma_L}{\sigma_T}$.  
The asymmetry $A_1$ is related to the longitudinal spin structure function $g_1$ by
\begin{equation}
    A_1 = \frac{g_1}{F_2/[2x(1+R)]} \simeq \frac{g_1}{F_1},
    \label{eq:Aone}
\end{equation}
where $F_2$ or $F_1$ denotes the spin-independent structure function.

In the inclusive DIS process, where only the scattered electrons are detected, the measured $F_1$ and $g_1$ structure
functions can be expressed at leading order (LO) in the parton model, for $Q$ being much smaller than the $Z$ boson mass, as
\begin{equation}
\label{eq:F1LO}
F_1(x,Q^2)=\frac{1}{2}\sum_{q=(u,d,s)} e_q^2\big[ q(x,Q^2)+\overline{q}(x,Q^2)\big] ,
\end{equation}
\begin{equation}
\label{eq:g1LO}
g_1(x,Q^2)=\frac{1}{2}\sum_{q=(u,d,s)} e_q^2\big[\Delta q(x,Q^2)+\Delta\overline{q}(x,Q^2)\big] ,
\end{equation}
where $q$ and $\Delta q$ denote the unpolarized and helicity parton distribution functions respectively.

On the other hand, in the SIDIS process, where a leading hadron such as $\pi^{\pm}$ or $K^{\pm}$ is also detected
in addition to the scattered electron, the measured asymmetry $A_1^h  \simeq \frac{g_1^h}{F_1^h}$
is related to the corresponding semi-inclusive
unpolarized and longitudinal spin structure functions, $F_1^h$ and $g_1^h$, which can be expressed at LO in the parton model, for $Q$ being much smaller than the $Z$ boson mass, 
as
\begin{equation}
\label{eq:F1hLO}
F_1^h(x,Q^2,z)=\frac{1}{2}\sum_qe_q^2\big[ q(x,Q^2) D^{q \to h}(Q^2,z)  \\
+\Delta\overline{q}(x,Q^2) D^{ \overline{q} \to h}(Q^2,z) \big],
\end{equation}
\begin{equation}
\label{eq:g1hLO}
g_1^h(x,Q^2,z)=\frac{1}{2}\sum_qe_q^2\big[\Delta q(x,Q^2) D^{q \to h}(Q^2,z)  \\
+\Delta\overline{q}(x,Q^2) D^{ \overline{q} \to h}(Q^2,z) \big] .
\end{equation}
Here $D^{q \to h}(Q^2,z)$ describes the fragmentation process from a quark $q$ to a hadron $h$, $z$ represents the momentum fraction 
of the final state hadron, whose four-momentum is denoted as $P_{h}$, with respect to the momentum of the produced quark. Experimentally it is defined as 
$z=\frac{P_{h} \cdot p}{q \cdot p}$. 

As one can tell from the LO expressions, the final-state hadron in the SIDIS processes offers different weights for different 
flavours of the initial state quark comparing to the inclusive DIS measurements. Hence, SIDIS processes provide a powerful way to separate single flavour distributions.
In the electron-proton collision, considering $\pi^{\pm}$ and $K^{\pm}$ SIDIS processes, there will be four sets of data.
In addition, using polarized $^3{\rm He}$ as an effective neutron target offers additional four sets of data. The LO expressions for $A_1^h$ of these eight data sets can be found in Appendix~\ref{app:A1}.

Beyond LO, factorization of short (high energy) and long (low energy) distance interactions in DIS and SIDIS allows to write the previous LO expressions in an all-order form. For instance, the spin dependent structure function $g_1^h$ can be written as
\begin{eqnarray}
\label{eq:g1hallorders}
g_1^h(x,z,Q^2) &=&\frac{1}{2}\sum_{f,f'=q,\bar{q},g} 
\int_x^1 \frac{d\hat{x}}{\hat{x}}\int_z^1 \frac{d\hat{z}}{\hat{z}}\, \Delta f \left(\frac{x}{\hat{x}},
\mu^2\right)D^{ f' \to h} \left(\frac{z}{\hat{z}},\mu^2\right)\,\Delta{\cal{C}}_{f'f}
\left(\hat{x},\hat{z},\frac{Q^2}{\mu^2},\alpha_s(\mu^2)\right) \nonumber\\
&\equiv&\,\frac{1}{2}\sum_{f,f'=q,\bar{q},g} \left[\Delta f\otimes \Delta{\cal{C}}_{f'f}\otimes D^{ f' \to h}\right](x,z,Q^2,\mu^2),
\end{eqnarray}
where with $\mu$ we collectively denote all factorization and renormalization scales, $\hat{x}$ and $\hat{z}$ are the partonic counterparts of the hadronic variables $x$ and $z$, and  $\Delta{\cal{C}}_{f'f}$ are spin-dependent coefficient functions.

Similar expressions can be written for inclusive $g_1$ by removing the fragmentation functions. For unpolarized cases, such as $F_1^h$ and $F_1$, one has to use
unpolarized parton distributions and unpolarized coefficient functions accordingly.

Moreover, using perturbation theory the coefficient functions can be expanded in terms of powers of the strong coupling constant $\alpha_s$. For example,
\begin{equation}
\label{eq:factcoef}
\Delta{\cal{C}}_{f'f}\,=\,\Delta C^{(0)}_{f'f}+\frac{\alpha_s(\mu^2)}{2\pi}\Delta C^{(1)}_{f'f}+{\cal O}(\alpha_s^2),
\end{equation}
and its LO expression yields,
\begin{equation}
   \Delta{\cal{C}}_{qq}(\hat{x},\hat{z})\,=\,\Delta{\cal{C}}_{\bar{q}\bar{q}}
(\hat{x},\hat{z})\,= \Delta C^{(0)}_{qq}(\hat{x},\hat{z})=\, e_q^2\,\delta(1-\hat{x})\delta(1-\hat{z})\,, 
\end{equation}
which, once being inserted into Eq.~(\ref{eq:g1hallorders}), trivially gives the LO expression of Eq.~(\ref{eq:g1hLO}) with $\mu^2=Q^2$. At LO, all other coefficient functions are vanishing.

The corresponding next-to-leading oder (NLO) expression to Eq.~(\ref{eq:g1hLO}) 
can be obtain from Eqs.~(\ref{eq:factcoef}) and~(\ref{eq:g1hallorders}):
\begin{eqnarray}
\label{eq:g1hNLO}
2g_1^h(x,z,Q^2)&=&\sum_q e^2_q \bigg\{\Delta q(x,Q^2)D^{ q \to h}(z,Q^2)+\Delta\bar{q}(x,Q^2)D^{ \bar q \to h}(z,Q^2)\nonumber\\
&&+\frac{\alpha_s(Q^2)}{2\pi}\left[\left(\Delta q\otimes D^{ q \to h}+
\Delta \bar{q}\otimes D^{ \bar q \to h}\right)\otimes \Delta C^{(1)}_{qq}\right.+\left(\Delta q+\Delta \bar{q}\right)\otimes \Delta C^{(1)}_{gq}\otimes D^{ g \to h}\nonumber\\[2mm]
&&\left. +\Delta g\otimes \Delta C^{(1)}_{qg}\otimes (D^{ q \to h}+D^{ \bar q \to h})\right](x,z,Q^2)\bigg\},
\end{eqnarray}
where we have set again $\mu^2=Q^2$.
Explicit expressions for the polarized and unpolarized coefficient functions can be found for SIDIS in ~\cite{Anderle:2012rq,deFlorian:1997zj,deFlorian:2012wk,Altarelli:1979kv,Nason:1993xx,Furmanski:1981cw,Graudenz:1994dq}
and for DIS in ~\cite{Furmanski:1981cw,Gluck:1995yr}.

In the following sections, the impact study on various helicity distributions at NLO taking advantage of the sets of SIDIS data at the EicC will be discussed in detail. 

\section{Description of the Pseudo-data}
\label{sec_data}
\begin{figure}[!t]
\begin{center}
\includegraphics[width=0.5\textwidth]{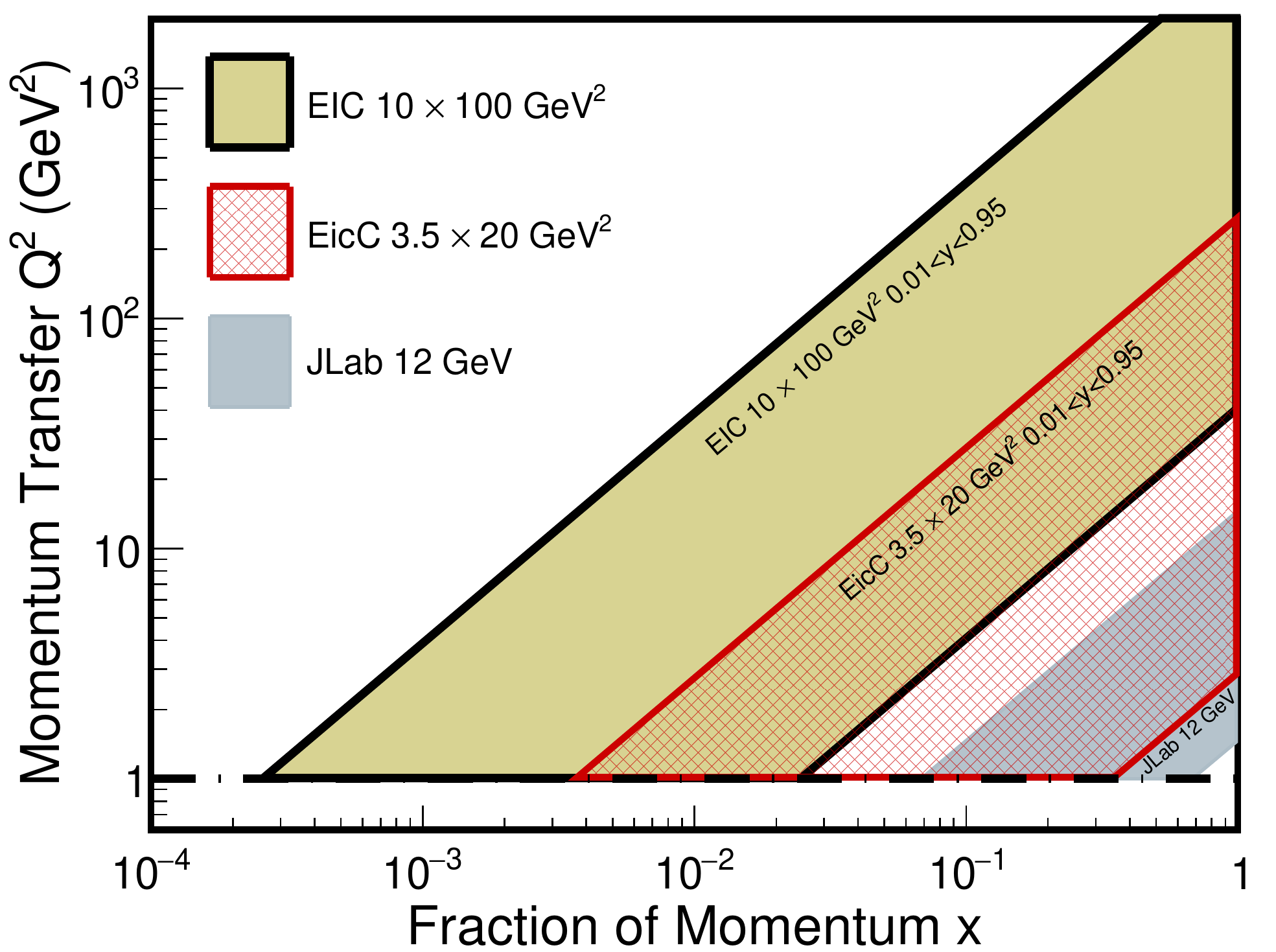}
\caption{\label{fig:Q2_VS_X}
Kinematic coverage of deep inelastic scattering process for the EicC and the US EIC (using 10 GeV $\times$ 100 GeV as an
example), as well as JLab-12 experiments.
}
\end{center}
\end{figure}

The EicC accelerator is optimized to provide a 3.5 GeV electron beam on a 20 GeV proton beam (40 GeV $^3{\rm He}$ beam). 
The pseudo-data were produced according to this design. 
The Q$^2$-$x$ coverage of the DIS process at the EicC, together with the coverage of an optional energy configuration at the US EIC and JLab-12 experiments,  are shown in Fig. \ref{fig:Q2_VS_X}. 
As mentioned above, the instantaneous luminosity at the EicC
is about $2 \times 10^{33} {\rm cm}^{-2} {\rm s}^{-1}$ per nucleon, which means that about 50~${\rm fb}^{-1}$ of integrated luminosity
will be accumulated with 10 months of running without considering beam delivery and detector efficiency.

The DJANGOH event generator ~\cite{CHARCHULA1994381} was employed to produce the pseudo-data. It has been widely used 
at HERA and then modified to accommodate the needs of the EIC community for 
various simulations. DJANGOH can simulate deep inelastic lepton-nucleon (nuclei) scattering
including both QED and QCD radiative effects. It is an interface of the Monte-Carlo
programs HERACLES ~\cite{KWIATKOWSKI1992155} and LEPTO ~\cite{INGELMAN1997108}. 
The HERACLES can treat the electron-proton scattering using either parametrized structure functions 
or PDFs in the framework of the quark-parton model. The 
LEPTO does the integration on electroweak cross-sections and, based on the cross-section,  
it simulates lepton-nucleon scattering with hadronic final states by using the 
JETSET library ~\cite{SJOSTRAND1987367}.

\begin{figure}[!t]
\begin{center}
\includegraphics[width=0.5\textwidth]{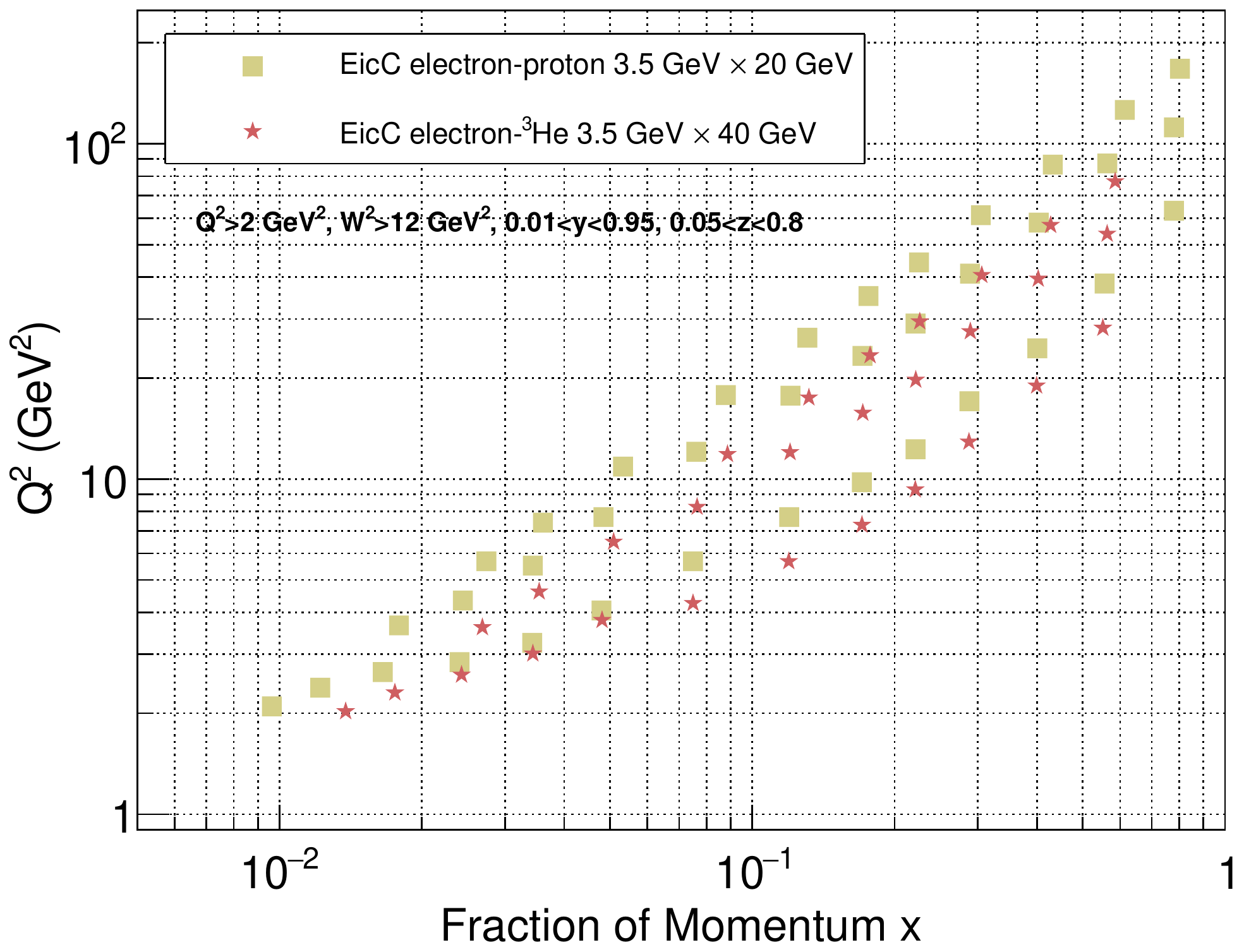}
\caption{\label{fig:Q2_VS_X_points}
The $Q^2$-x coverage of the pseudodata points for electron-proton and electron-$^3$He collisions
after kinematics cuts. In electron-proton (or electron-$^3$He) collision, there are 5 sets of data: inclusive DIS, $\pi^{\pm}$ and $K^{\pm}$ SIDIS.
}
\end{center}
\end{figure}

Once the pseudo-data were produced, the following cuts were applied: 
$Q^2 > 2~GeV^2$, $W^2 > 12~GeV^2$, $0.05<y<0.8$, and  $0.05<z<0.8$. Afterwards, the data were
binned in $x-Q^2$ two dimensions, as shown in Fig. \ref{fig:Q2_VS_X_points}. In each $x-Q^2$ bin, a log-likelihood was defined as
\begin{equation}
    L = \log\left( \prod_{events} \frac{yield(x, Q^2)}{\mathcal{N}}\right),
\end{equation}
where 
\begin{equation}
    yield = (1+\lambda P_e P_p D(y) A_1) \cdot \sigma_{0} \cdot \mathcal{L} \cdot \mathcal{A},
\end{equation}
and the normalization
\begin{equation}
\mathcal{N} = \int dx \int dQ^2 [1+ \lambda P_e P_p D(y) A_1] \cdot \sigma_{0} \cdot \mathcal{L} \cdot \mathcal{A}.
\end{equation} 
Here $\lambda$, with value $\pm1$, denotes the different spin combinations in Eq.~(\ref{eq_asy}), $\mathcal{A}$ is the detector acceptance, and $\mathcal{L}$ is the integrated luminosity. If the acceptance is the same for $\lambda = \pm 1$ states, the uncertainty for 
the $A_1$ measurement in a particular bin, after maximizing the log-likelihood, is given by
\begin{equation}
\sigma_{A_1} = \sqrt{ \frac{1}{\sum_{events} \lambda^2 P_e^2 P_p^2 D(y)^2} }.
\label{equation:A1_projection}
\end{equation}

For electron-proton collisions, the obtained uncertainty projection on $A_{1p}$, where $p$ denotes proton data, can be used for the following impact study. While for e-$^3{\rm He}$ collisions, after the uncertainty
projection on $A_{1^3He}$ is obtained, the dilution factor needs to be considered
in order to get $A_{1N}$ projection, where $N$ denotes neutron.

The ground state of $^3{\rm He}$ nuclear wave-function is dominated by the S-state, in which the proton spins cancel each other and the
nuclear spin is mostly carried by the neutron ~\cite{PhysRevC.65.064317}. Hence, $^3{\rm He}$ can be used as an effective 
polarized neutron source. Neutron asymmetries can be obtained from $^3{\rm He}$ asymmetries using the effective nucleon 
polarization ~\cite{PhysRevC.42.2310,PhysRevC.64.024004,Zhao:2015brn} by
\begin{equation}
    A_{1^3He} = P_n (1- f_p) A_{1N} + P_p f_p A_{1p},
\label{eq:he3_correction}
\end{equation}
with dilution factor $f_p = \frac{2\sigma_p}{\sigma_{^3He}}$, neutron effective polarization $P_n = 0.86^{+0.036}_{-0.02}$, and  proton 
effective polarization $P_p = -0.028^{+0.009}_{-0.004}$. The dilution factor was calculated bin by bin for individual SIDIS channels 
using a dedicated electron-proton simulation using DJANGOH with proton beam energy set to be 40/3 GeV in order to match the electron-$^3{\rm He}$ collisions.
To extract neutron results out of $^3$He data, various nuclear effects have to be taken into account, including spin depolarization, 
nuclear binding and Fermi motion of nucleons, the off-shellness of the nucleons, presence of non-nucleonic degrees
of freedom, and nuclear shadowing and anti-shadowing, etc.
Since we are merely interested in the uncertainty propagation of data in this paper, we shall apply Eq.~(\ref{eq:he3_correction}) as an effective method to extract neutron results
out of $^3$He data without including
the presence of non-nucleon degrees of freedom and nuclear shadowing and anti-shadowing effects, etc. 
Considering the precision of the data
expected in the future electron-ion colliders, one has to use the convolution approach, instead of the effective method, to include all 
other nuclear effects in order to obtain the correct central value of the neutron results \cite{PhysRevC.65.064317}. 
This would require dedicated theoretical and experimental effort in the EIC era.

In the following discussion, the data sample for electron-proton and electron-$^3{\rm He}$ (effective neutron) collisions are both assumed to be 50~${\rm fb}^{-1}$, respectively.
In this scenario, while discussing the impact of different data subsets, without mentioning proton or neutron data explicitly, for example ``EicC(50~${\rm fb}^{-1}$)DIS'' or ``EicC(50~${\rm fb}^{-1}$)SIDIS'', it means 100~${\rm fb}^{-1}$ of data (50~${\rm fb}^{-1}$ e-p + 50~${\rm fb}^{-1}$ e-$^3{\rm He}$ collisions). 

\section{Description of the impact study using ePump}
\label{sec_fitting}

In this section, we discuss how to quantitatively study the impact of new (pseudo) data on updating the existing PDF sets. 
After briefly reviewing various methods for this type of study, we discuss how to convert a Monte Carlo PDF set to a Hessian PDF set. 
Following that, the Hessian profiling method will be discussed and applied to the pseudo-data to carry out the impact study.  

The two commonly-used methods for extracting PDFs and their uncertainties from a global analysis of high-energy scattering data 
are the Monte Carlo method, used by NNPDF~\cite{Ball:2012cx}, and the Hessian method, used in CT14HERA2~\cite{Dulat:2015mca,Hou:2016nqm}, for example.
In the Monte Carlo method,
a statistical ensemble of PDF sets is provided, which are assumed to approximate the probability distribution of possible PDFs, as constrained from the global analysis of the data.
In the Hessian method, a smaller number of error PDF sets are provided along with the central set which minimizes the $\chi^2$-function in a global analysis. These error PDF sets correspond to the plus and minus eigenvector directions in the space of PDF parameters,
which are used to approximate the $\chi^2$-function near its global minimum.

An understanding of uncertainties due to PDFs is crucial to precision studies of the standard model, as well as to searches for new physics beyond the standard model at lepton-hadron and hadron-hadron colliders.  In turn, new measurements of standard model processes can be used to constrain the uncertainties of PDFs.  The most complete method for obtaining constraints from the new data on the PDFs would be to add the new data into the global analysis package and to do a full re-analysis on the PDFs.  However, this is impractical for most users of PDFs.
A technique for estimating the impact of new data on the PDFs, without performing a full global analysis, is extremely useful.
In the context of the Monte Carlo PDFs, the PDF reweighting method has become commonplace.  This involves applying a weight factor, which is dependent on the
new data and the theory predictions, to each of the PDFs in the ensemble~\cite{Giele:1998gw, Ball:2010gb,Ball:2011gg} when performing ensemble averages. Because the weight factor for some of the PDFs in the ensemble may be small, the effective number of PDFs in the ensemble is reduced. Therefore, the number of initial PDF replicas in the ensemble must be increased to get sufficient statistics in the reweighted averages. However, this may not be always sufficient to guarantee a successful outcome of the reweighting method. In the case of particularly significant data improvements the effective number of replicas surviving the reweighting procedure can drop down to a few dozens or less, and any statistical value of the reweighted replica set is therefore lost. For instance, this is what was observed in~\cite{Aschenauer:2012ve} when attempting to assess the impact on DSSV14 PDFs~\cite{deFlorian:2019zkl} of semi-inclusive deep inelastic scattering off helium at the future USA Electron-Ion Collider. In that specific case, they have observed the failure of the reweighting method and stressed the need for a new fit.  

To overcome this possible limitation, it is also possible to estimate the impact of new data directly using the so-called ``Hessian profiling''~\cite{Paukkunen:2014zia,Camarda:2015zba,Schmidt:2018hvu,Hou:2019gfw} method to update the existing Hessian PDF sets.
The advantage of this Hessian updating method over the Monte Carlo reweighting method is that it directly works with the (smaller set of) Hessian PDFs, and it is a simpler
and a much faster way to estimate the effects of the new data. 
This method directly calculates the minimum of the updated $\chi^2$ function within the Hessian approximation.
 
In this work, the software package {\tt ePump} (error PDF Updating Method
Package)~\cite{Schmidt:2018hvu,Hou:2019gfw}, which can update any given set of Hessian PDFs obtained from an earlier global analysis, was used. 
Although the DSSV14 PDFs were presented as Monte Carlo sets, we can apply the package {\tt MC2Hessian}~\cite{Carrazza:2015aoa} to produce an equivalent Hessian set, which will be named as DSSV14H in this work. The details of this conversion will be discussed below.  After that,
one is able to use the {\tt ePump} package to estimate the impact of new (pseudo) data on updating the DSSV14H Hessian PDFs.
A few similar studies, but for unpolarized PDFs, can be found in Refs.~\cite{Willis:2018yln,Fu:2020mxl,Deng:2020sol}.

\subsection{A Hessian representation for Monte Carlo PDFs}
\label{sec_mc2hessian}

In this subsection, a brief overview of the methodology employed by the {\tt MC2Hessian} package to generate a reliable Hessian representation from a PDF Monte Carlo (MC) replica set will be presented, followed by its application in our specific case for the DSSV14 PDF set. 

PDFs extracted using Monte Carlo methods are given in terms of an ensemble of functions, called ``replicas'', which form a discrete representation of the probability distribution describing the PDF functional space for a given set of experimental data. Although the probabilistic interpretation of PDFs' ``best-fit'' and ``uncertainties'' as the mean and standard deviation of the replica distribution is in this case straightforward, Monte Carlo methods usually do not require optimizing the end number of replicas describing a specific PDF set. As a consequence, PDFs are often given in terms of a large number of replicas which may be strongly correlated with each other. In such a case, it has been shown~\cite{Carrazza:2015hva} that it is possible to find an equivalent representation of the PDFs using a smaller subset of the original replicas. This also implies that, for a sufficiently large number of replicas $\{ f^{(k)}_\alpha
\}_{k=1,\ldots,N_{\rm rep}}$, where $\alpha=1,\ldots,N_{\rm pdf}$
runs over the type of quarks, antiquarks, and the gluon PDFs, one may be able to describe the Monte Carlo sample as a linear combination of a suitably chosen subset of replicas $\{
\eta^{(i)}_\alpha \}_{i=1,\ldots,N_{\rm eig}} \subset
\{f^{(k)}_\alpha\}$:

\begin{equation}
  f_\alpha^{(k)} \approx f^{(k)}_{H,\alpha} \equiv f^{(0)}_\alpha +
  \sum_{i=1}^{N_{\rm eig}} a_i^{(k)} (\eta^{(i)}_\alpha -
  f^{(0)}_\alpha)\, ,
  \label{eq:hessian}
\end{equation}
where $f^{(0)}_\alpha$ denotes the average value of the original replica set,
$a_i^{(k)}$ are constant coefficients, and
 $f^{(k)}_{H,\alpha}$ is the linear representation
of the original replica $f_\alpha^{(k)}$.

As one can see more in detail hereinafter, given such replica basis, it is possible to produce a Hessian representation of the original PDF set in the space of linear expansion coefficients $a_i^{(k)}$. However, the deviation of the Hessian representation from the original MC sample ends up being proportional to the deviation from the gaussianity of the starting probability distribution. This might be the case for specific kinematic regions (such as small-$x$ and large-$x$) where limited experimental data are available and the PDF uncertainties are determined mainly by theoretical constraints. Nonetheless, for most cases, where PDF uncertainties are driven by copious experimental data, gaussianity is a reasonable approximation and the described strategy turns out to be a reliable way to convert MC PDF sets into Hessian PDF sets. From a practical point of view, this is achieved by choosing the optimal $\{\eta^{(i)}_\alpha \}_{i=1,\ldots,N_{\rm eig}}$ and determining the parameters $\{a_i^{(k)}\}$.

In order to get the coefficients $\{a_i^{(k)}\}$, starting from the covariant matrix in the PDF functional space
\begin{equation}
\label{eq:covmat}
{\rm cov}^{\rm pdf}_{ij,\alpha\beta} \equiv \frac{N_{\rm rep}}{N_{\rm
    rep} -1 } \left( \la f^{(k)}_{\alpha}(x_i,Q^2_0)\cdot
f^{(k)}_{\beta}(x_j,Q^2_0)\ra_{\rm rep} -
\la f^{(k)}_{\alpha}(x_i,Q^2_0) \ra_{\rm rep}
\la f^{(k)}_{\beta}(x_j,Q^2_0)\ra_{\rm rep} \right) \, ,
\end{equation}
where the averages are calculated over the original $N_{\rm rep}$ replicas,
one can define a figure of merit
\begin{equation}
\label{eq:chi2def}
\chi^{2(k)}_{\rm pdf} \equiv \sum_{i,j=1}^{N_x}
\sum_{\alpha,\beta=1}^{N_{\rm f}}\Bigg( \lc
f_{H,\alpha}^{(k)}(x_i,Q^2_0) - f_{\alpha}^{(k)}(x_i,Q^2_0)\rc \cdot
\lp {\rm cov^{pdf}}\rp^{-1}_{ij,\alpha\beta} \cdot \lc
f_{H,\beta}^{(k)}(x_j,Q^2_0) - f_{\beta}^{(k)}(x_j,Q^2_0)\rc \Bigg) \,,
\end{equation}
where ${N_x}$ runs over the number of sampling of a discretized $x$-grid.

The coefficients $\{a_i^{(k)}\}$ are obtained by minimization of Eq.~(\ref{eq:chi2def}) using Singular Value Decomposition techniques. The Hessian representation of the original replica set is obtained by computing the following co-variant matrix,

\begin{eqnarray}
\label{eq:covmata}
   {\rm cov}^{(a)}_{ij} = \frac{N_{\rm rep}}{N_{\rm rep}-1} \lp \la   a_i^{(k)} a_j^{(k)}\ra_{\rm rep}  -
   \la   a_i^{(k)}\ra_{\rm rep}
   \la  a_j^{(k)}\ra_{\rm rep} \rp \, , \qquad
   i,j=1,\ldots,N_{\rm eig}\, .
\end{eqnarray}
 and calculating the Hessian matrix, defined as the diagonalized inverse matrix $\left({\rm cov}^{(a)}_{ij}\right)^{-1}$. If we define $v_{ij}$ to be the rotation matrix used to diagonalize $\left({\rm cov}^{(a)}_{ij}\right)^{-1}$ and $\lambda_i$ the set of obtained eigenvalues, the PDF uncertainties can be expressed as
  \begin{equation}
\sigma^{\rm PDF}_{H,\alpha}(x,Q^2) = \sqrt{ \sum_{i=1}^{N_{\rm
   eig}}\left[\sum_{j=1}^{N_{\rm
   eig}}\frac{v_{ij}}{\sqrt{\lambda_i}}\left(\eta^{(j)}_\alpha(x,Q^2)-f^{(0)}_\alpha(x,Q^2)
\right)\right]^2}\, ,
\label{eq:eigs}
\end{equation}
whereas the $N_{\rm eig}$ symmetric Hessian eigenvectors describing the original Monte Carlo sample are given by
\begin{equation}
\label{eq:eigenHessian}
\widetilde{f}_\alpha^{(i)}(x,Q^2)=f^{(0)}_\alpha(x,Q^2)+\sum_{j=1}^{N_{\rm
   eig}}\frac{v_{ij}}{\sqrt{\lambda_i}}\left(\eta^{(j)}_\alpha(x,Q^2)-f^{(0)}_\alpha(x,Q^2)
\right) \, .
\end{equation}
Using Eq.~(\ref{eq:eigenHessian}), Eq.~(\ref{eq:eigs}) yields
\begin{equation}
\label{eq:sigma}
\sigma^{\rm PDF}_{H,\alpha}(x,Q^2) = \sqrt{ \sum_{i=1}^{N_{\rm eig}}
\lp \widetilde{f}_{\alpha}^{(i)}(x,Q^2) -  f_{\alpha}^{(0)}(x,Q^2)   \rp^2} \, .
\end{equation}
A successful determination of $\{a_i^{(k)}\}$ implies that Eq~(\ref{eq:sigma}) should yield similar results as the one-sigma PDF of the Monte Carlo representation defined as
\begin{equation}
\label{eq:sigmaMC}
\sigma^{\rm PDF}_{\alpha}(x,Q^2) = \sqrt{  \la
  \lp {f}^{(k)}_{\alpha}(x,Q^2)\rp^2\ra_{\rm rep} -  \la
  {f}^{(k)}_{\alpha}(x,Q^2)\ra_{\rm rep}^2 } \, .
\end{equation}

To determine the optimal set of $\{\eta^{(i)}_\alpha \}_{i=1,\ldots,N_{\rm eig}}$, {\tt MC2Hessian} utilizes a Genetic Algorithm (GA) to optimize an ``estimator'' defined as

\begin{equation}
  \label{eq:estimator}
  \textrm{ERF}_{\sigma} = \sum_{i=1}^{N_x} \sum_{\alpha=1}^{N_{\rm pdf}} \abs{ \frac{\sigma^{\rm PDF}_{H,\alpha}(x_i,Q^2_0)
      - \sigma^{\rm PDF}_{\alpha}(x_i,Q^2_0)}{\sigma^{\rm PDF}_{\alpha}(x_i,Q^2_0)} } \,,
\end{equation}
for a fixed value of $N_{\rm eig}$. A detailed discussion of the specific GA is beyond the scope of this paper and we refer the reader to the original work~\cite{Carrazza:2015aoa}.

The package has an additional optimization parameter defined as 
\begin{equation}
  \label{eq:epsestimator}
  \epsilon_\alpha(x_i,Q^2_0) = \frac{\left|\sigma_{\alpha}(x_i,Q^2_0)
      - \sigma^{68}_{\alpha}(x_i,Q^2_0)\right|}{\sigma^{68}_{\alpha}(x_i,Q^2_0)}  \, ,
\end{equation}
where $\sigma_{\alpha}(x_i,Q^2_0)$ and $\sigma_{\alpha}^{68}(x_i,Q^2_0)$
are respectively the one-sigma and 68\% confidence level intervals
for the $\alpha$-th PDF
which allows to discard points on the $x$-grid for which the gaussian approximation deviates more than a threshold value $\epsilon$, i.e.  $\epsilon_\alpha(x_i,Q^2_0)< \epsilon$.

For the purpose of this paper, the {\tt MC2Hessian} has been applied to the DSSV14 PDF set~\cite{deFlorian:2019zkl}, which, from its original analysis, is given in the form of $N_\text{rep}=1000$ Monte Carlo replicas. 
Among the consistency checks of the PDF Monte Carlo extraction, the DSSV collaboration has performed a comparison between the provided Monte Carlo sample and a version of the same analysis with error bands produced using the Lagrange Multiplier procedure. Similar to the Monte Carlo procedure, this method allows dropping the requirement of a linearized error analysis,  typical of the Hessian representation. However, uncertainties bands are defined in terms of a tolerated increase in the $\chi^2$, denoted ``tolerance'' $\Delta \chi^2$. For normal (Gaussian) errors a 68\% Confidence Level (CL) band would correspond to a tolerance $\Delta \chi^2=1$. In the context of PDF fits, a deviation from this standard textbook value is usually employed to cope with neglected uncertainties which cannot be quantified and included in the analysis, such as possible tensions among data sets included in a global analysis. It is interesting to notice that in their comparison, they achieve a good agreement between the two extracted sets by setting $\Delta\chi^2\sim 10-15$. As the resulting one-sigma variance from the Monte Carlo replica method has a solid probabilistic interpretation, the corresponding comparable error band for the Lagrange Multiplier method with $\Delta\chi^2\sim 10-15$ shares the same interpretation. 

\begin{figure}[!t]
\begin{center}
\includegraphics[width=0.4\textwidth]{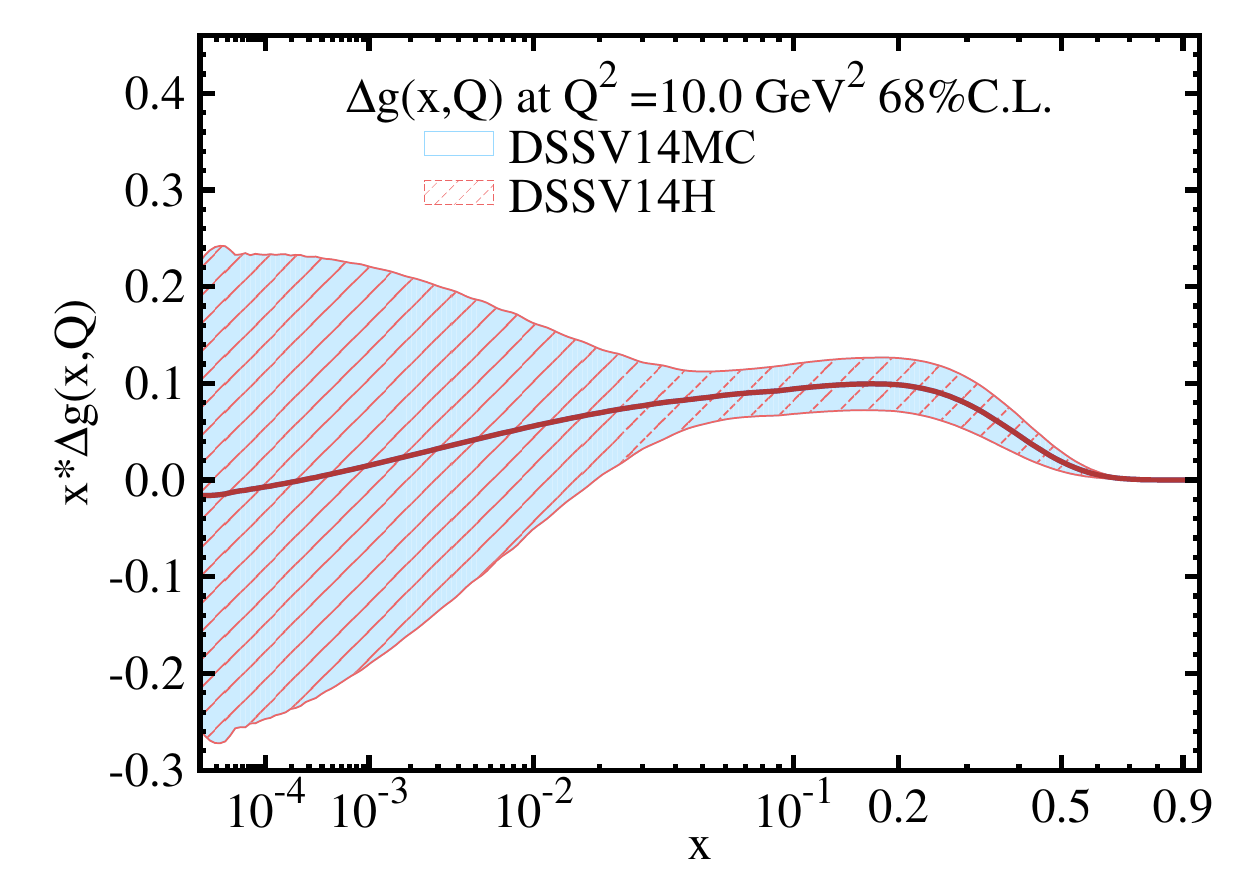}
\includegraphics[width=0.4\textwidth]{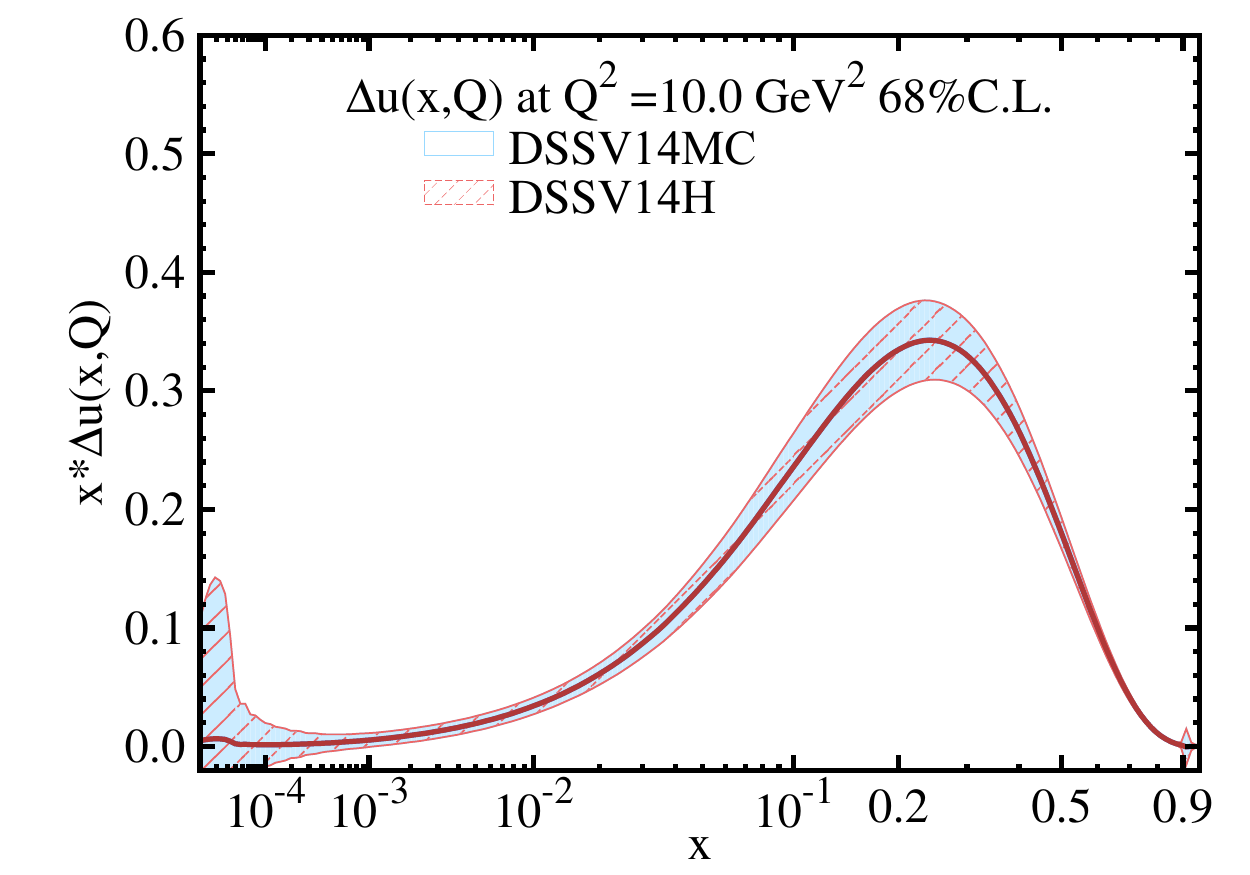}
\includegraphics[width=0.4\textwidth]{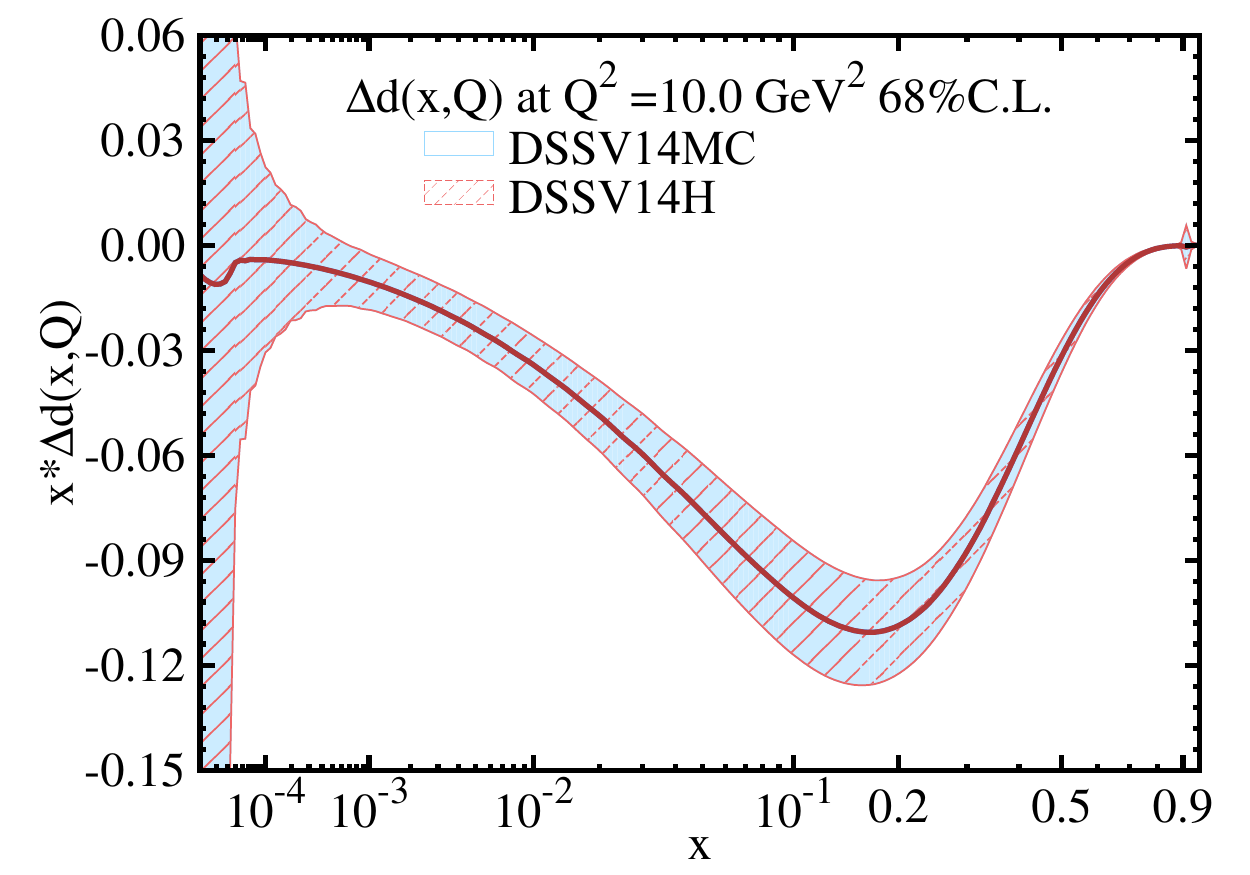}
\includegraphics[width=0.4\textwidth]{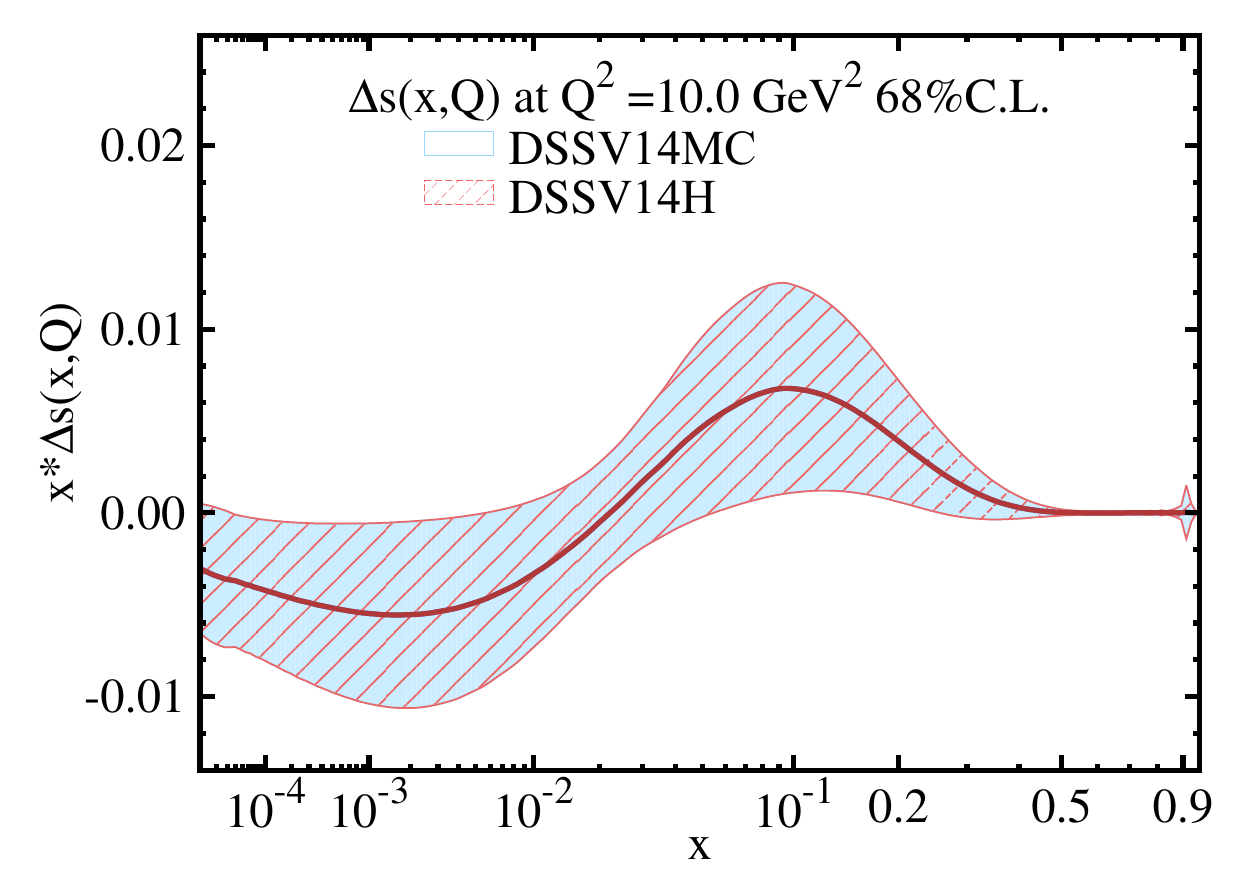}
\caption{\label{fig:consistent}
PDFs plots of DSSV14 representative distributions in different formats. The DSSV14MC is the original DSSV14 in form of 1000 Monte Carlo replicas; the DSSV14H is the DSSV14 in form of symmetric Hessian sets with 52 eigenvector pairs. The PDFs uncertainties of DSSV14 in different forms agree with each other. 
}
\end{center}
\end{figure}

Following the logic of the methodology described above, this may suggest that the deviation from the Gaussian distribution could end up spoiling our attempt to use {\tt MC2Hessian} to convert DSSV14 into a Hessian representation.
However, we have performed the conversion for different values of $N_{\rm eig}$ and $\epsilon$ at $Q_0^2=1 \text{GeV}^2$, and have found great agreement in the range $0.001<x<0.9$, useful to our analysis, for a value of $N_{\rm eig}=52$. 
This leads to a total of 104 error PDF sets, with symmetric error in both positive and negative eigenvector directions away from the central PDF set. Below, we refer to this set of PDFs as the DSSV14H Hessian PDFs. 
Moreover, no limitation on the parameter $\epsilon$ was found, which means no $x$-grid points are eliminated during the conversion according to the gaussianity condition $\epsilon_\alpha(x_i,Q^2_0)< \epsilon$. Fig.~\ref{fig:consistent} shows a direct comparison between the two representations; one is the original DSSV14 MC replica set, and another is the derived DSSV14H Hessian set. 
The difference between representations is below the per-mil level which is more than sufficient for the scope of this work. Hereafter, we shall only consider DSSV14H Hessian  PDFs, which will also be referred as DSSV14 interchangeably.

\subsection{A brief review of the Hessian profiling method}
\label{sec:hpm}

The error PDFs and the PDF-induced uncertainty of theoretical calculations can be effectively calculated by using the Hessian method. In order to find the Hessian eigenvector pairs of PDF sets ({\it i.e.}, error PDFs) after the inclusion of new data sets, 
a full global fit is needed so that one can evaluate variations of PDFs around the best-fit with respect to PDF parameters. Since a full global fit might be complicated and time-consuming, one may desire a faster and simpler approach to estimate the impact of a new data set. Paukkunen and Zurita~\cite{Paukkunen:2014zia} introduced a method that utilizes the Hessian eigenvector set to study the impact of the new data input. This method has been implemented in the software package \texttt{ePump} ~\cite{Schmidt:2018hvu}, which we used in this study. In this subsection, we briefly review some details of this method.

Consider a new data set for the measurement of the observable $X$ with $N_{pt}$ data points. Let's denote its experimental values as $X_i^E$, the inverse of its covariance matrix for the correlated experimental errors as $C_{ij}^{-1}$, and the corresponding theoretical prediction values $X_i^T$, which depend on $N$ diagonalized PDF parameters $\{z^{\pm}_r\} = \{0, \dots, z_r = \pm 1, \dots, 0 \} $. With the inclusion of this new data set, the variation of the total $\chi^2$ becomes,

\begin{equation}
 \Delta \chi^2_{\text{new}} = \chi^2_{old} + \sum_{j,k=1}^{N_{pt}} \Big{(}X_j^E - X_j^T(\textbf{z})\Big{)} C_{jk}^{-1} \Big{(}X_k^E - X_k^T(\textbf{z})\Big{)},
\end{equation}
where $\chi^2_{old} =  T^2 \sum_{r=1}^N z_r^2$ and $T$ is the tolerance parameter. In practice, the deviation in a particular eigenvalue direction $z_r$ is limited by the Dynamical Tolerance, $T^{\pm}_r$, at the given confidence level. So the parameters take the values $z^{\pm}_r = \pm T^{\pm}_r/T$. But for simplicity, we will ignore the Dynamical Tolerance dependence in the discussion.  For more detail, refer to Refs.~\cite{Schmidt:2018hvu,Hou:2019gfw}.

The new total $\chi^2$ can be rewritten by expanding the $X_j^T(\textbf{z})$ up to the linear term,

\begin{equation}
\label{eq:UpdateNewchi2}
 \Delta \chi^2_{\text{new}} = \sum_{j,k=1}^{N_{pt}} \Big{(}X_j^E - X_j^{T,0}\Big{)} C_{jk}^{-1} \Big{(}X_k^E - X_k^{T,0}\Big{)} + T^2 \bigg{(} \sum_{m=1}^N z^2_m + \sum_{m,n =1}^N z_m M^{mn} z_n - 2 \sum_{m=1}^N A^m z_m  \bigg{)},
\end{equation}
where we define a vector $A$ and a matrix $M$ such that,

\begin{eqnarray}
 M^{mn} &=& \sum_{j,k=1}^{N_{pt}} X_j^{T,m} C_{jk}^{-1} X_j^{T,n}, \\
 A^m &=& \frac{1}{T} \sum_{j,k=1}^{N_{pt}} \Big{(}X_j^E - X_j^{T,0}\Big{)} C_{jk}^{-1} X_k^{T,m}.
\end{eqnarray}
With new information of PDFs being introduced by the new data set, the PDFs parameters are then driven to the updated values. By minimizing Eq. (\ref{eq:UpdateNewchi2}), the updated central values of parameters can be found,

\begin{equation}
\label{eq:z0_update}
 z^0_n \xrightarrow{\text{update}} z^0_n = \sum_{m=1}^N \big{(} I + M \big{)}^{-1}_{mn} A^m.
\end{equation}
Then, the $r$-th new eigenvector and its corresponding eigenvalue are

\begin{eqnarray}
 \sum_{n=1}^N M^{mn}U^{r}_n &=& \lambda^{r} U^{r}_m, \\
 \sum_{m=1}^N U^{r}_m U^{s}_m &=& \delta^{rs}.
\end{eqnarray}
With the inclusion of new information, the Hessian eigenvalue directions are also updated from the set of old bases $\textbf{z}$ to a new set of bases $\textbf{z}'$. By diagonalizing the quadratic terms of the Eq. (\ref{eq:UpdateNewchi2}), the new parameters are

\begin{equation}
\label{eq:update_new_z}
 z'_r = \sqrt{1+\lambda_r} \sum_{m=1}^N U^r_m z_m.
\end{equation}
Consider another observable $Y=Y(z)$ whose PDF-induced uncertainty is constructed by the Hessian eigenvector sets. After the inclusion of the new data set $X_i$, the central value of $Y(z)$ is updated to
\begin{equation}
\label{eq:UpdateCenVal}
 Y_{\text{new}}^0 = Y_{\text{old}}^0 + \sum_{s=1}^{N} z^0_{s} \Delta Y^s,
\end{equation}
where $\Delta Y^s = (Y^{s,+}-Y^{s,-})/2$. The extreme values of $Y$ for the new $r$-th eigenvector can be calculated in a similar manner as in Eq.~(\ref{eq:update_new_z}),

\begin{equation}
\label{eq:UpdateErrObs}
 Y_{\text{new}}^{\pm r} = Y_{\text{new}}^0 + \frac{1}{\sqrt{1+\lambda^{r}}} \sum_{s=1}^N U_s^{r} \big{(} Y^{s, \pm}- Y^0 \big{)}.
\end{equation}
Notice that we can of course choose the $Y$ to be the PDFs $f(x,Q^0)$. In this case, Eq. (\ref{eq:UpdateCenVal}) and Eq.  (\ref{eq:UpdateErrObs}) stand for the updated central value and the error set of PDFs.
This procedure is referred as ePump-updating, and is implemented in the ePump package. 

To quantitatively summarize in a single value the change in the best-fit PDFs after the new data has been added to the global fit, the measure $d^0$ was  introduced in Ref.~\cite{Schmidt:2018hvu} as 
\begin{eqnarray}
	(d^0)^2&=&\sum_{r=1}^N\left(\frac{T}{T_r}z^0_r\right)^2 \, ,
	\label{eq:distance1}
\end{eqnarray}
where, again, the dynamical tolerance $T_r$
limits the deviation in the particular ($r$-th) eigenvector direction.
To be precise, $d^0$ is the length of the shift of the best-fit point in parameter space, relative to the 90\% confidence level (C.L.) boundary of the original PDFs.  Thus, $d^0=1$ means that the new best-fit touches the 90\% C.L. boundary, while a value of $d^0\ll 1$ implies a very small change to the best-fit PDFs. One should note that $d^0$ only reflects the change in the best-fit PDFs, so that it is still possible for the new data to produce a significant reduction in the PDF error bands, even if $d^0$ is small.
A value $d^0 > 1 $ would indicate
that either there is tension between the new data and the original data, or else the uncertainties in the original global analysis were
under-estimated~\cite{Schmidt:2018hvu}.
	
\subsection{A brief review of the data set rediagonalization}
\label{sec:hmo}

In the previous subsection, it has been discussed in detail that the PDF uncertainty can be effectively expressed in terms of a Hessian set, and that the impact of the measurement for a new observable is also easily accessible with the updating method. 
These methods are frequently used in the analyses of the PDF uncertainty with respect to experimental errors or kinematical cuts. Although they are straightforward conceptually and well-applicable numerically, repeated exercises for a Hessian set with a large number of error PDFs would still be time-consuming. 
It would be more convenient if one can reproduce the majority of the PDF dependence for given observables with a reduced Hessian set, so that it is not necessary to repeatedly evaluate all of the error PDFs, but a smaller number of them. The members of this optimized Hessian set are chosen in such a way that the combination of them recovers the PDF uncertainty for the observables to any desired precision.

The idea of this optimization method is based on the data set diagonalization procedure by Pumplin~\cite{Pumplin:2009nm}. Noting that the representation of the diagonalized parameters $\textbf{z}$ is not unique, one could take the advantage of this freedom to rotate the diagonalized parameters $\textbf{z}$ into a new set of parameters $\textbf{z}'$ where the PDF sensitivity for a given data set is maximized on a certain direction. Note that the given data set may or may not be included in the original global fit, and that the optimized eigenvectors after the rotation contain exactly the same information as the original eigenvectors do.

Our goal is to find a direction on which the variation for a set of observables $X_{i}(\textbf{z})$, where $i$ runs from 1 to $N_{pt}$, from its best-fit values $X_{i}^0 = X_{i}(\textbf{0})$ is maximized. Therefore, we define the following function,

\begin{equation}
\label{eq:OptFunc}
 S(\textbf{z}, \lambda) = \sum_{i=1}^{N_{pt}} \frac{1}{T^2} \bigg{(} X_{i}(\textbf{z}) - X_{i}^0 \bigg{)}^2 - \lambda \bigg{(} \frac{\Delta \chi^2(\textbf{z})}{T^2} - 1 \bigg{)},
\end{equation}
where $\lambda$ is the Lagrange multiplier. To simplify the expression, let's again take the usual approximation and expand $X_{i}(\textbf{z})$ up to the linear terms,

\begin{equation}
\label{eq:OptEq1}
 S(\textbf{z}, \lambda) = \sum_{r,s=1}^N z_r M_{rs} z_s - \lambda \bigg{(} \sum_{i=1}^N z^2_r -1 \bigg{)},
\end{equation}
where

\begin{equation}
 M_{rs} = \sum_{i = 1}^{N_{pt}} X^r_{i} X^s_{i},
\end{equation}
and as defined before the $X^r_{i} = (X^{r,+}_i - X^{r,-}_i)/2$. The matrix $M_{ij}$ is normalized in such a way that $\text{Tr}\{M\} = N_{pt}$. The extreme values of $S(\textbf{z}, \lambda)$ appear when the equality $\partial S/\partial z_r = 0$ holds for all Hessian indices $r$. Therefore the optimized parameters can be found by solving the eigenequation,

\begin{equation}
 \big{(} M_{rs} - \lambda \delta_{rs} \big{)} z_s = 0.
\end{equation}
If we rediagonalize the matrix $M$ by

\begin{eqnarray}
 \sum_{r=1}^N M_{rs}U^t_r &=& \lambda^t U^t_s, \\
 \sum_{t=1}^N U^t_r U^t_s &=& \delta_{rs},
\end{eqnarray}
we can find a new set of parameter $\textbf{z}'$ in the rediagonalized space,

\begin{equation}
z'_r = \sum_{s=1}^N \sqrt{\lambda^r} U^r_s z_s,
\end{equation}
The rediagonalized error PDF is calculated as usual,

\begin{equation}
 f^{\pm r} = f(\textbf{0}) + \sum_{s=1}^N U_s^r \bigg{(} f({z'_s}^{\pm}) - f(\textbf{0}) \bigg{)}.
\end{equation}

For the old parameters $\textbf{z}$, we do not discriminate among $N$ Hessian eigenvalue directions, since when diagonalizing the Hessian matrix, we have already normalized the orthogonal bases by their corresponding eigenvalues. But for the rediagonalized parameters $\textbf{z}'$, each of new eigenvalue directions $z'_r$ is associated with its corresponding eigenvalue $\lambda^r$ of the error matrix $M$ for the interested observable $X$.
The eigenvalue $\lambda^r$ provides useful information that reflects how much this rediagonalized direction $z'_r$ is sensitive to the given observable $X$. Since the matrix $M$ is normalized in a way that $\text{Tr}\{M\} = N_{pt}$, we have $\sum_{r} \lambda^r = N_{pt}$. Therefore, one is able to quantify how many data points of the set $X_i$ are particularly constraining a specific direction $z'_r$. If a rediagonalized error PDF has a small eigenvalue $\lambda^r$, one can draw a conclusion that no data points in the set $X_i$ are sensitive to this rediagonalized direction $z'_r$. Thus, one can ignore the error PDFs in this direction without a significant loss of accuracy. 
This procedure of the data set rediagonalization is referred as the ePump-optimization procedure, and is implemented in the ePump package.


\subsection{Updating the PDFs with pseudo data}
\label{sec_tables}

In our analysis, the quantitative study to update the DSSV14H Hessian PDFs with the EicC new (pseudo) data is done 
by using the above detailed Hessian updating method, via {\tt ePump}.
In order to perform the analysis, \texttt{ePump} requires two sets of inputs: data templates and theory templates. The data templates consist of the new experimental data values and
their statistical and systematic uncertainties, including correlations, exactly as what would be included in a standard global analysis.
The theory templates consist of the corresponding theory predictions for the same observables, evaluated using the central PDF 
and each of the DSSV14H Hessian eigenvector PDFs.  Note that any number of new data sets can be included in the update by \texttt{ePump}.
The output of \texttt{ePump} consists of an updated central and Hessian eigenvectors PDFs, which
approximate the result that would be obtained from a full global re-analysis that includes the new data.  
As an additional benefit, \texttt{ePump} can also directly output the updated predictions and uncertainties for any other observables 
of interest (such as the cross-section in the signal region), without the necessity to recalculate them using the updated PDFs.
More details about the use of \texttt{ePump} can be found in  Refs.~\cite{Schmidt:2018hvu,Hou:2019gfw}.

The theory predictions have been generated according to the NLO formulae discussed in Section~\ref{sec_DIS_SIDIS} expressed in the standard $\overline{MS}$ factorization scheme. To include data from both  proton and neutron targets in the analysis, we shall apply SU(2) proton-neutron isospin symmetry and impose $\Delta u^{\;\text{neutron}}=\Delta d^{\;\text{proton}}$, $\Delta \bar u^{\;\text{neutron}}=\Delta \bar d^{\;\text{proton}}$, $\Delta d^{\;\text{neutron}}=\Delta u^{\;\text{proton}}$ and $\Delta \bar d^{\;\text{neutron}}=\Delta \bar u^{\;\text{proton}}$. We have also set all renormalization and factorization scales $\mu^2=Q^2$. Factorization scale and scheme dependence investigations are well beyond the scope of this analysis. However, it may be a very interesting factor to include in the error analysis of future global extractions of PDFs, once the real EicC data will become available. 
Another source of theoretical errors that we defer to future studies is the fragmentation function uncertainties. 
In the case of parton-to-pion fragmentation functions, it has already been shown ~\cite{deFlorian:2009vb} that the inclusion of such systematic errors produces effects of at most a few percent level. However, due to the lower rate of production of kaons in respect to pions, and a consequent lower precision of kaon SIDIS data, uncertainties for parton-to-kaon fragmentation functions are generally larger compared to the respective pion ones (see for example ~\cite{deFlorian:2017lwf} and references therein). Analyses such as~\cite{Aschenauer:2019kzf} suggest that future electron-ion collider SIDIS data will have a remarkable impact on the kaon fragmentation functions. On the other hand, it is also argued that in order to use the future rich information of the high precision SIDIS data to extract parton and fragmentation distribution functions with reliable uncertainties, a simultaneous fitting of PDFs and FFs is needed in order to disentangle the highly correlated set of parameters describing them~\cite{Borsa_2017,Ethier_2017,Moffat:2021dji}.
In fact, traditional methods of global fitting of FFs using SIDIS data fix a specific PDF set as a baseline and account for their error by propagating them into the FFs themselves. This introduces a non-trivial correlated double-counting effect when such FFs are used to extract PDFs and their uncertainties.
For example, in~\cite{Borsa_2017} when performing a reweighting of PDFs using SIDIS data, it has been shown how explicitly choosing to include (or not to include) current kaon FF uncertainties~\cite{deFlorian:2017lwf} extracted with traditional methods results in an under (or over) estimation of the PDF uncertainties. 
Solving this conundrum will definitely be a task of the more sophisticated simultaneous PDFs and FFs fitting machinery once real SIDIS data with precision comparable to the inclusive case are available.
For the time being and for the scope of this article we want to concentrate on the effect of EicC future data solely on helicity PDFs. Hence, we choose not to include FFs uncertainties in our analysis and to use only the central ``best-fit'' values of DSS~\cite{deFlorian:2007ekg}.
Our results may be biased by this assumption and have to be taken as the ``best possible outcome'' in the future case where pion and, in particular, kaon FFs will be known at very high precision.

The data templates have been constructed using the uncertainty calculations of the pseudo-data according to Sec.~\ref{sec_data}, cf. Eq.~(\ref{equation:A1_projection}).
However, the fit using \texttt{ePump} requires a central value of the experimental observable as well. 
The central value of the asymmetry A$_1$, cf. ~Eq.(\ref{eq:Aone}), for each data point, was  
taken from the theory tables after a smearing procedure with a Gaussian distribution centered at 0 and a standard deviation equal to the estimated A$_1$ uncertainty of the pseudo-data. This ensures a reasonable estimation of the central value of A$_1$ while not affecting the $\chi^2$ artificially during the fit. 

Lastly, the tolerance value for the {\tt ePump} updating has been set to be $\Delta\chi^2=10$, which is of the same order of magnitude as the tolerance used in the DSSV14 analysis when studying the uncertainties via means of the Lagrange multiplier's method.

In total, ten pairs of theory-data templates have been prepared for this analysis: two for DIS process, one for electron-proton collision and one for electron-neutron collision, and eight for SIDIS process which corresponds to each combination of the two possible nucleon targets (proton or neutron) and the four observed final state hadrons ($\pi^\pm$ or $K^\pm$). In the original DSSV14 analysis they also included data with charged hadrons as the final state. For the purpose of this study, we concentrate only on the final states of known flavour content as this is very helpful to investigate the impact of specific data sets on the flavour content of the proton (neutron) target.

\subsection{Updated PDFs and their moments}
\label{sec_discussion}

\begin{figure}[t]
\begin{center}
\includegraphics[width=0.48\textwidth]{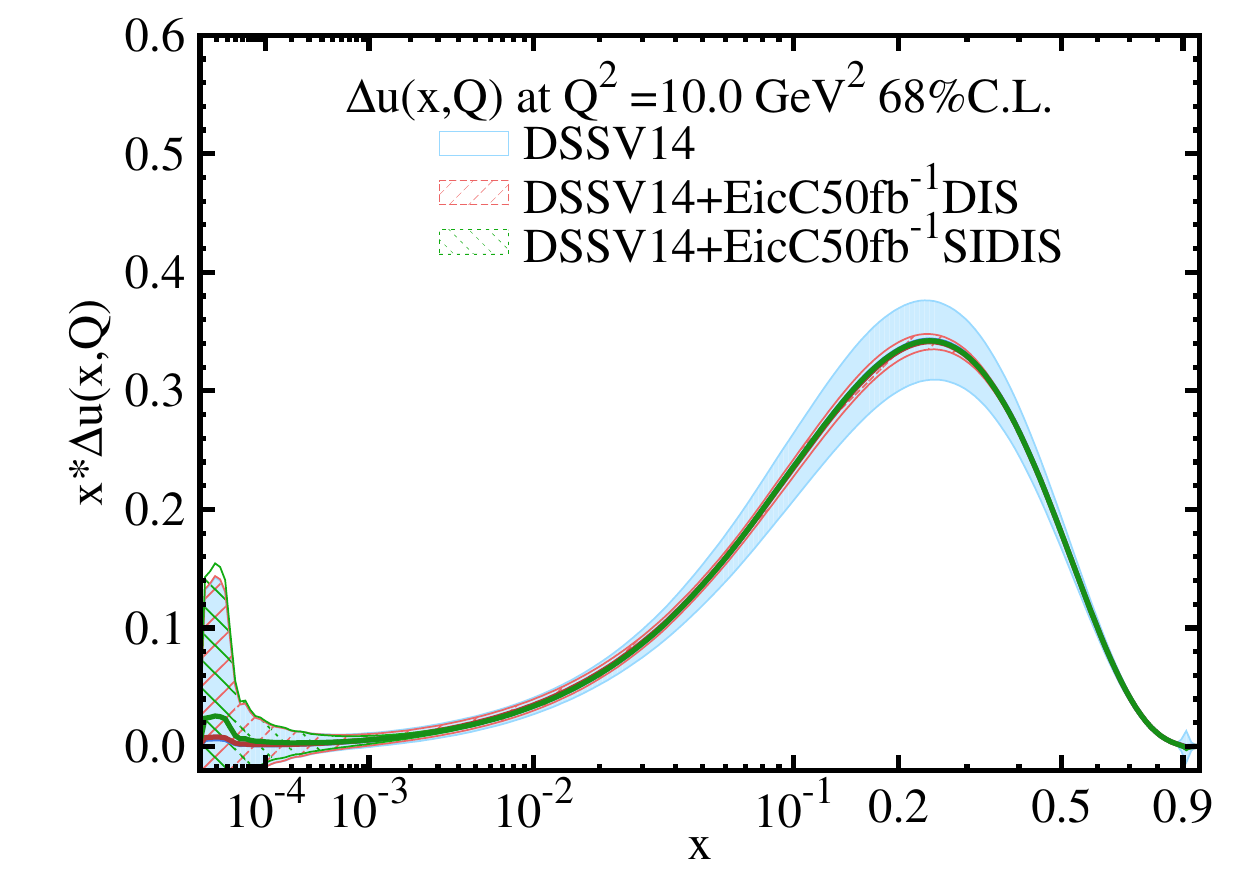}
\includegraphics[width=0.48\textwidth]{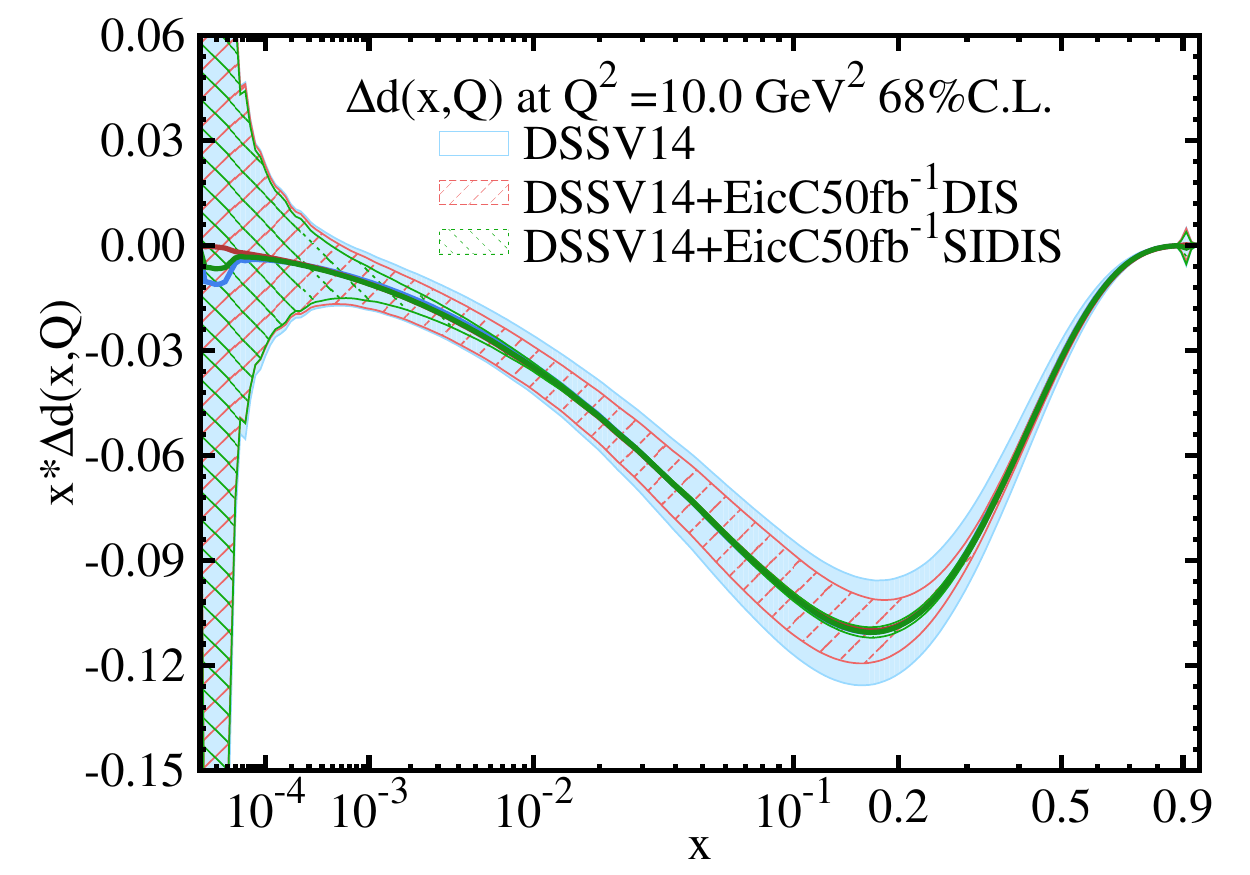}
\includegraphics[width=0.48\textwidth]{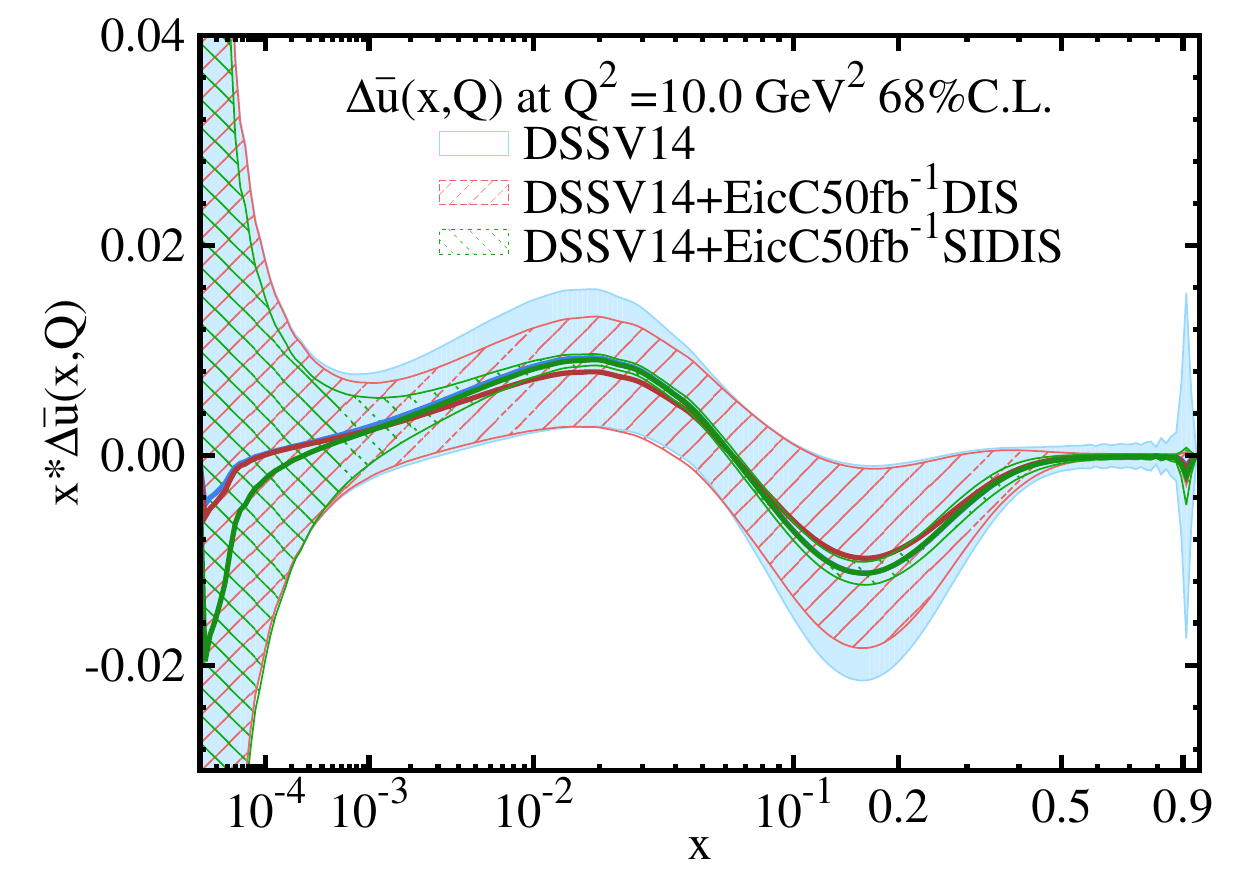}
\includegraphics[width=0.48\textwidth]{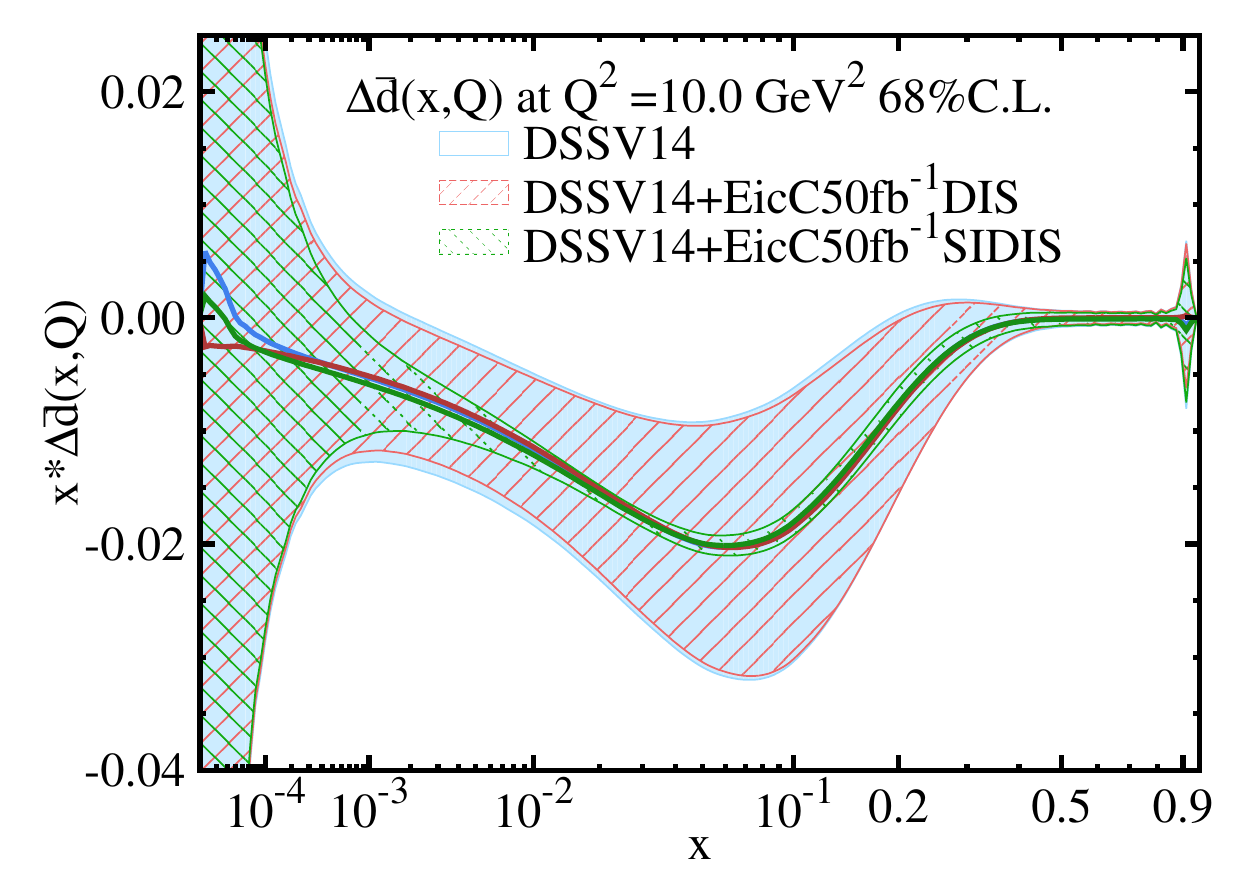}
\includegraphics[width=0.48\textwidth]{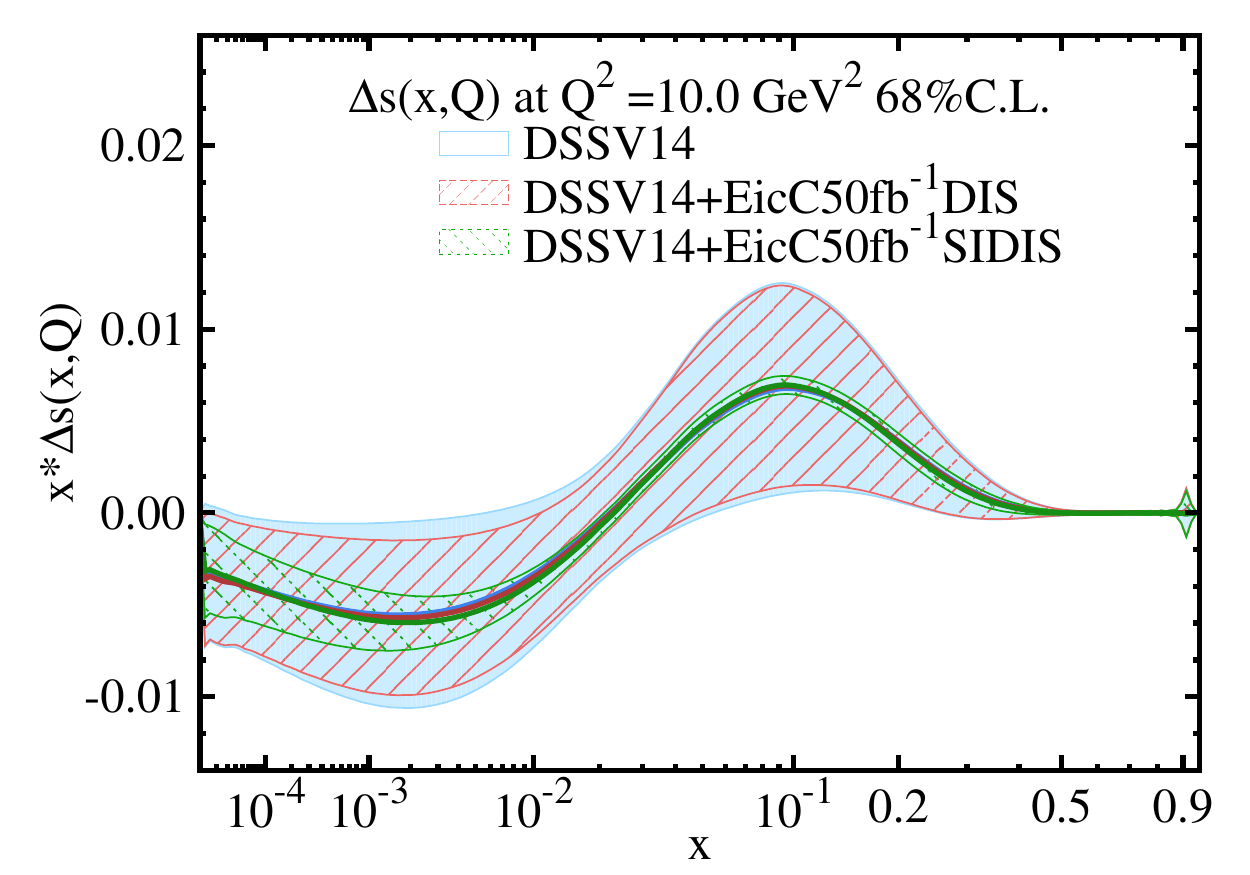}
\includegraphics[width=0.48\textwidth]{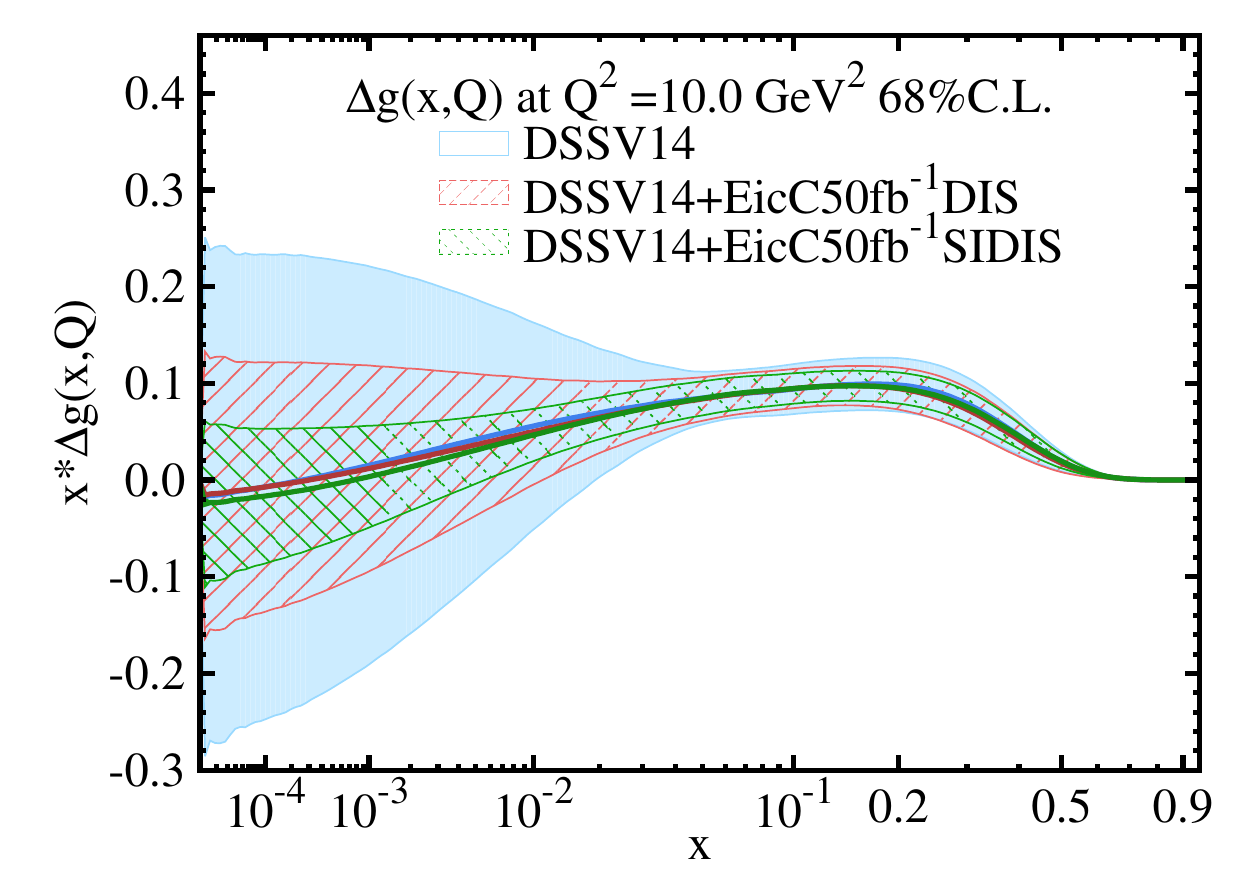}
\caption{\label{fig:updated_dist}
Results on the uncertainty band of polarized quark and gluon distributions after a next-to-leading
order fit by including EicC pseudo-data. The light blue band represents the original DSSV14 global fit.
The red (green) band shows the results by adding EicC DIS (SIDIS) pseudo-data.
}
\end{center}
\end{figure}

In the following, the results of the {\tt ePump} updating procedure will be presented. All the results are presented 
at $Q^2=10$ GeV$^2$ with uncertainties given at 68\% CL.    

\begin{figure}[p]
\begin{center}
\subfigure[DIS pseudo data]{\label{fig:PDFup_DIS-neu-pro}\includegraphics[width=0.42\textwidth]{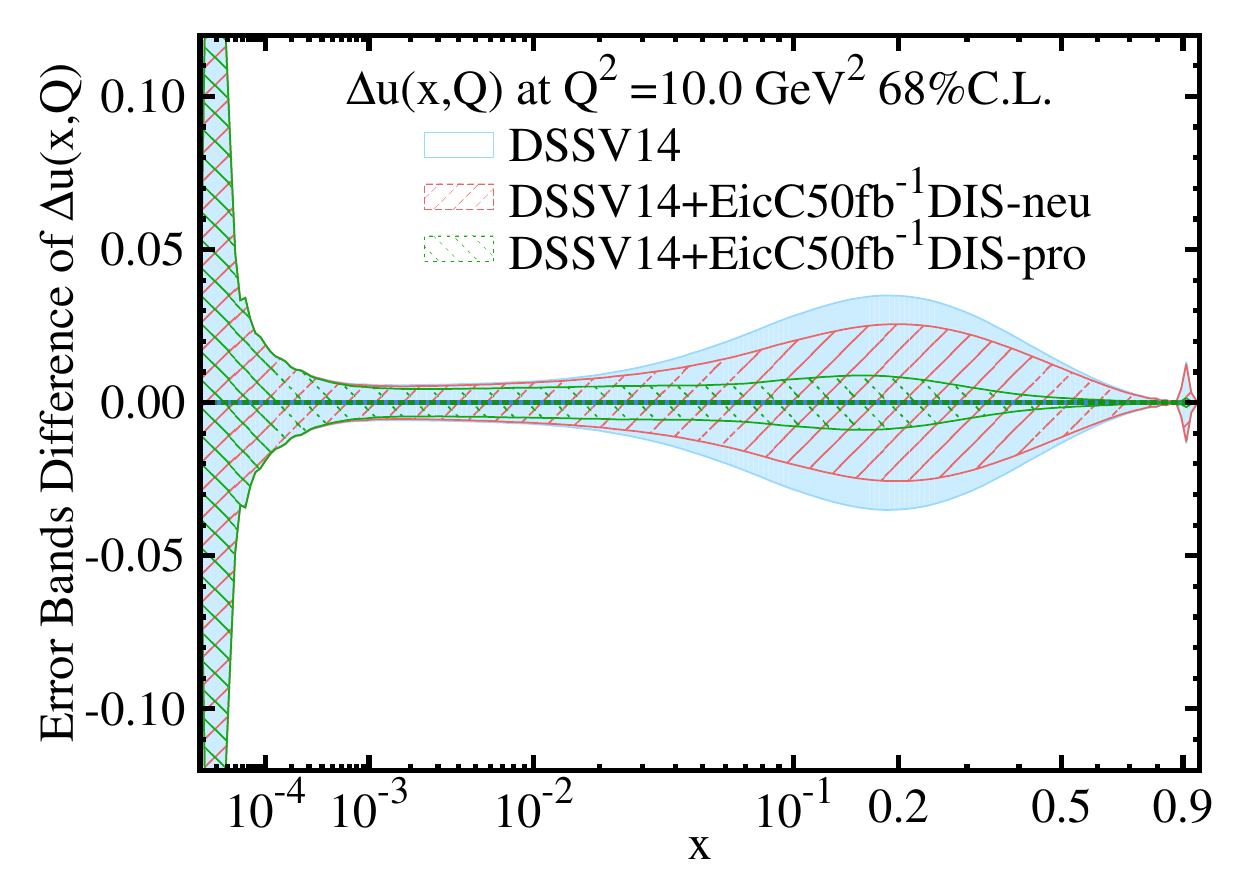}
}\\
\includegraphics[width=0.42\textwidth]{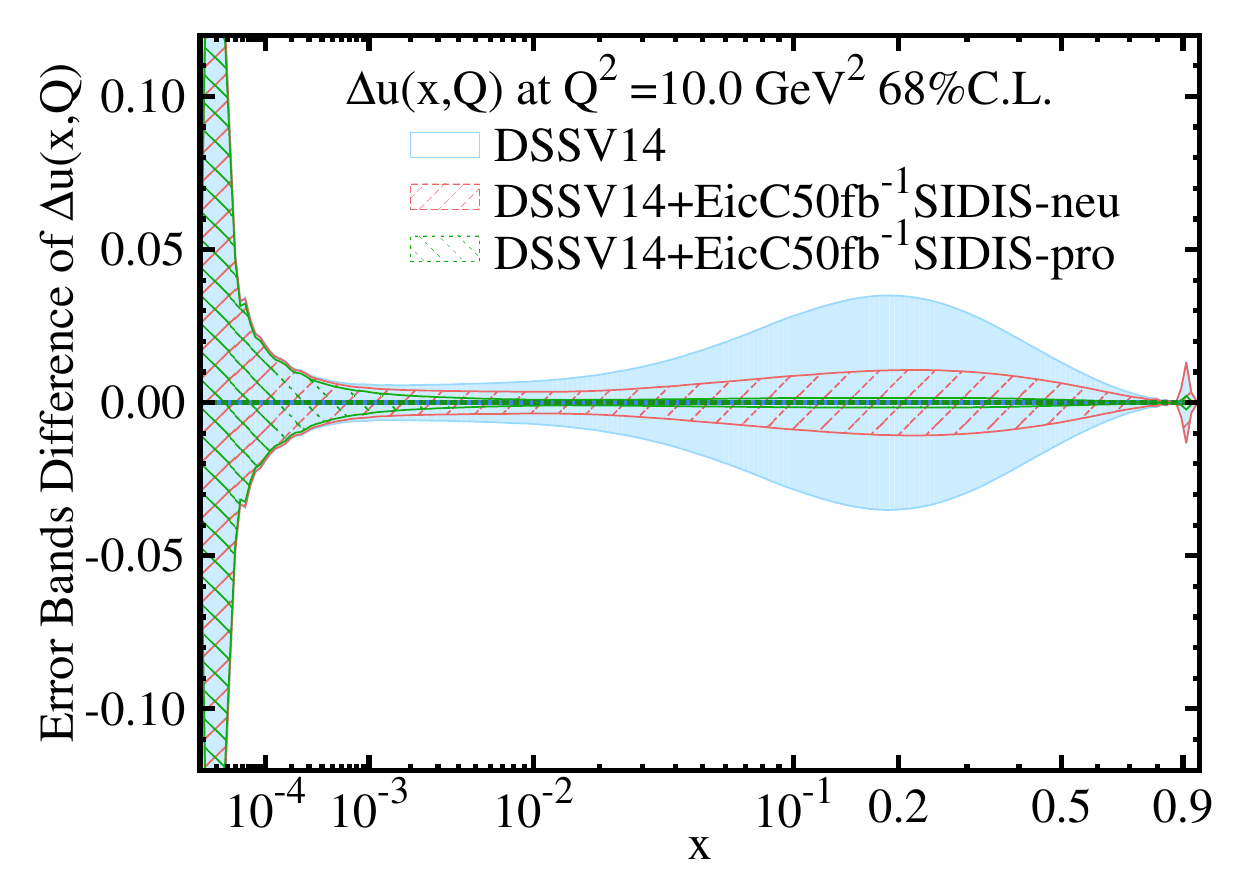}
\includegraphics[width=0.42\textwidth]{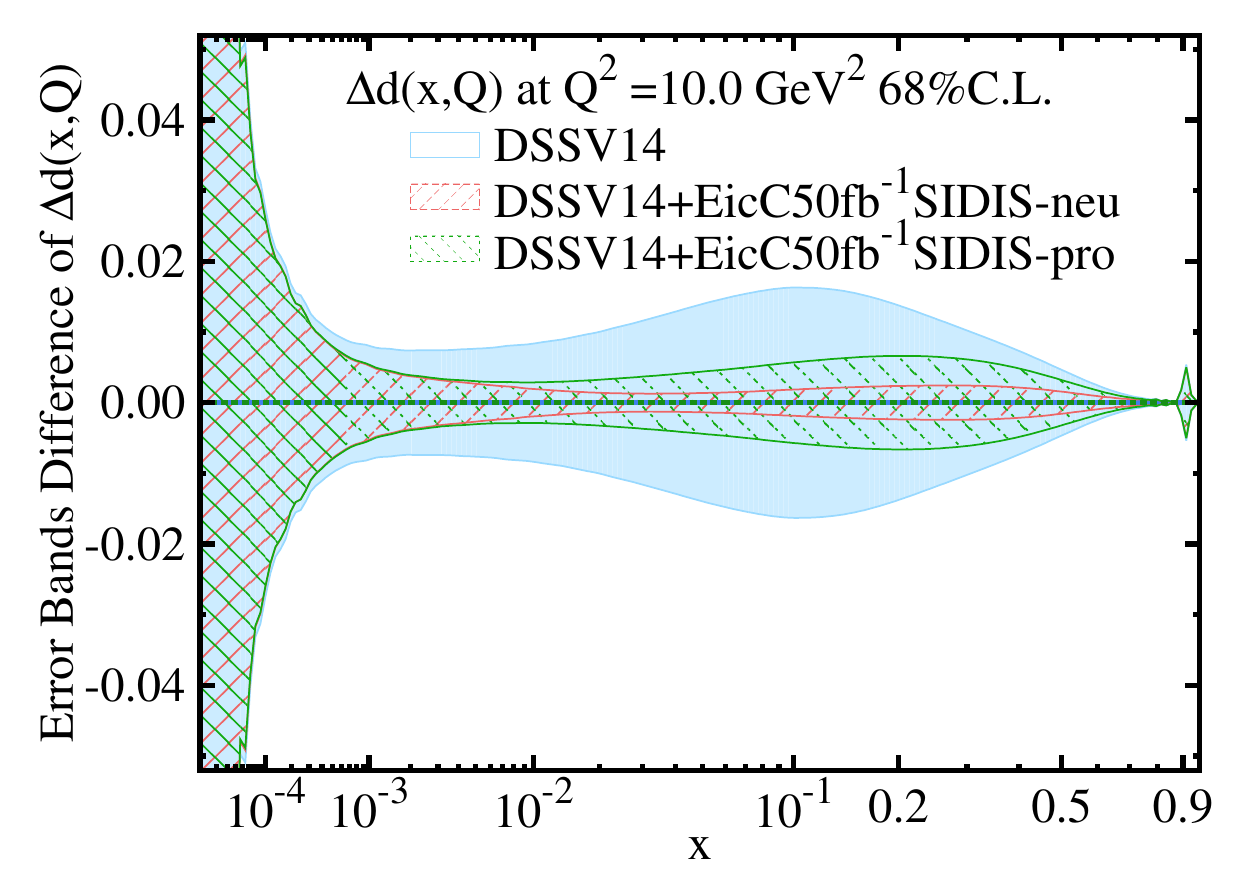} \\
\includegraphics[width=0.42\textwidth]{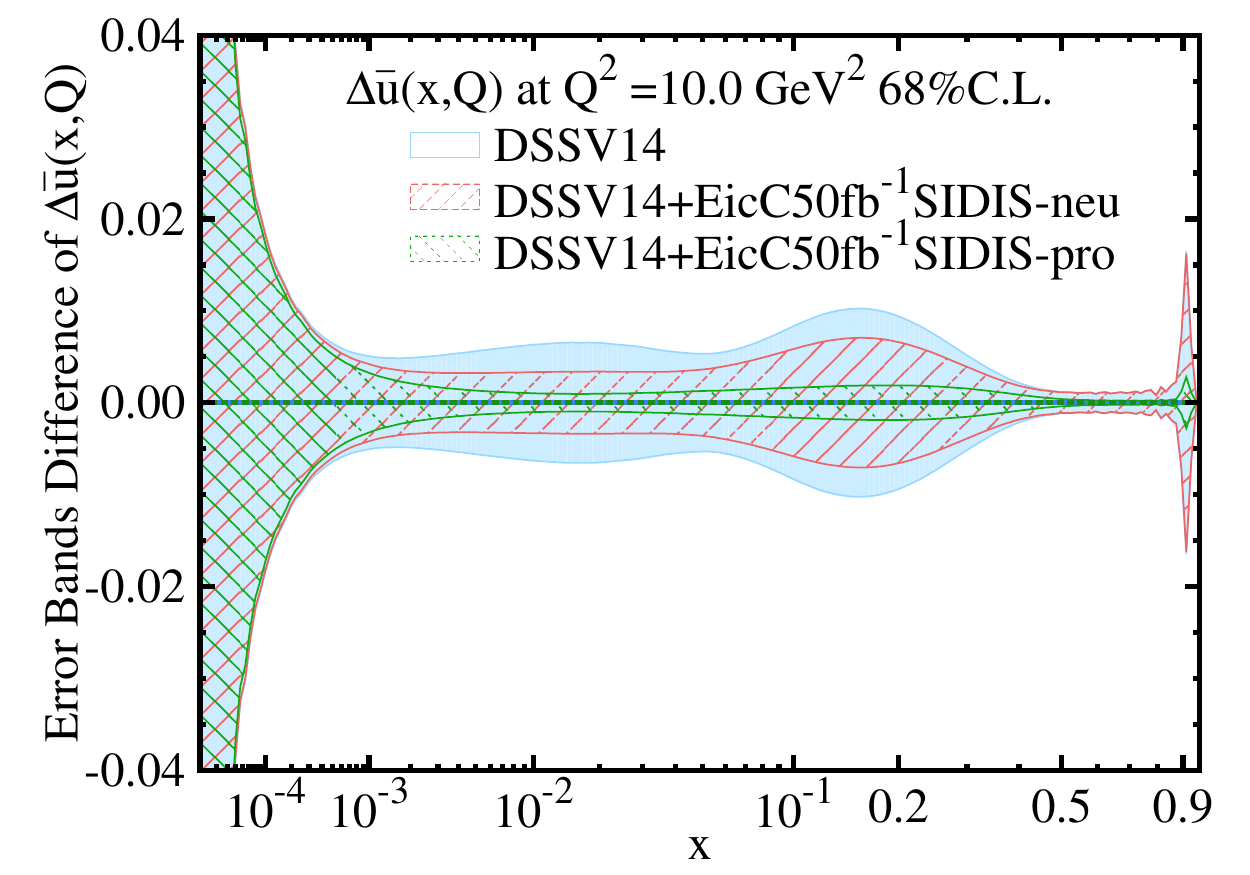}
\includegraphics[width=0.42\textwidth]{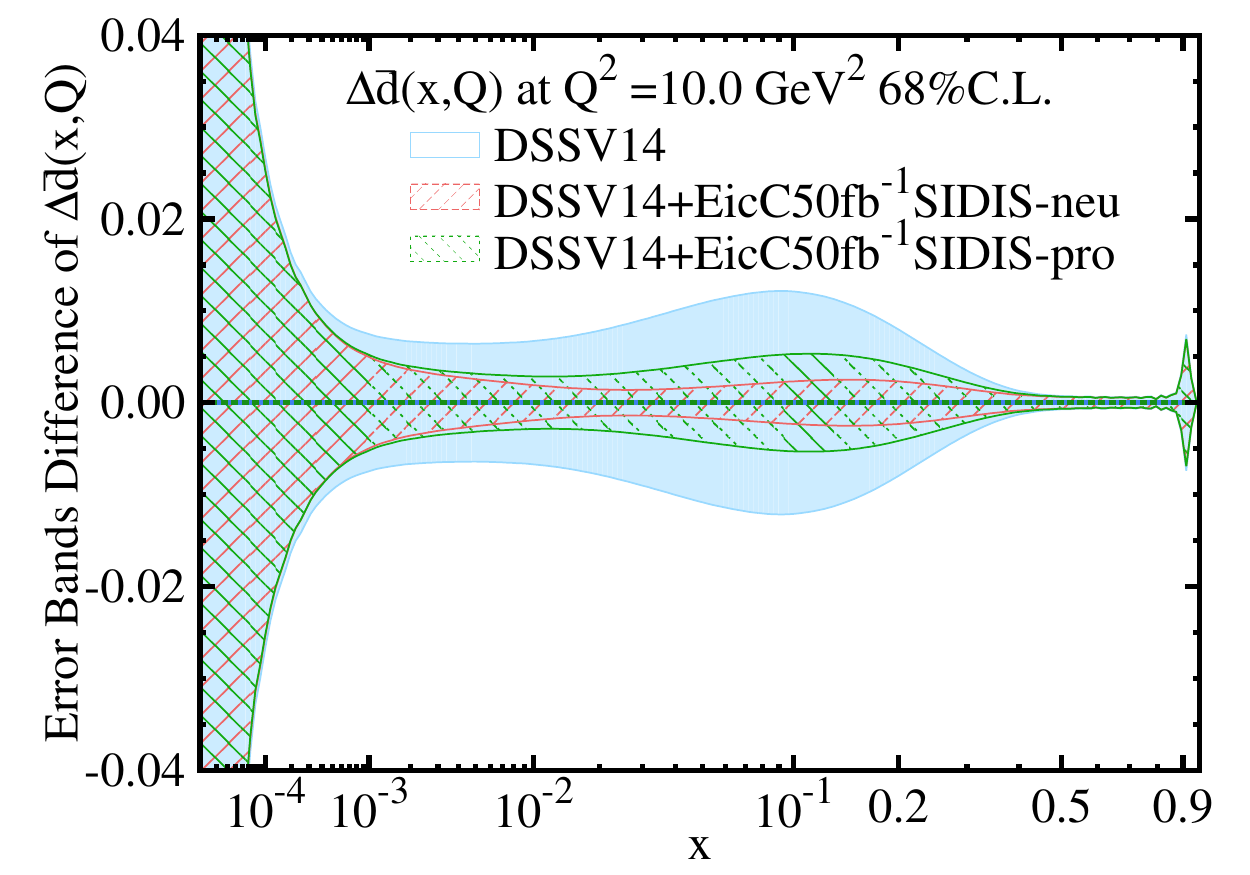} 
\subfigure[SIDIS pseudo-data]{\label{fig:PDFup_SIDIS-neu-pro}
\includegraphics[width=0.42\textwidth]{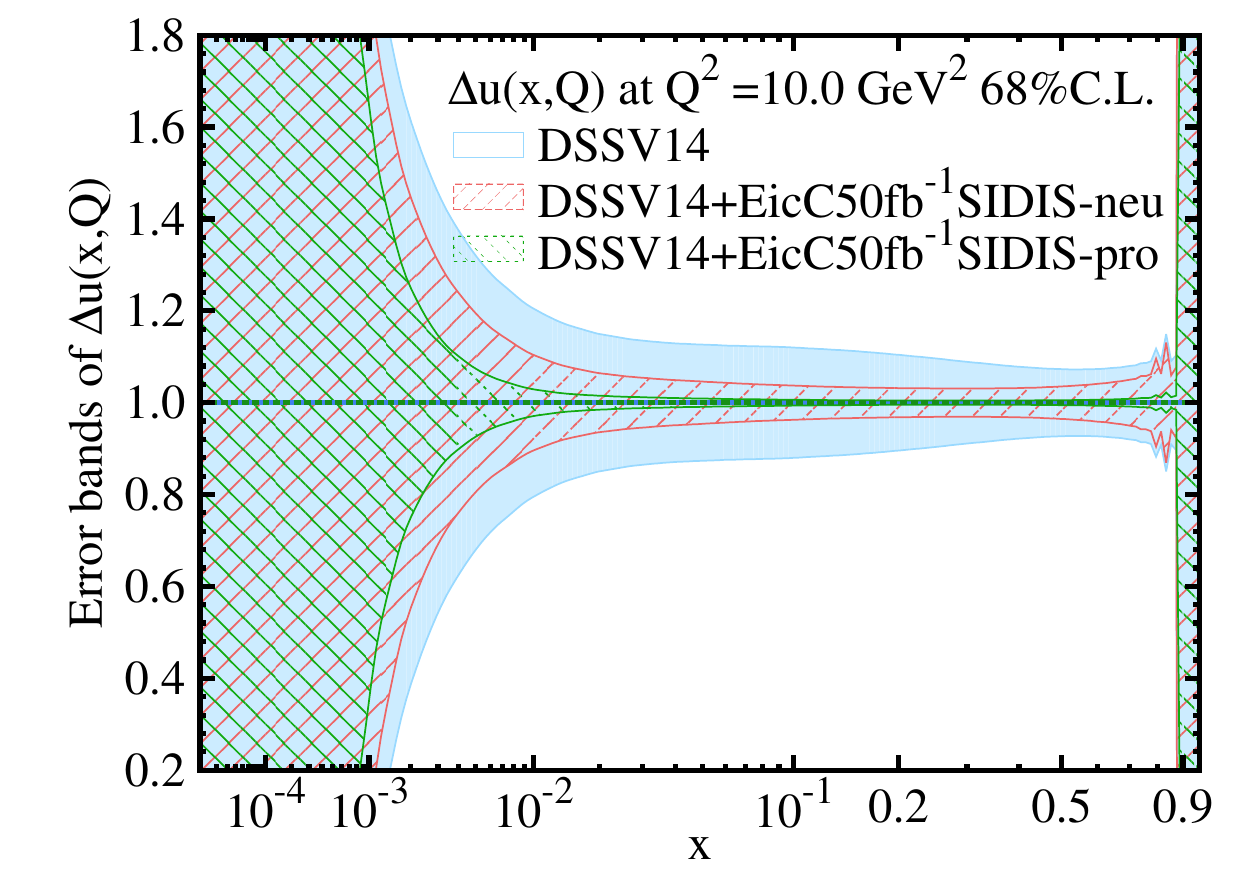}
\includegraphics[width=0.42\textwidth]{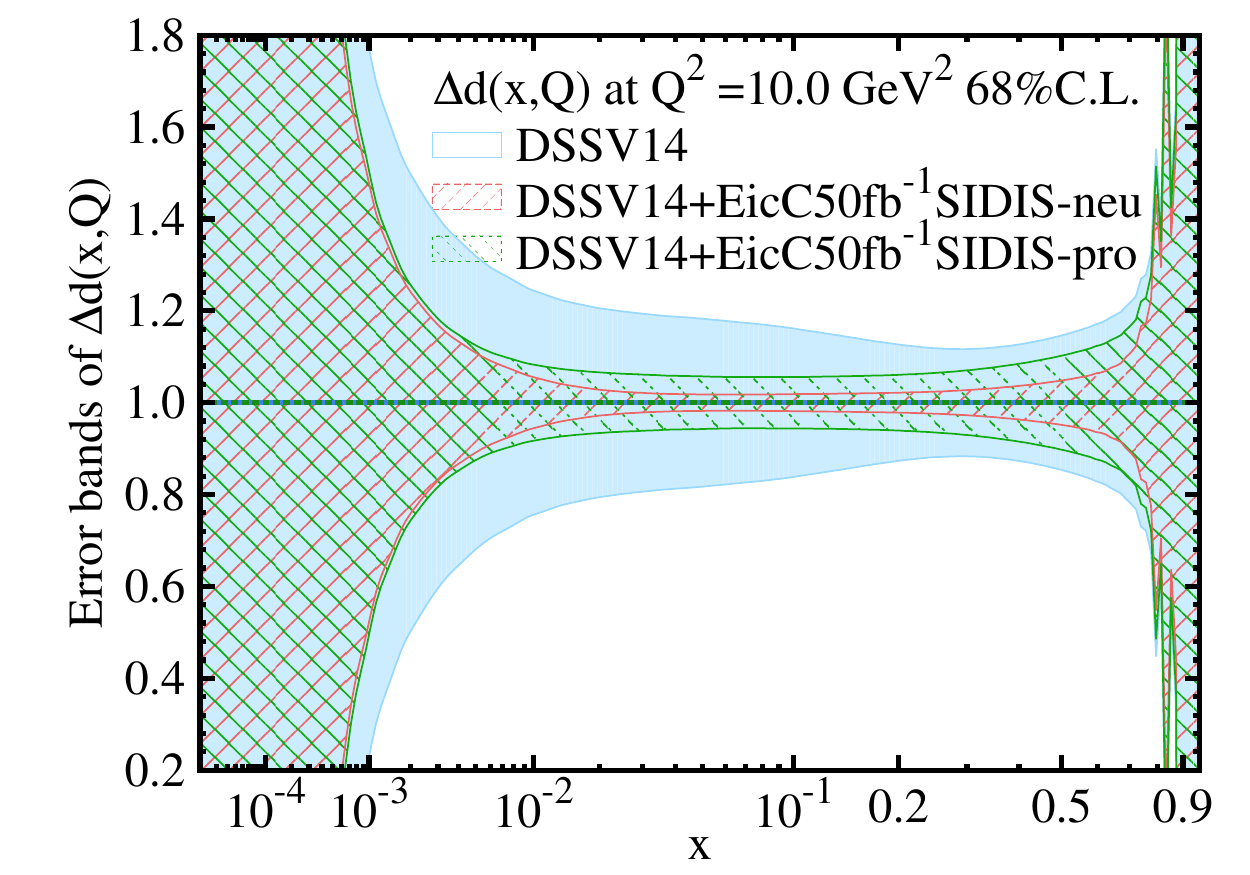}
}
\caption{\label{fig:PDF_SIDIS-neu-pro}
Results on the uncertainty band of polarized quark and sea quark distributions after a next-to-leading
order fit by including EicC pseudo-data. The light blue band represents the original DSSV14 global fit.
The red (green) band shows the results by adding EicC neutron (proton) pseudo-data.
The error bands in the first 5 plots are presented in the format of absolute differences comparing to the central values in Fig. \ref{fig:updated_dist},
while in the last row 
the error bands of $\Delta u$ and $\Delta d$ are presented as ratios to central values for better visualization of percentage uncertainties.
}
\end{center}
\end{figure}

Fig.~\ref{fig:updated_dist} shows the impact of DIS and SIDIS EicC pseudo-data on the parton helicity distributions, separately. 
As a general remark, all plots show a larger constraining power of some degree of SIDIS data (green areas) compared to DIS data (red areas). The small impact difference between the two types of data on the $\Delta u$ distribution is to be expected as $\Delta u$ is already the most constrained flavour distribution by the already available high precision proton inclusive data.
 On the other hand, the largest difference between the effects of including DIS versus SIDIS data can be seen in the sea quark distributions ($\Delta\bar u$, $\Delta \bar d$ and $\Delta s$) over the whole $x$-range spanned by the pseudo-data, down to about $x=0.005$. The ability of the EicC machine to pin down the sea distributions through the SIDIS process is a core feature around which the accelerator is being designed. Our result shows the benefit of using EicC SIDIS data to determine the sea quark distributions for which the impact of DIS pseudo-data only accounts with a minor or zero reduction of the PDF  uncertainties.
For $\Delta u$ and $\Delta d$, the largest uncertainty reduction happens around $x=0.2$ which is the region where the valence quarks are expected to contribute the most to the proton and neutron flavour content.

The EicC is not specifically planned to investigate the small-$x$ gluon distribution. A better machine suited for this purpose will be the future Electron-Ion Collider planned in the USA (EIC). Nonetheless, we show in Fig.~\ref{fig:updated_dist} that both DIS and SIDIS EicC pseudo-data are able to reduce the uncertainties on $\Delta g$ for $10^{-3}\lesssim x\lesssim 4\times10^{-2}$. The reduction of the uncertainties below the pseudo-data range $x\lesssim10^{-3}$ that we can observe in both $\Delta s$ (the green grid part) and $\Delta g$ (the red and green grid parts) can be sourced back to assumptions made in the original DSSV14 analysis. Among those, the most stringent ones are the initial parametrization bias of the helicity PDFs, and, in particular for $\Delta s$, the hyperon $\beta$-decay constraints that will be discussed later on in this section. 

To study the difference of the impact between the proton and neutron target pseudo-data,
in Fig.~\ref{fig:PDF_SIDIS-neu-pro} the uncertainty bands for the first 5 plots are presented as a difference with their respective central values, meaning that the uncertainty bands for the original DSSV14H (light-blue), the DSSV14H including proton pseudo-data (green) and the DSSV14H including neutron pseudo-data (red) are always centered along the zero axis.  
In the last row of Fig.~\ref{fig:PDFup_SIDIS-neu-pro}, the error bands of $\Delta u$ and $\Delta d$ are presented as ratios to central values for better visualization of percentage uncertainties.

For both DIS (Fig.~\ref{fig:PDFup_DIS-neu-pro}) and SIDIS (Fig.~\ref{fig:PDFup_SIDIS-neu-pro}), the proton data are able to constrain the $\Delta u$ distribution better compared to the neutron data. This is consistent with a quark model picture in which the proton content is dominated by the $u$ quarks whereas the neutron content by the $d$ quarks. This can be observed more in detail in Fig.~\ref{fig:PDFup_SIDIS-neu-pro}, where this behavior remains the same for $\Delta \bar u$ and is accordingly inverted for the $\Delta d$ and $\Delta \bar d$ distributions. More significantly, it shows the importance of including both SIDIS neutron and proton target data in order to efficiently constrain the up and down quark and anti-quark distributions.

\begin{figure}[t]
\begin{center}
\includegraphics[width=0.48\textwidth]{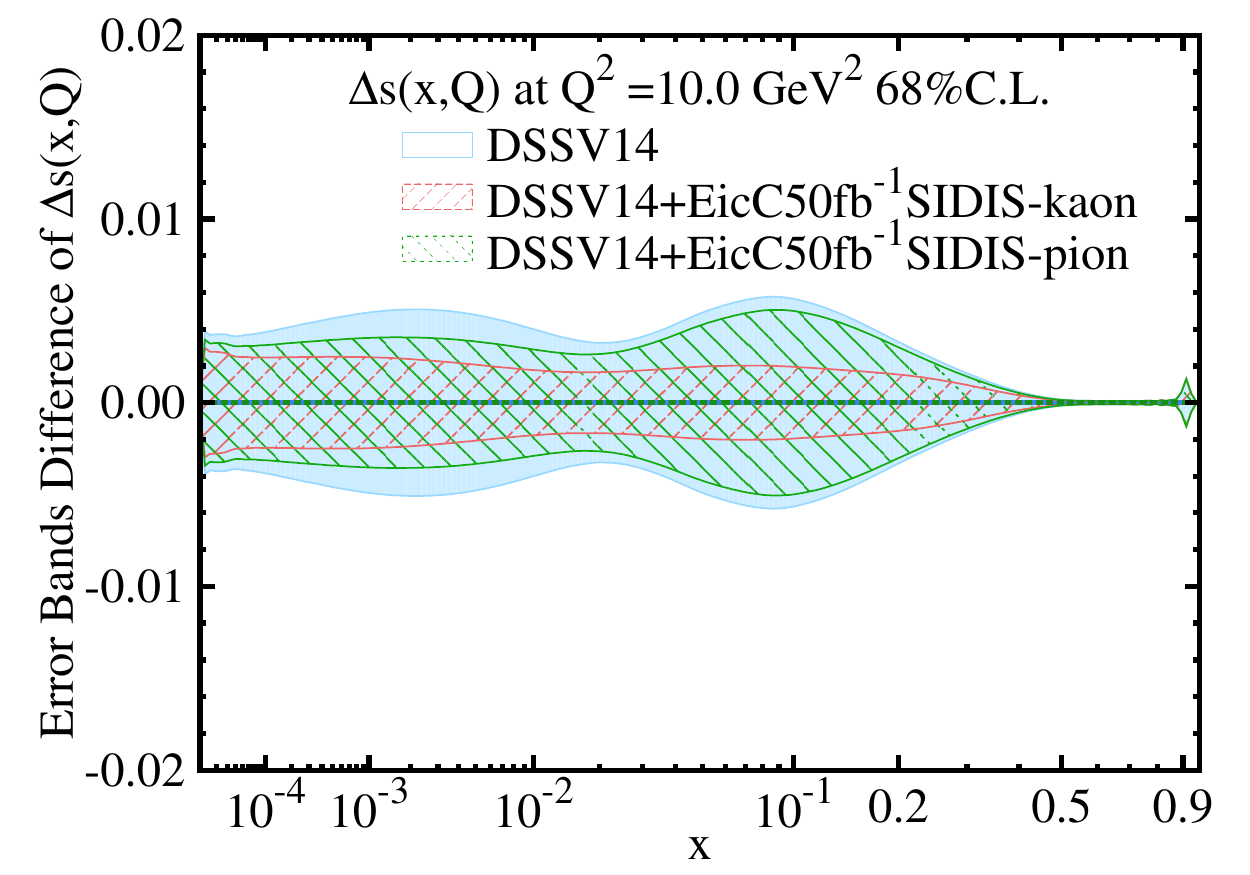}
\caption{\label{fig:PDFup_SIDIS-kaon-pion}
Results on the uncertainty band of polarized strange quark distribution after a next-to-leading
order fit by including EicC pseudo-data. The light blue band represents the original DSSV14 global fit.
The red (green) band shows the results by adding EicC SIDIS $K$ (SIDIS $\pi$) pseudo-data.
}
\end{center}
\end{figure}

In Fig.~\ref{fig:PDFup_SIDIS-kaon-pion} the impact of SIDIS pion pseudo-data (green) versus SIDIS kaon pseudo-data (red) is shown in the same type of difference plot for the $\Delta s$ distribution. Since the flavour content of kaon mesons is dominated by strange quarks, identifying kaons in the final state effectively ``tags'' the strange quarks scattering out of the target. Hence, the SIDIS kaon data are able to better constrain $\Delta s$ in respect to SIDIS data with non-strange final state hadrons such as pions. Moreover, comparing with Fig.~\ref{fig:updated_dist} we notice that the $\Delta s$ is further constrained after including the pion data on top of the kaon data. This is due to the correlation between $\Delta u$, $\Delta d$ and $\Delta s$ introduced by the relation imposed in the DSSV14 analysis that we discuss further down in Eq.~(\ref{eq:su2_su3_1}).


As shown in Figs. \ref{fig:updated_dist}-\ref{fig:PDFup_SIDIS-kaon-pion}, the EicC pseudo-data can effectively constrain the polarized PDFs, namely, their error bands have been significantly reduced. As for the changes of PDFs' central values after the updates, we have calculated the measure $d^0$ defined in Eq. (\ref{eq:distance1}) and list them in 
Table \ref{tab:d0}. 
The fact that none of those  $d^0$ values is greater than one indicates  that the ePump-updating provides a reasonable fit, cf. Sec.~\ref{sec:hpm}. This result is expected by the construction of the pseudo-data, discussed in Sec.~\ref{sec_data}.
Although this measure is a powerful tool to quantify the shift of the central value due to a new set of data, in the contest of pseudo-data it cannot act as a physically meaningful prediction of the shift on the best-fit that will result when real experimental data are used. Pseudo-data are constructed such that they embed a faithful estimate of the future EicC data uncertainties but they have an unknown degree of deviation from the future actual experimental data central values. For this reason, any definitive statement on central-value shifts has to be postponed for when updating will be possible with the EicC real experimental data.

\begin{table}[]
    \centering
    \begin{tabular}{ll|ll}
    \texttt{ePump} update & $d^0$ & \texttt{ePump} update & $d^0$ \\ \hline\hline
    EicC SIDIS & 0.88 & EicC DIS & 0.50 \\
    EicC SIDIS proton & 0.89 & EicC DIS proton & 0.45 \\
    EicC SIDIS neutron & 0.64 & EicC DIS neutron & 0.34 \\
    EicC SIDIS pion & 0.68 & EicC SIDIS kaon & 0.59
    \end{tabular}
    \caption{The $d^0$ measures, defined in the Eq. (\ref{eq:distance1}), for various \texttt{ePump} update analyses.}
    \label{tab:d0}
\end{table}


\begin{figure}[t]
\begin{center}
\subfigure[After updating with DIS and SIDIS pseudo-data.]{\label{fig:bubble_5E-3_1}\includegraphics[width=0.42\textwidth]{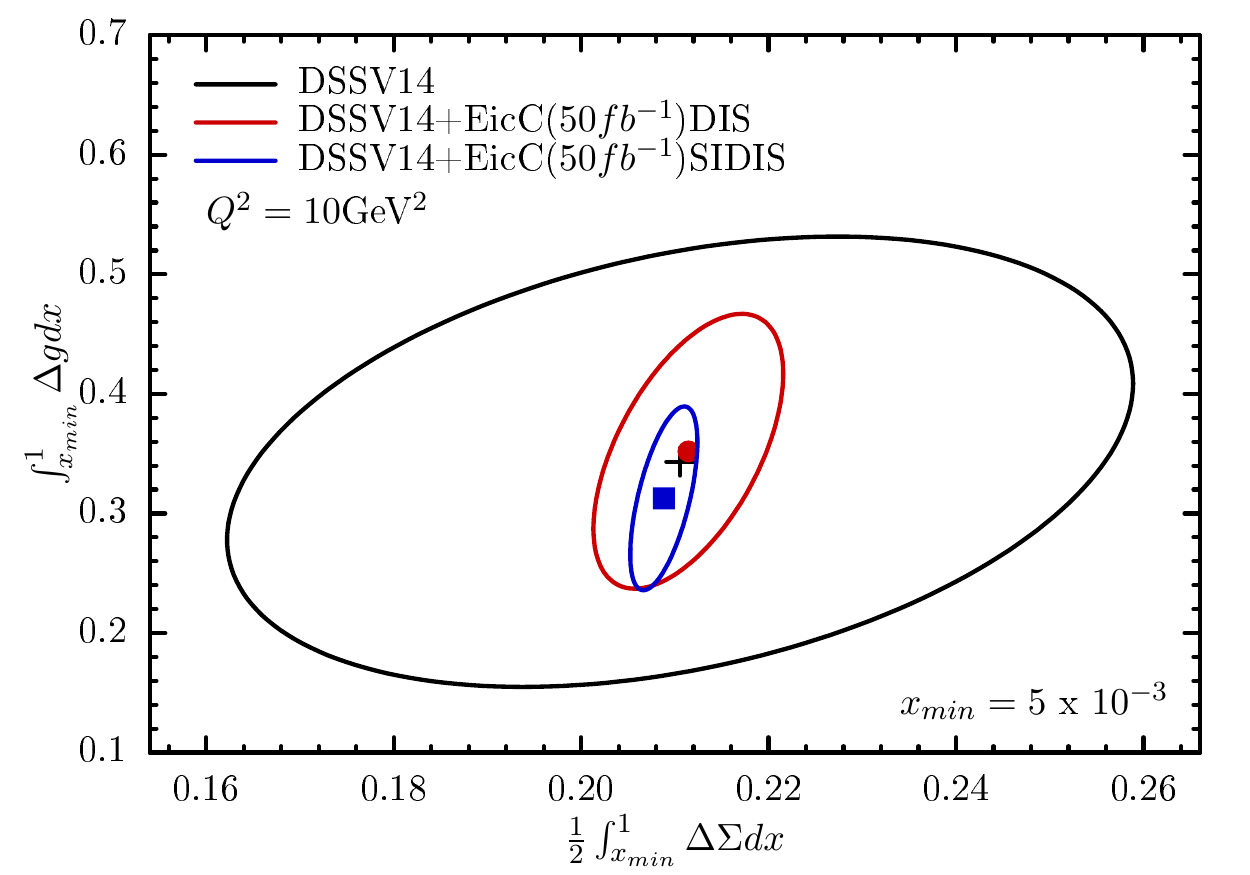}}
\subfigure[After updating with DIS electron-proton and electron-${}^3$He collision pseudo-data.]{\label{fig:bubble_5E-3_2}\includegraphics[width=0.42\textwidth]{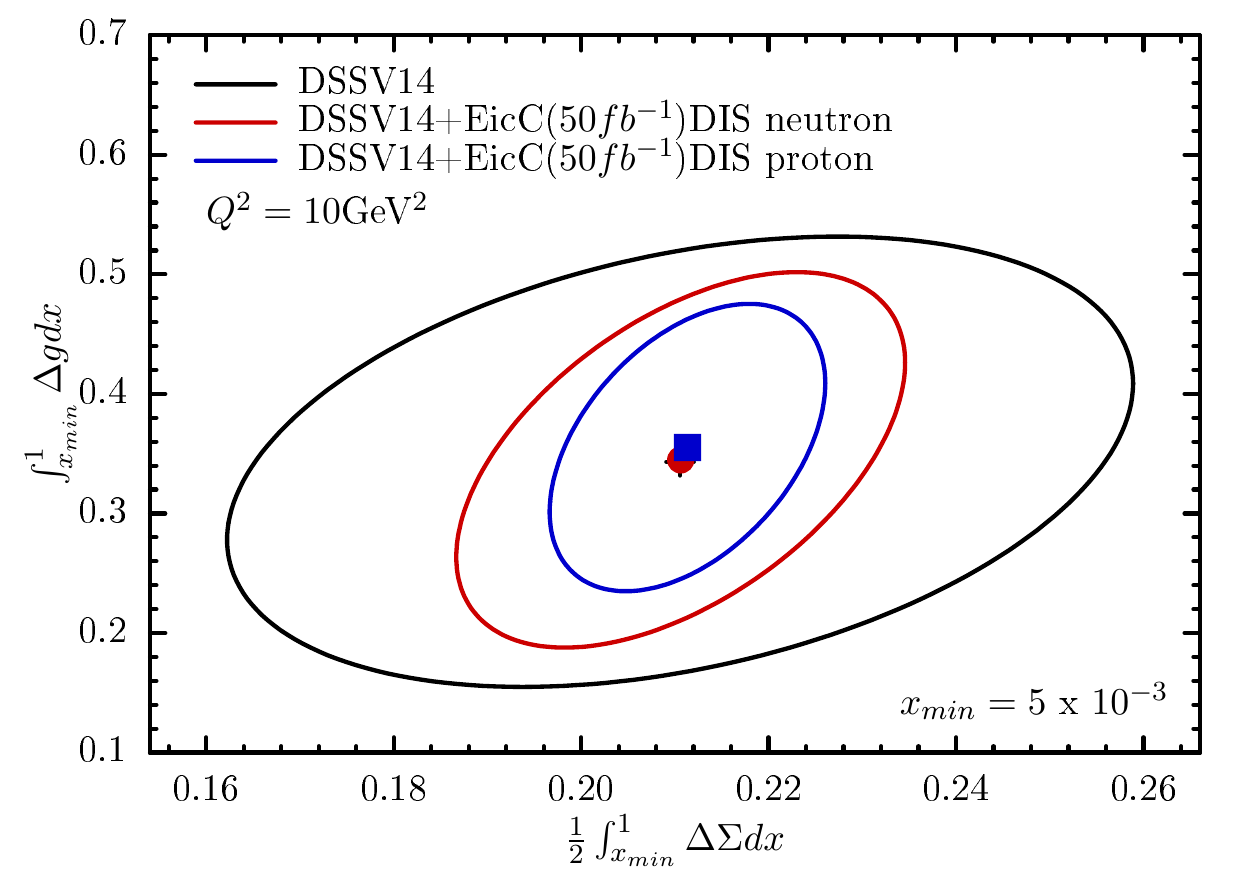}}
\subfigure[After updating with SIDIS electron-proton and electron-${}^3$He collision pseudo-data where the identified hadrons are pions or kaons.]{\label{fig:bubble_5E-3_3}\includegraphics[width=0.42\textwidth]{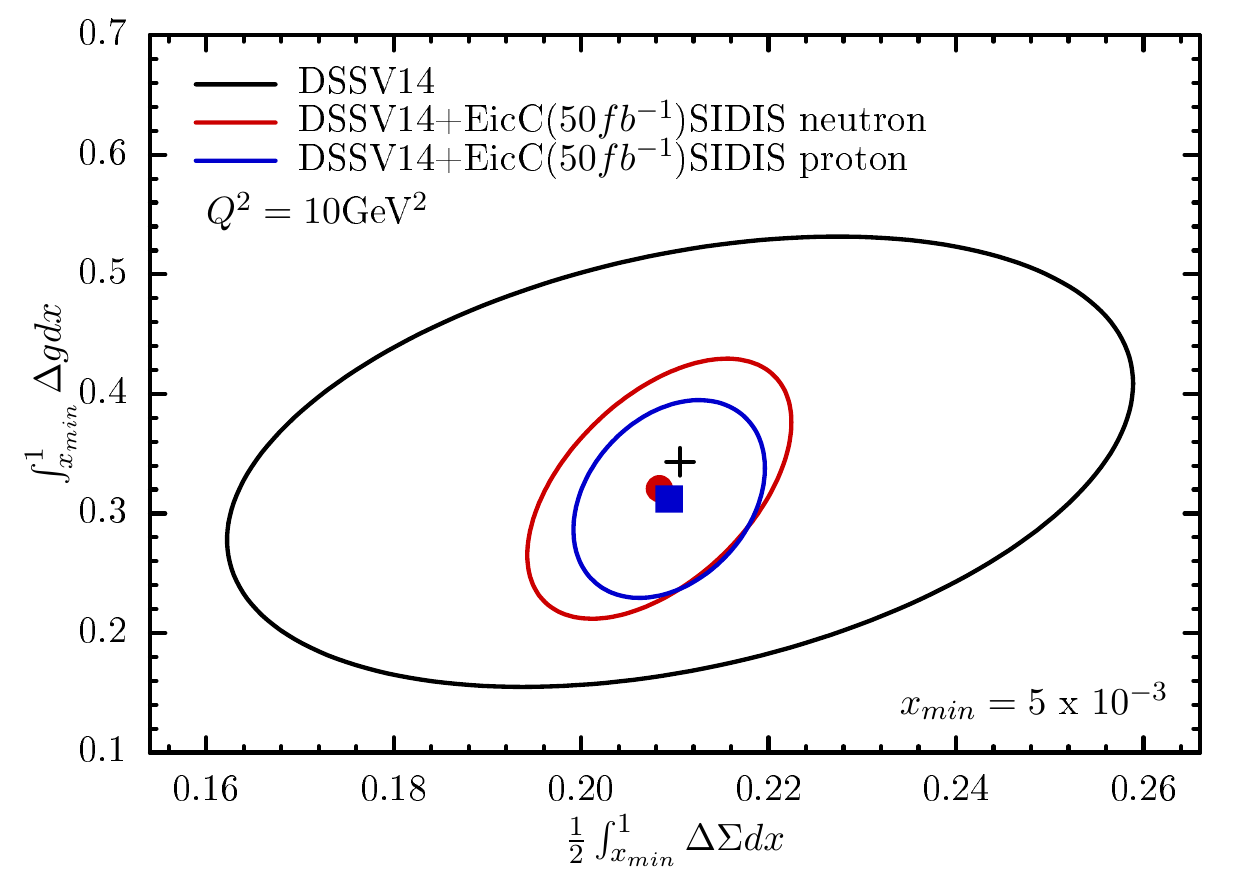}}
\subfigure[After updating with SIDIS pseudo-data for identified pions and kaons in the final state.]{\label{fig:bubble_5E-3_4}\includegraphics[width=0.42\textwidth]{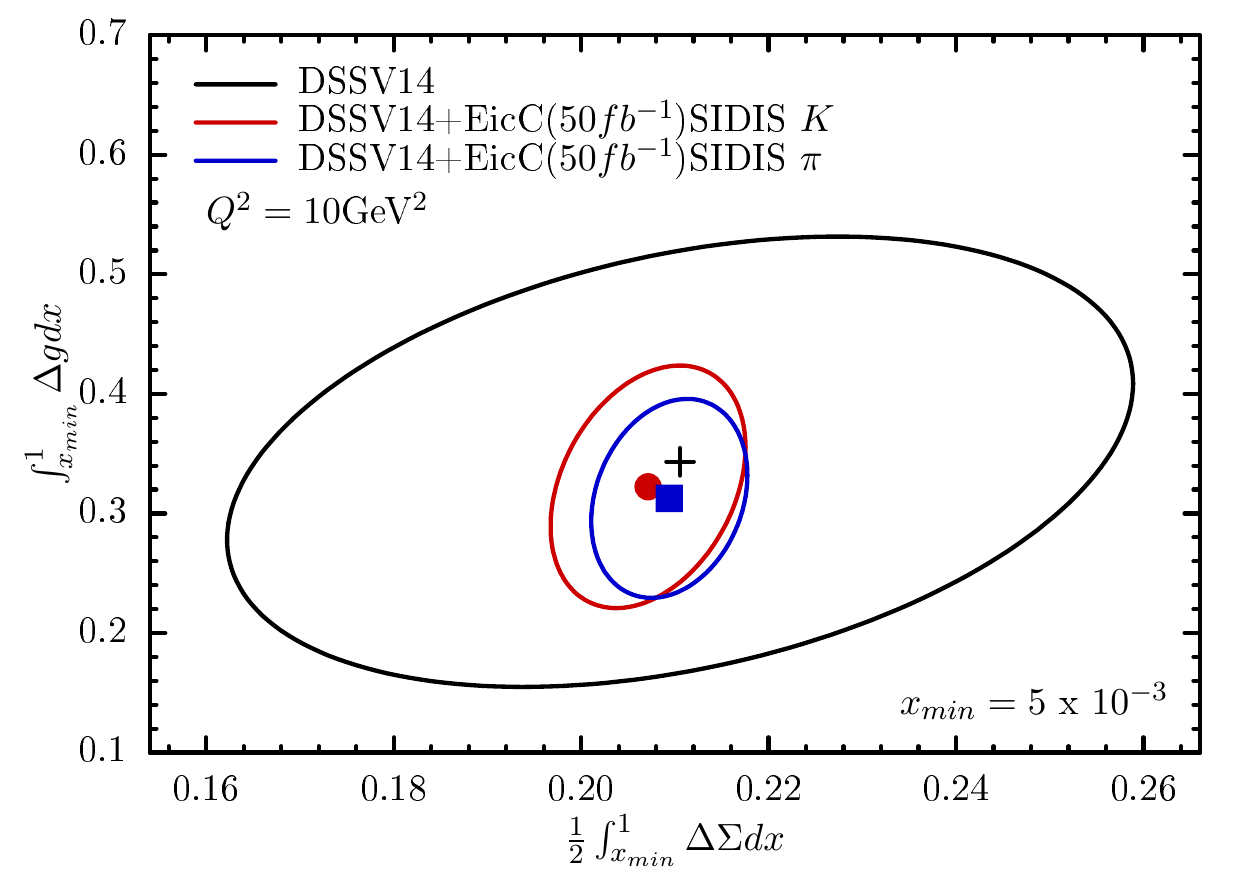}}
\caption{\label{fig:bubble_5E-3}
Correlation contour between $\frac{1}{2}\int_{x_{min}}^{1} \Delta \Sigma dx$ and $\int_{x_{min}}^{1} \Delta g dx$. The black contour shows the correlation with DSSV14 PDFs. The red and blue contours show the correlation with DSSV14 PDFs after ePump updating by using different EicC DIS and SIDIS pseudo-data. In all data sets the integrated luminosity for EicC pseudo-data is 50 $fb^{-1}$ for e-p collisions and 50 $fb^{-1}$ for e-$^3$He collisions. See text for detailed discussion.
}
\end{center}
\end{figure}

Quantities of particular interest in the field are the moments of the singlet combination $\Delta\Sigma$, i.e. the sum over all flavour PDFs (see Appendix~\ref{app:A1}), and the gluon distribution. More specifically, their first moment, i.e. their integral over the parton momentum fraction, has a simple interpretation as the net quark and gluon contribution to the proton spin. 

The impact of the EicC pseudo-data on such quantities is shown in Fig.~\ref{fig:bubble_5E-3}, where the correlated uncertainties of the truncated first moment of the gluon and singlet distribution are depicted in a two-dimensional plot at $Q^2=10\text{ GeV}^2$. The lower truncation of the integral is set to $x_{min}=0.005$, which corresponds to the theoretical lower momentum fraction accessible to the future EicC machine and, consequently, below which pseudo-data have not been generated for this analysis. Trying to investigate the spin contribution to the proton from quarks and gluons with smaller momentum fraction by stretching our analysis beyond this lower threshold, would return biased results only constrained by the original DSSV14 assumptions such as the choice of the initial parametrization form or the continuity, integrability, and positivity (i.e. $|\Delta q|<q$) requirements of the helicity distributions. However, lower values of $x_{min}$ will be accessible from the future EIC and analyses using EIC pseudo-data down to $x_{min}\sim 10^{-5}$ such as~\cite{Aschenauer:2012ve} have been performed. In that specific study, they have also extended their integration from $x_{min}\sim 10^{-5}$ down to $x_{min}\sim 10^{-6}$ and observed that the integrals tend to saturate quite early, suggesting a picture where very low-$x$ partons become unpolarized. Independently from whether or not this feature will be confirmed by the future EIC measurements at very low-$x$, the actual allowed central value of the two integrals will be proportional to the precision at which the distributions are known in the full $x$-range. In this respect, the EicC acts as a complementary machine to the EIC by being able to better determine distributions in the sea-quark region ($x\gtrsim10^{-2}$). The red and blue contours in Fig.~\ref{fig:bubble_5E-3} show the allowed values of the contribution to the integrals, together with their central values, for $x>5\times10^{-3}$ according to different EicC pseudo-data sets. The black cross and black contour are, respectively, the central value and the allowed values of the contribution to the integrals for the same $x$-region for the actual DSSV14. As can be observed in all plots, the uncertainty area of the DSSV14 is predicted to be well reduced by the \texttt{ePump} updating after including EicC pseudo-data.

More in detail, Fig.~\ref{fig:bubble_5E-3_1} explicitly shows the higher constraining power of SIDIS pseudo-data (blue region) in respect to the DIS ones (red region). The shifts on the respective central values are a consequence of the slight shifts observed for the ``sea-quark'' distribution $\Delta\bar u$. Moreover, the effect of SIDIS pseudo-data on the central value is sizeably bigger than the effect observed for DIS pseudo-data.
As already discussed above, the particular value of the shift has no real physical meaning in this analysis and depends on the specifically chosen iteration of the gaussian smearing used to produce the pseudo-data. Nonetheless, it cannot be excluded that real future EicC data may change the shape of the gluon and single flavour distribution for the region $x>5\times10^{-3}$.

Figs.~\ref{fig:bubble_5E-3_2} and~\ref{fig:bubble_5E-3_3} show the effect on the integrals for EicC proton and neutron DIS and SIDIS pseudo-data, separately. Both electron-proton and electron-neutron collisions have been generated with an integrated luminosity of $50~fb^{-1}$. However, the neutron data sample has been obtained from electron-${}^3$He collision data through the dilution procedure described in Sec.~\ref{sec_data}. The effect of the additional uncertainties introduced in the neutron data sample, by the required neutron and proton effective polarization values, can be directly observed for both DIS and SIDIS as a lower constraining power of the neutron pseudo-data (red regions) in respect to the proton pseudo-data (blue regions). Moreover, the central values updated with proton and neutron (SI)DIS pseudo-data do not shift between each other by a significant amount. This is an expected result dictated by the underlying SU(2) proton-neutron isospin symmetry imposed when calculating the theoretical tables for this analysis. Future global fitting will be able to lift this assumption by exploiting the ability for precise SIDIS data to discriminate flavours over a large kinematical range. Actually, the DSSV collaboration already allows deviations from exact SU(2) and SU(3) flavour symmetries in their analyses in the form of two additional fitting parameters $\epsilon_{\text{SU(2)}}$ and $\epsilon_{\text{SU(3)}}$. They are inserted in the fitting procedure in order to relax the constraints coming from the hyperon semi-leptonic $\beta$-decay and its implicit flavour symmetry assumptions, normally imposed in polarized PDFs extractions based on solely DIS data~\cite{deFlorian:2007ekg}.
More specifically, this translates in various first moments being related by
\begin{eqnarray}
\label{eq:su2_su3_1}
\Delta\Sigma^1_u-\Delta\Sigma^1_d&=&(F+D)\;[1+\epsilon_{\text{SU(2)}}] , \nonumber \\
\Delta\Sigma^1_u+\Delta\Sigma^1_d-2\Delta\Sigma^1_s&=&(3F-D)\;[1+\epsilon_{\text{SU(3)}}] , 
\end{eqnarray}
where $F$, $D$ are constants parametrizing the $\beta$-decay rates~\cite{Airapetian:2007mh} at the input scale $\mu_0=1$ GeV of the DSSV analysis, and
\begin{equation}
\label{eq:su2_su3_2}
    \Delta\Sigma^1_q \equiv \int^1_0\;[\Delta q+\Delta\bar q](x,\mu_0)\;dx , \qquad q\in\{u,d,s\}.
\end{equation}

As Eqs.~(\ref{eq:su2_su3_1}) and~(\ref{eq:su2_su3_2}) show, the precision at which the $\epsilon_{\text{SU(2)}}$ and $\epsilon_{\text{SU(3)}}$ parameters can be determined are tied together with the accuracy at which the integrals can be calculated over the full $0<x<1$ span. 
Due to the lower kinematical limit lying at $x_{min}\sim 10^{-3}$, data from the planned EicC machine wouldn't be sufficient by itself to impose very strict constraints to the parameters. As for $x\lesssim10^{-3}$, the integrals in Eq.~(\ref{eq:su2_su3_1}) would be strongly biased by the PDF initial parametrization form and only loosely constraint by general helicity PDFs requirements such as the continuity, integrability, and positivity of the distributions.
However, precise determination of the deviations from flavour SU(2) and SU(3) from global fitting will be possible and highly improved by taking into account both future EIC and EicC SIDIS data which, combined, will span over a larger $x-Q^2$ area in respect to SIDIS world data currently available with unprecedented precision. A study of the effect of DIS and SIDIS EIC pseudo-data was presented in~\cite{Aschenauer:2012ve}.

The effect of data samples with identified pions and kaons in the final state is shown in Fig.~\ref{fig:bubble_5E-3_4}. The larger area delimited by the SIDIS kaon pseudo-data comes from the fact that larger statistical uncertainties are associated with the kaon pseudo-data as kaons are produced with a lower rate in respect to pions. In the same plot, we can observe a slightly different shift of the central values produced by the two data sets. Since in the case of the SIDIS process different PDFs are weighted with different fragmentation functions, the different flavour content of the identified final state hadrons is responsible for the dissimilar shift after \texttt{ePump} updating. The ability for some specific data set to constrain particular flavour distributions will be discussed in detail in Sec.~\ref{sec_optimize}.

\begin{figure}[t]
\begin{center}
\subfigure[Net quark contribution to the proton spin as a function of $x_{min}$]{\label{fig:PDFint_log_1}\includegraphics[width=0.42\textwidth]{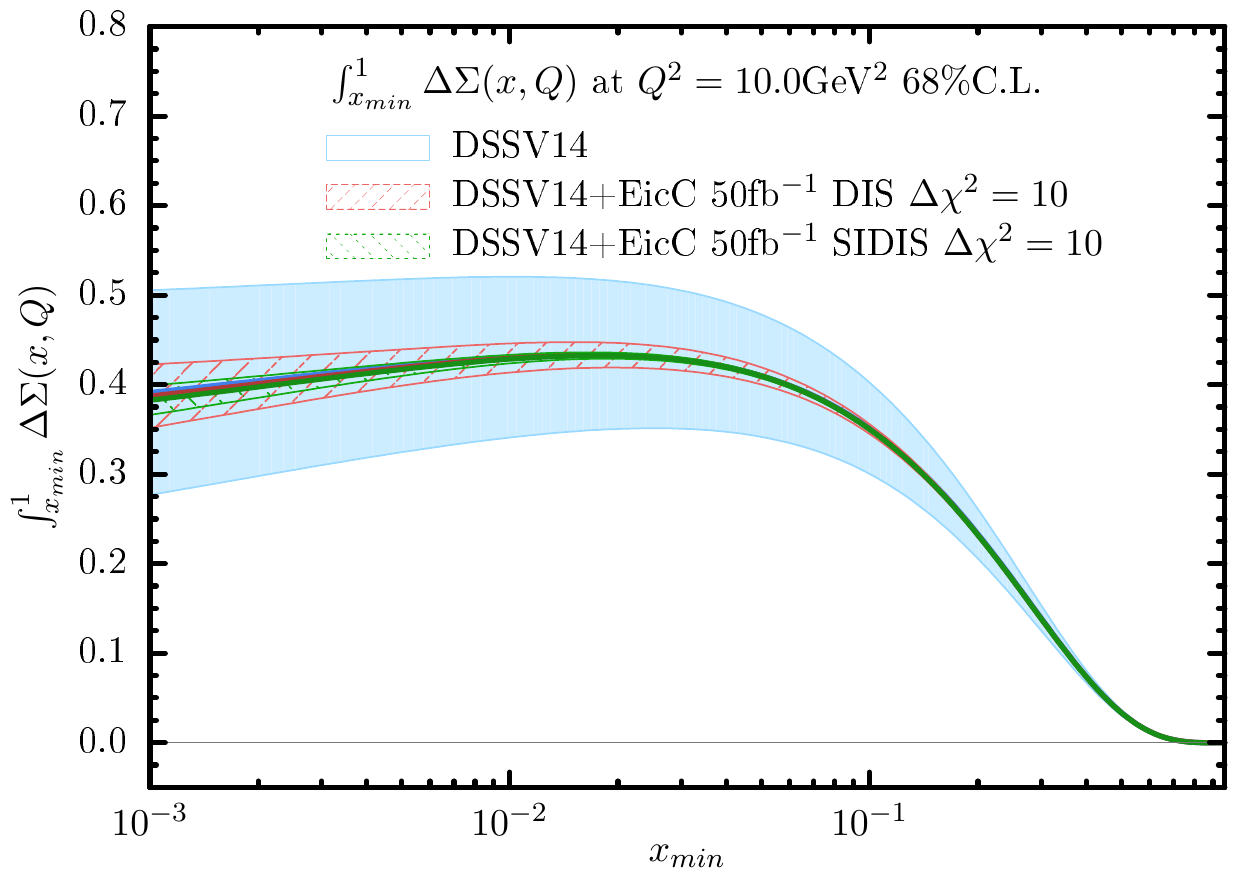}}
\subfigure[Net gluon contribution to the proton spin as a function of $x_{min}$]{\label{fig:PDFint_log_2}\includegraphics[width=0.42\textwidth]{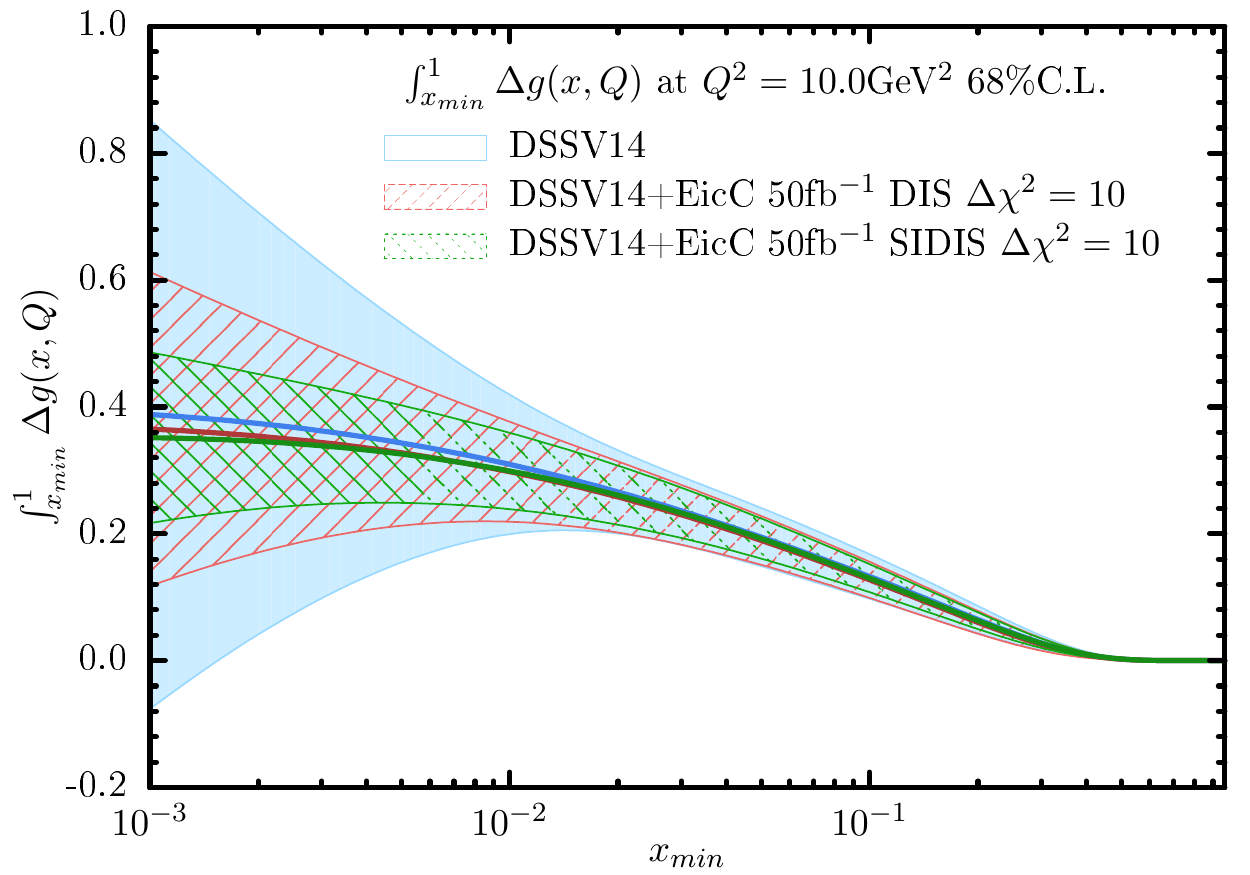}}
\subfigure[Difference between the proton spin 1/2 and the net spin contribution of partons as a function of $x_{min}$]{\label{fig:PDFint_log_3}\includegraphics[width=0.42\textwidth]{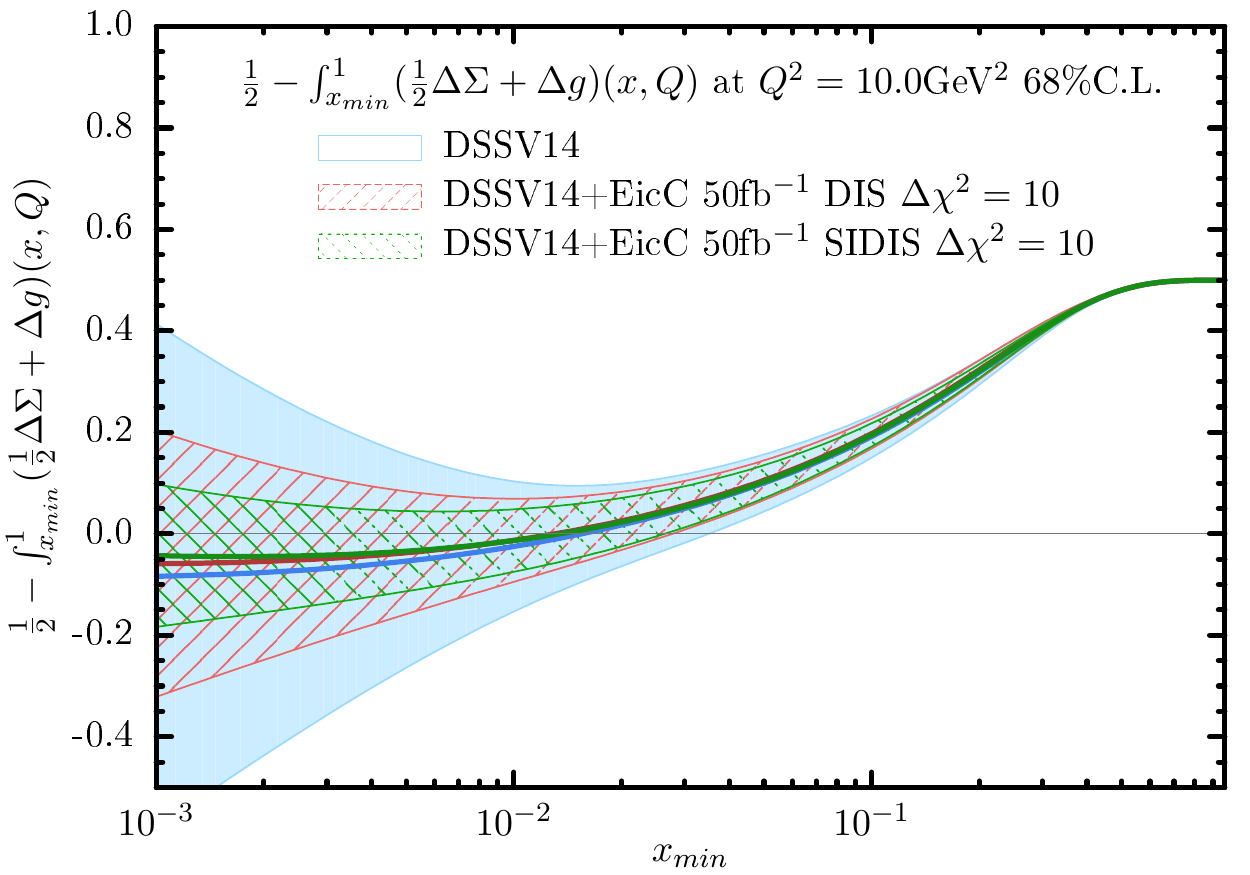}}
\caption{\label{fig:PDFint_log}
Central values and uncertainty limits for the net spin contribution of gluons, quarks and other sources as a function of their minimum momentum fraction $x_{min}$. The limit imposed by the original DSSV14 analysis is given in light-blue. The results including EicC DIS (SIDIS) pseudo-data are shown in red (green).
}
\end{center}
\end{figure}

In addition to the quark and gluon contributions to the proton spin, the remaining missing part is related to the quark and gluon orbital momentum. (For more on the subject, see ~\cite{Ji:2020ena} and references therein). In Fig.~\ref{fig:PDFint_log} we show the net contributions of quarks and gluons to the proton spin and their cumulative difference with the actual proton spin value 1/2 as a function of the lower bound $x_{min}$ used to compute the truncated moments. In all plots one can observe a clear reduction of uncertainties when including DIS and SIDIS EicC pseudo-data, with SIDIS data having the larger impact. The tendency for the central values of the integrals to saturate at low-$x$ shown in Figs.~\ref{fig:PDFint_log_1} and~\ref{fig:PDFint_log_2} is compatible with a picture where partons carrying very small momentum fraction $x$ are mostly unpolarized. However, contributions from partons with lower momentum fraction than $x\sim 10^{-3}$ may still contribute to the proton spin, in which case the above picture could result to be incorrect. At the moment, the huge uncertainties associated with the $\Delta g$ distribution at low-$x$ is still the main limiting factor in order to state a more definitive conclusion on the matter. The US EIC will be the perfect machine to precisely pin down the low-$x$ $\Delta g$. On the other hand, the EicC is planned to explore that complementary part of the phase space particularly suited for a better determination of the ``sea-quark'' sector. This becomes apparent if we observe the significant uncertainties' reduction on the quark spin contribution in Fig.~\ref{fig:PDFint_log_1}, which extends from $x_{min}\sim 10^{-3}$ up to high $x_{min}$ values. In contrast, in Fig.~\ref{fig:PDFint_log_2} we observe a much lower impact to the uncertainties, which is almost entirely relegated to the $x_{min}\lesssim 10^{-2}$ region of the plot.

Nonetheless, this exercise shows, once again, the importance of including the information coming from the SIDIS process when it comes to precision extraction of both gluon and quarks helicity distribution functions and their moments. 

Finally, Fig.~\ref{fig:PDFint_log_3} shows the evolution of the missing contribution to the proton spin as we consider contributions from partons with smaller and smaller momentum fraction $x$ in the computation of the integrals. 
The central value of the quantity shown in the plot seems to saturate asymptotically for small-$x$, with the SIDIS data having a greater constraining effect on the uncertainties in respect to DIS data.
In the assumption that all missing proton spin comes exclusively from the quark and gluon orbital momentum and that partons with $x\lesssim10^{-3}$ are mostly unpolarized (i.e. their spin contribution turns out to be negligible), 
the uncertainties of the plots at $x\sim 10^{-3}$ precisely represent the room left by the EicC data to the quark and gluon orbital momentum contribution to the proton spin. Deviation from this picture will become apparent as soon the EIC will be able to precisely fix the very low-$x$ region, and in particular the $\Delta g$ distribution, and the EicC will constrain with unprecedented accuracy the remaining middle and high-$x$ range.

\subsection{Optimizing the PDFs with pseudo data and results}
\label{sec_optimize}

\begin{figure}[p]
\begin{center}
\subfigure[ The EV1 to $\Delta u$]{ \label{fig:opt_ev_1:a} 
\includegraphics[width=0.475\columnwidth]{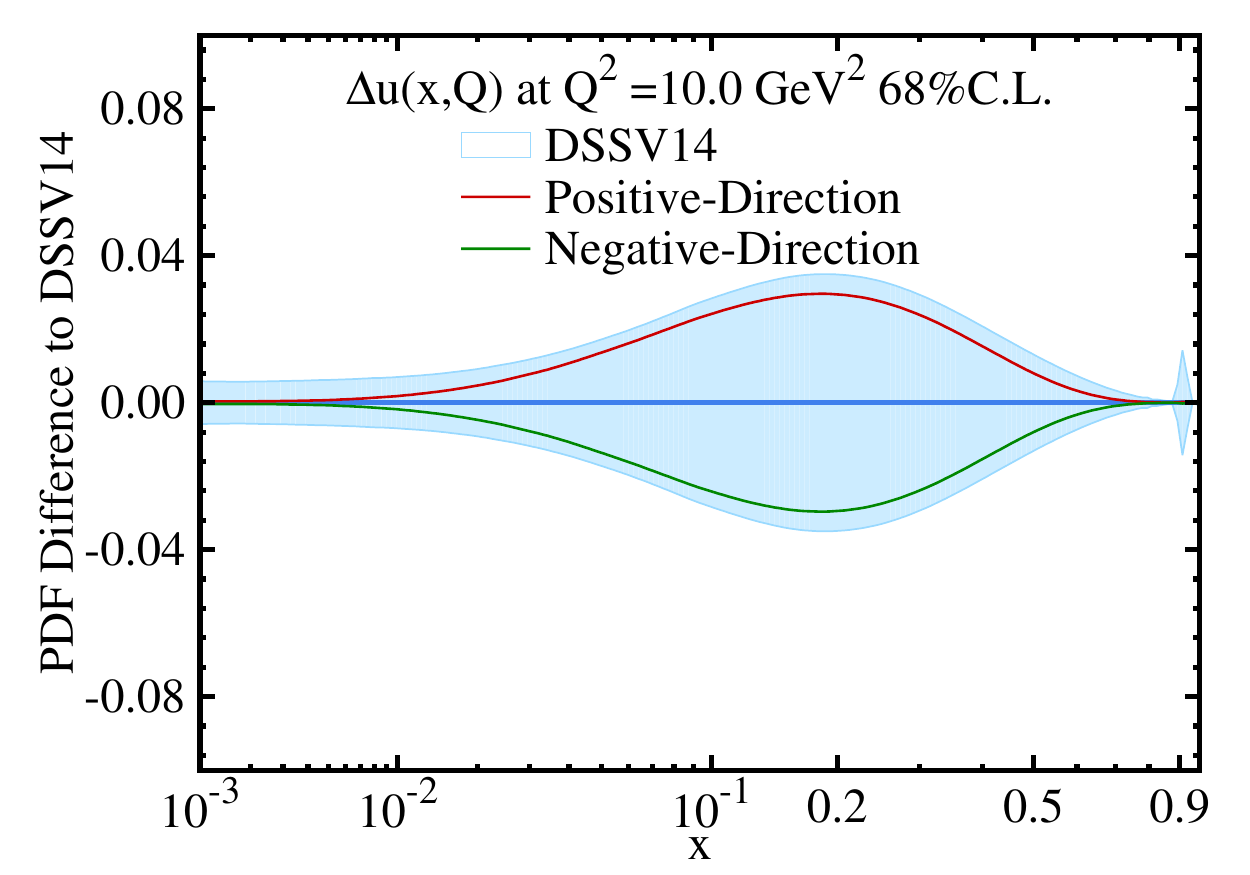}} 
\subfigure[ The EV1 to $\Delta d$]{ \label{fig:opt_ev_1:b} 
\includegraphics[width=0.475\columnwidth]{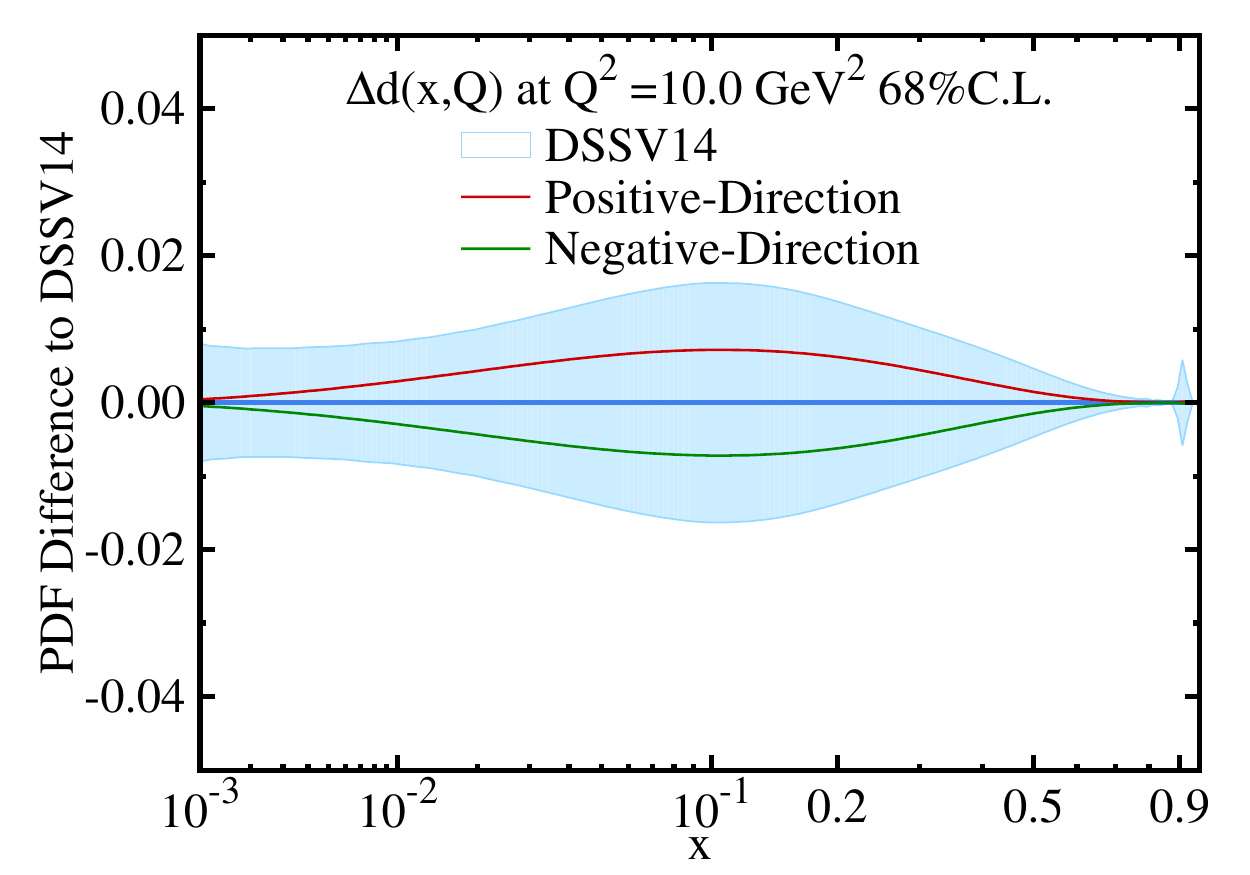}} 
\subfigure[ The EV2 to $\Delta u$]{ \label{fig:opt_ev_1:c} 
\includegraphics[width=0.475\columnwidth]{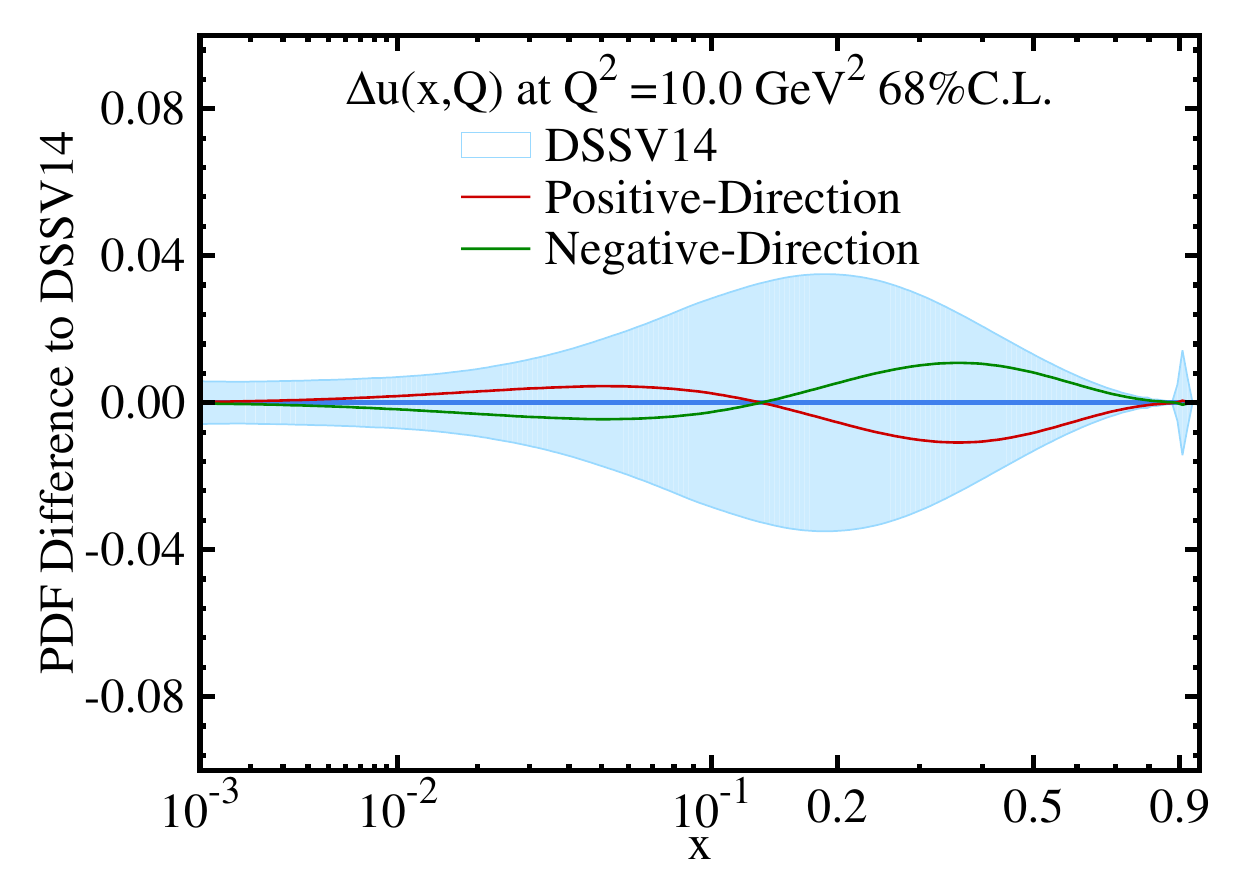}} 
\subfigure[ The EV3 to $\Delta d$]{ \label{fig:opt_ev_1:d} 
\includegraphics[width=0.475\columnwidth]{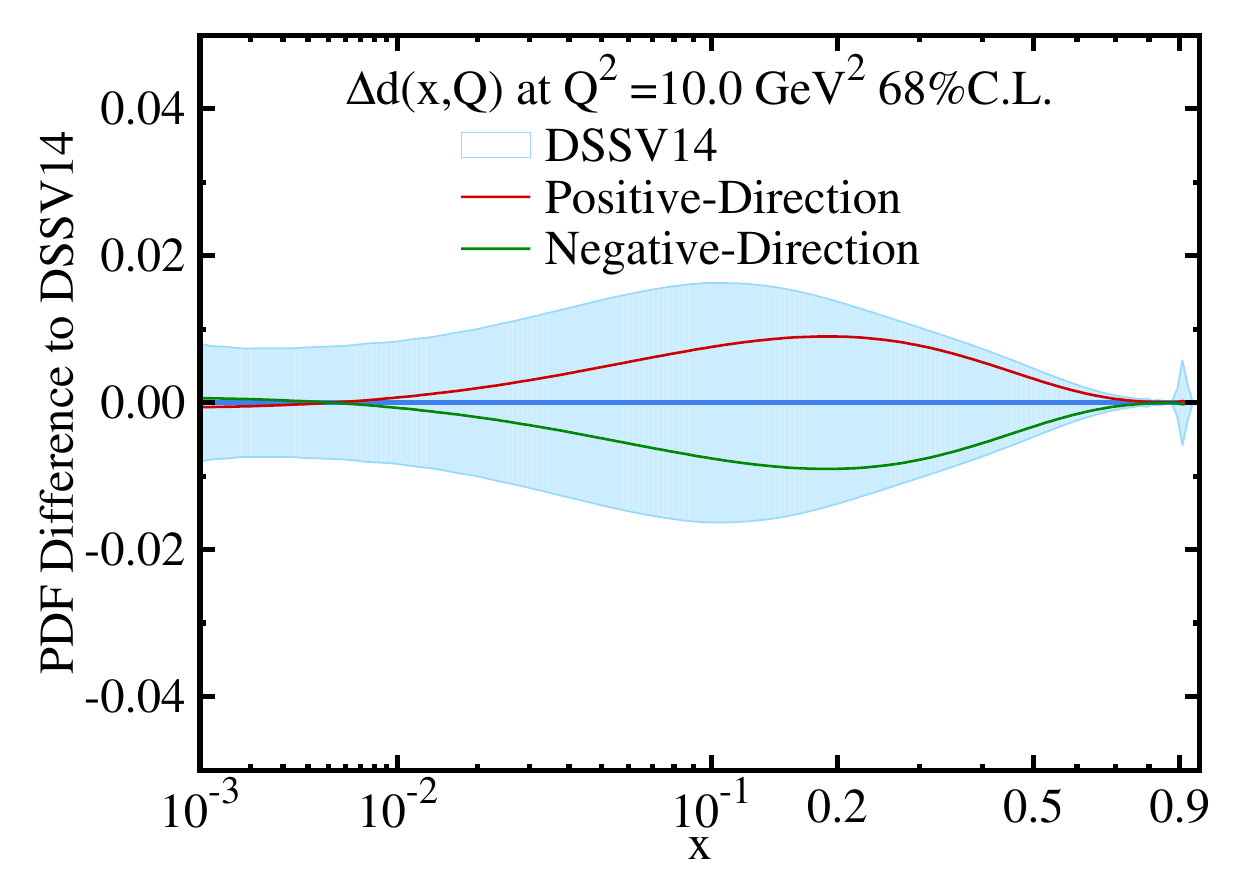}} 
\subfigure[ The EV4 to $\Delta s$]{ \label{fig:opt_ev_1:e} 
\includegraphics[width=0.475\columnwidth]{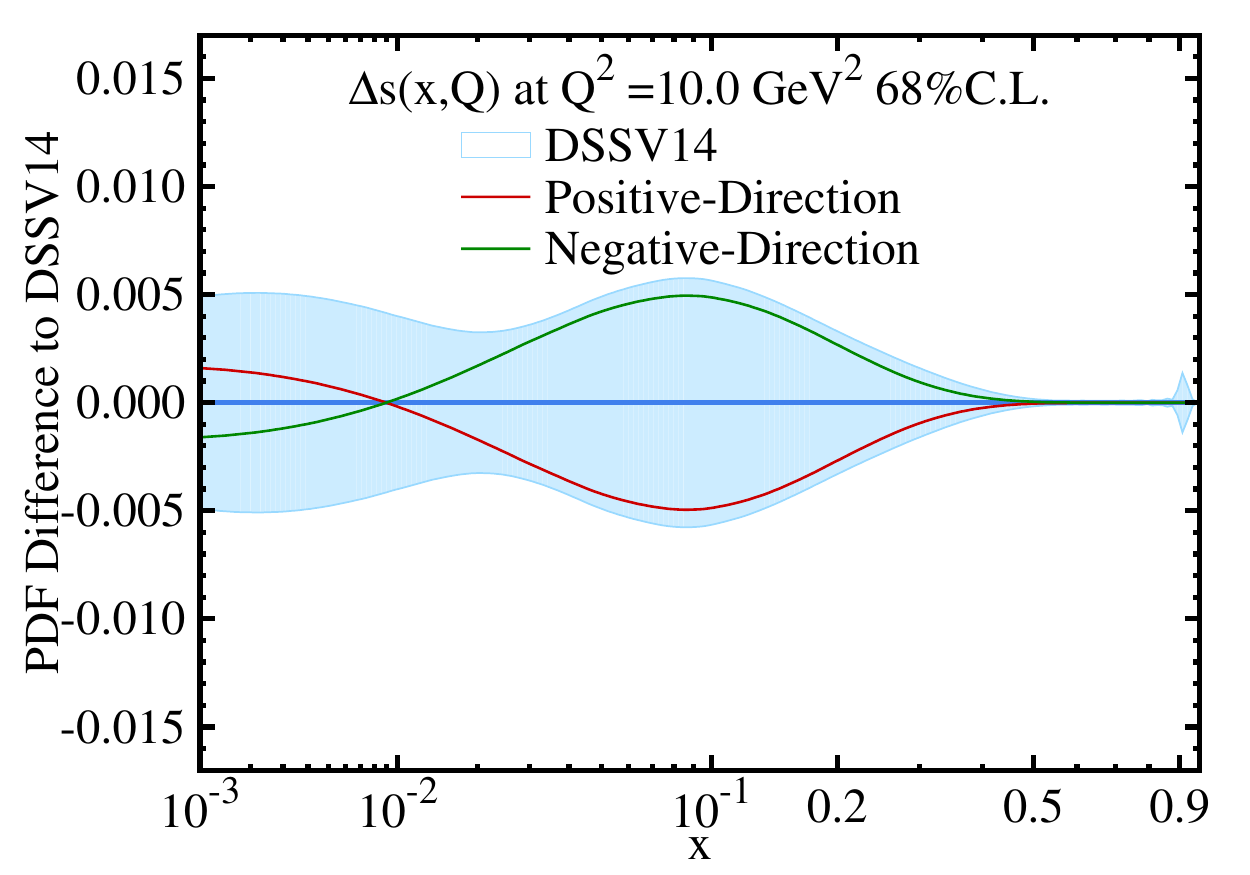}}
\subfigure[Optimization only with SIDIS Kaon data]{ \label{fig:opt_ev_1:f} 
\includegraphics[width=0.475\columnwidth]{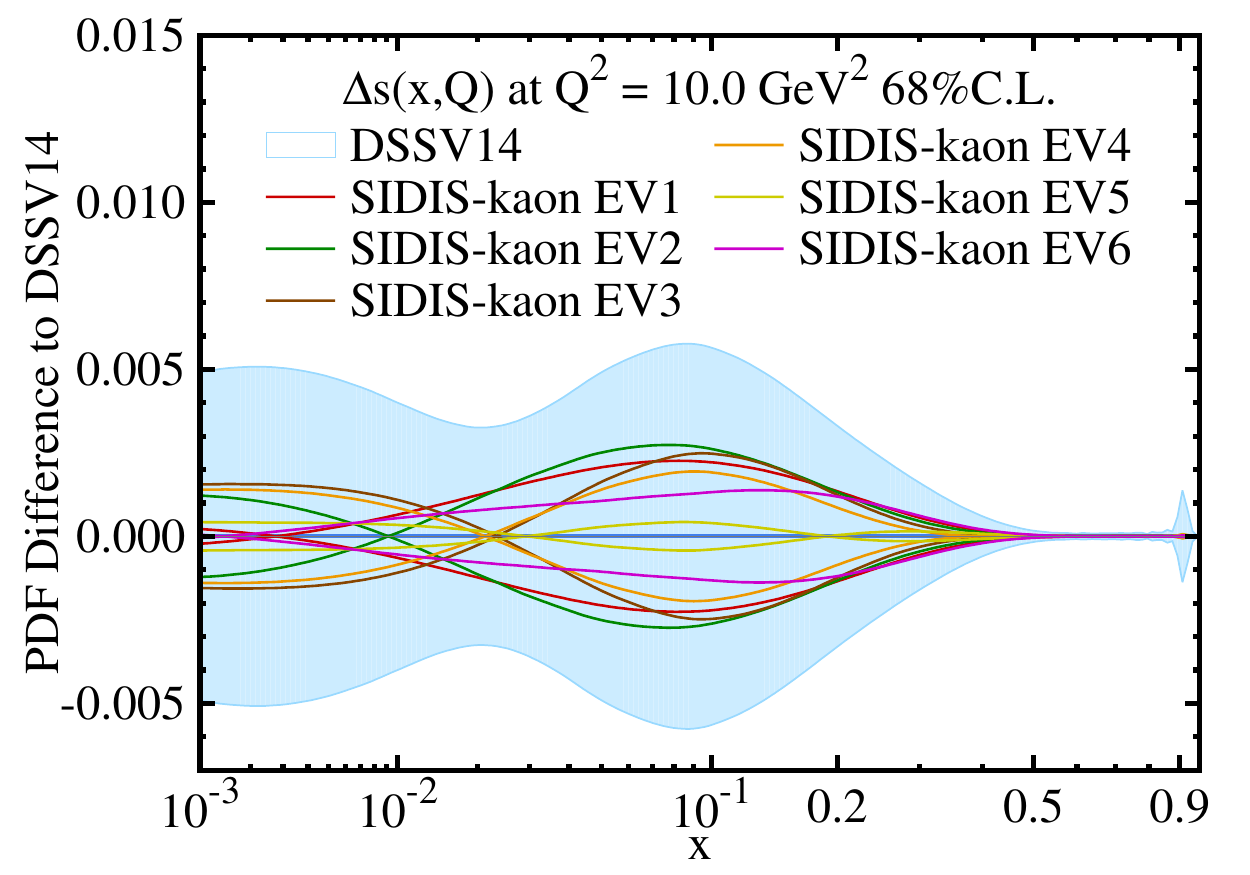}}
\caption{\label{fig:opt_ev_1}
In Fig. \ref{fig:opt_ev_1:a}-\ref{fig:opt_ev_1:e}, the first four pairs of the optimized eigenvector PDFs are compared to the central values of the DSSV14 PDFs at $Q^2 = 10.0$ GeV$^2$.
For each eigenvector pair only the most impacted flavours are shown. The positive and negative directions of the eigenvector pairs are shown with red and green lines respectively. The light-blue areas are the original DSSV14 error bands.
In Fig. \ref{fig:opt_ev_1:f}, only kaon data are considered in the \texttt{ePump} optimization.
}
\end{center}
\end{figure}

\begin{figure}[p]
\begin{center}
\subfigure[ The EV6 to $\Delta d$]{ \label{fig:opt_ev_2:b} 
\includegraphics[width=0.475\columnwidth]{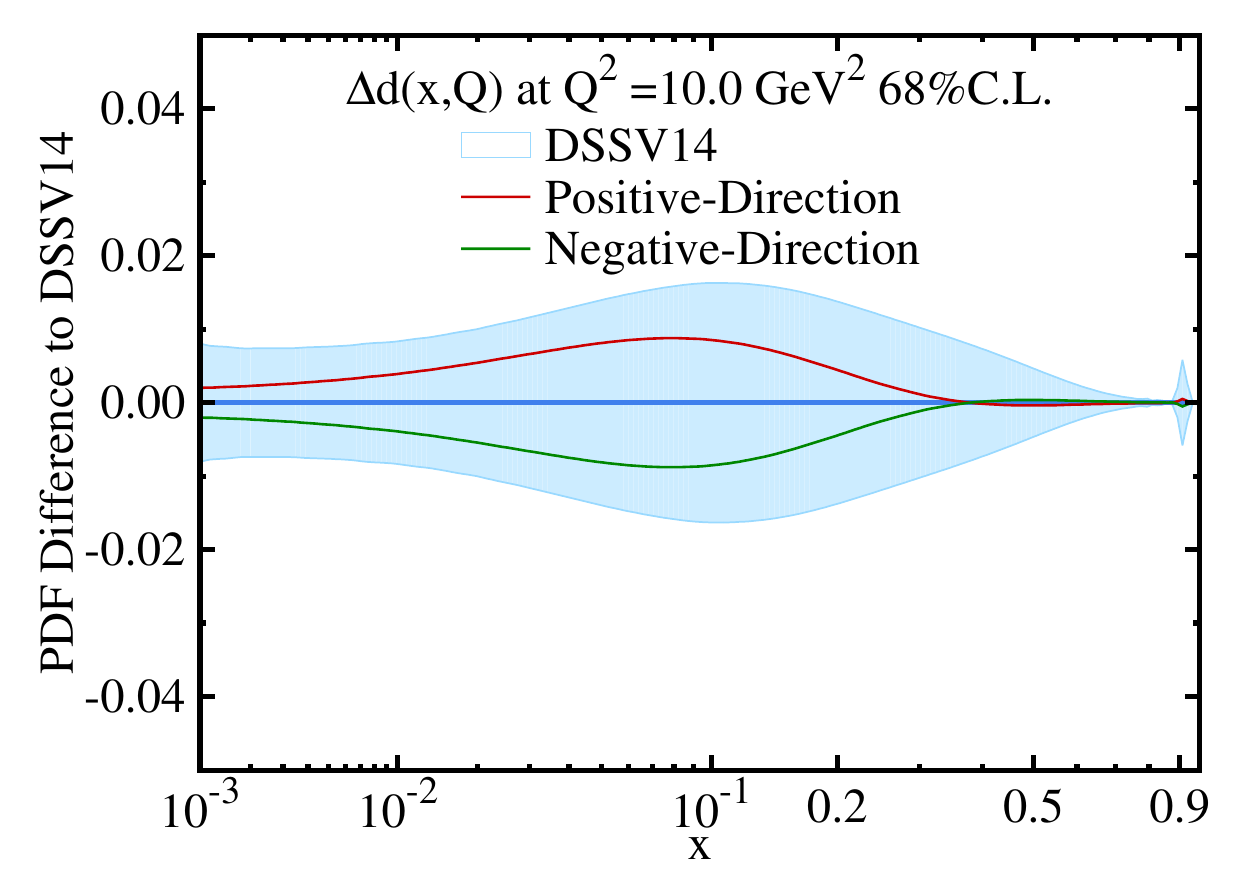}}
\subfigure[ The EV6 to $\Delta \bar{d}$]{ \label{fig:opt_ev_2:a}
\includegraphics[width=0.475\columnwidth]{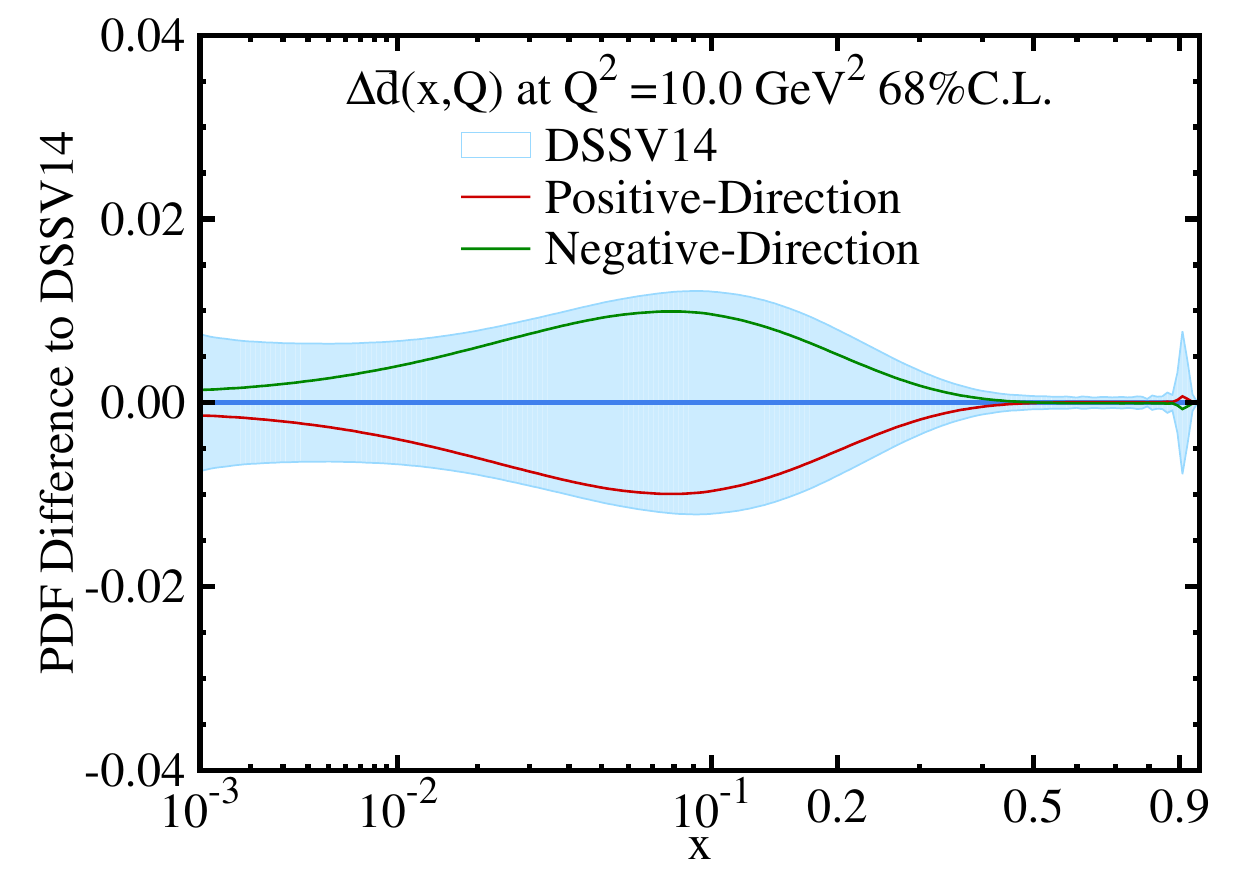}}
\subfigure[ The EV10 to $\Delta \bar{d}$]{ \label{fig:opt_ev_2:c}
	\includegraphics[width=0.475\columnwidth]{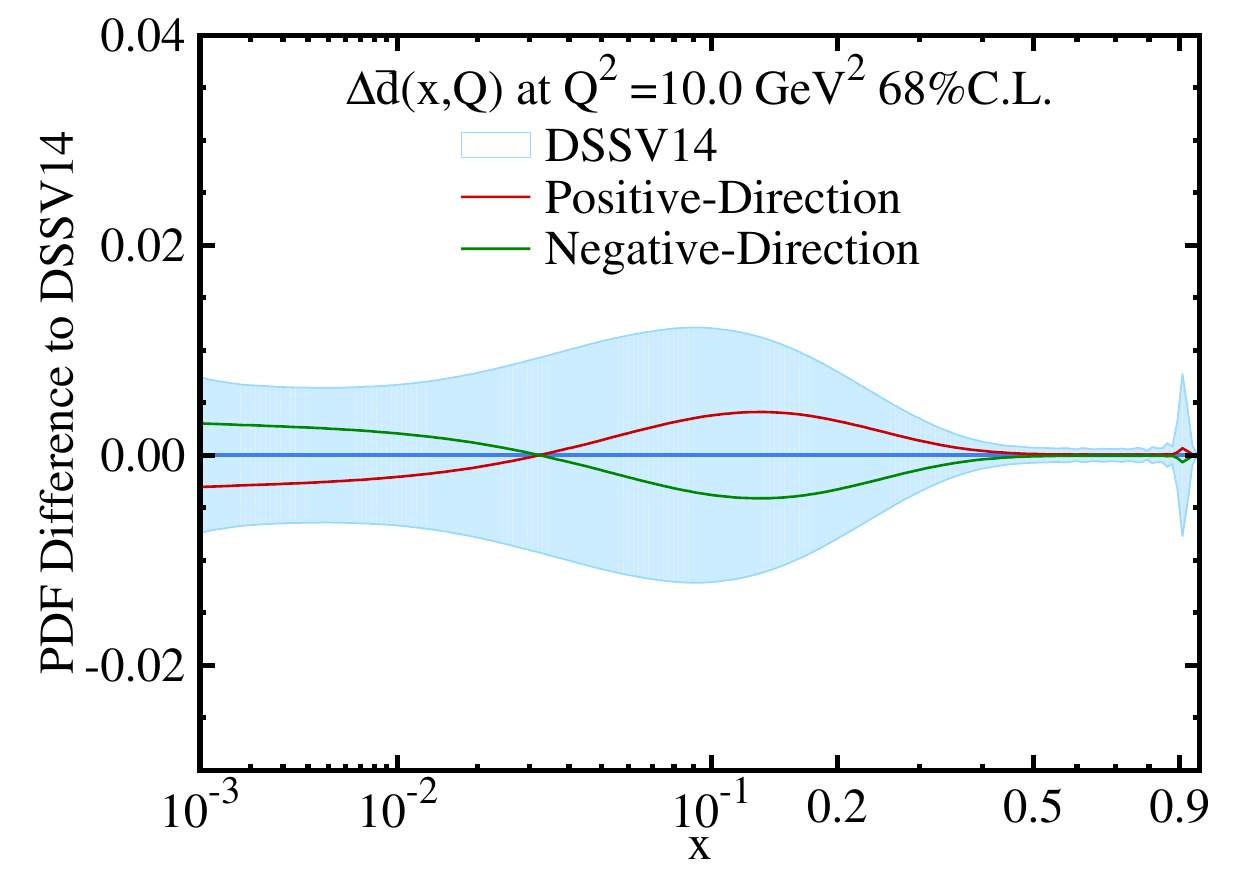}}
\subfigure[ The EV10 to $\Delta \bar{u}$]{ \label{fig:opt_ev_2:d}
\includegraphics[width=0.475\columnwidth]{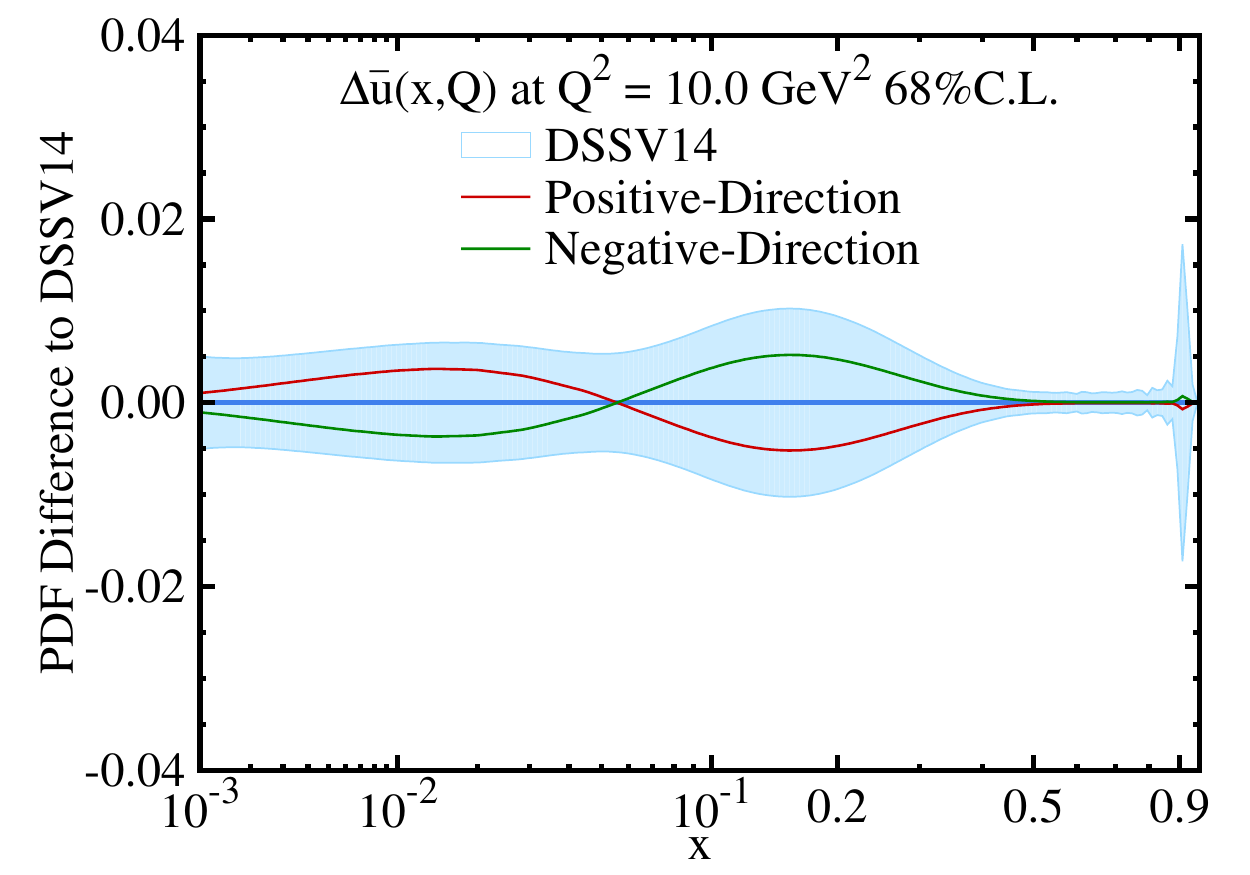}}
\subfigure[ The EV13 to $\Delta \bar{u}$]{ \label{fig:opt_ev_2:e}
\includegraphics[width=0.475\columnwidth]{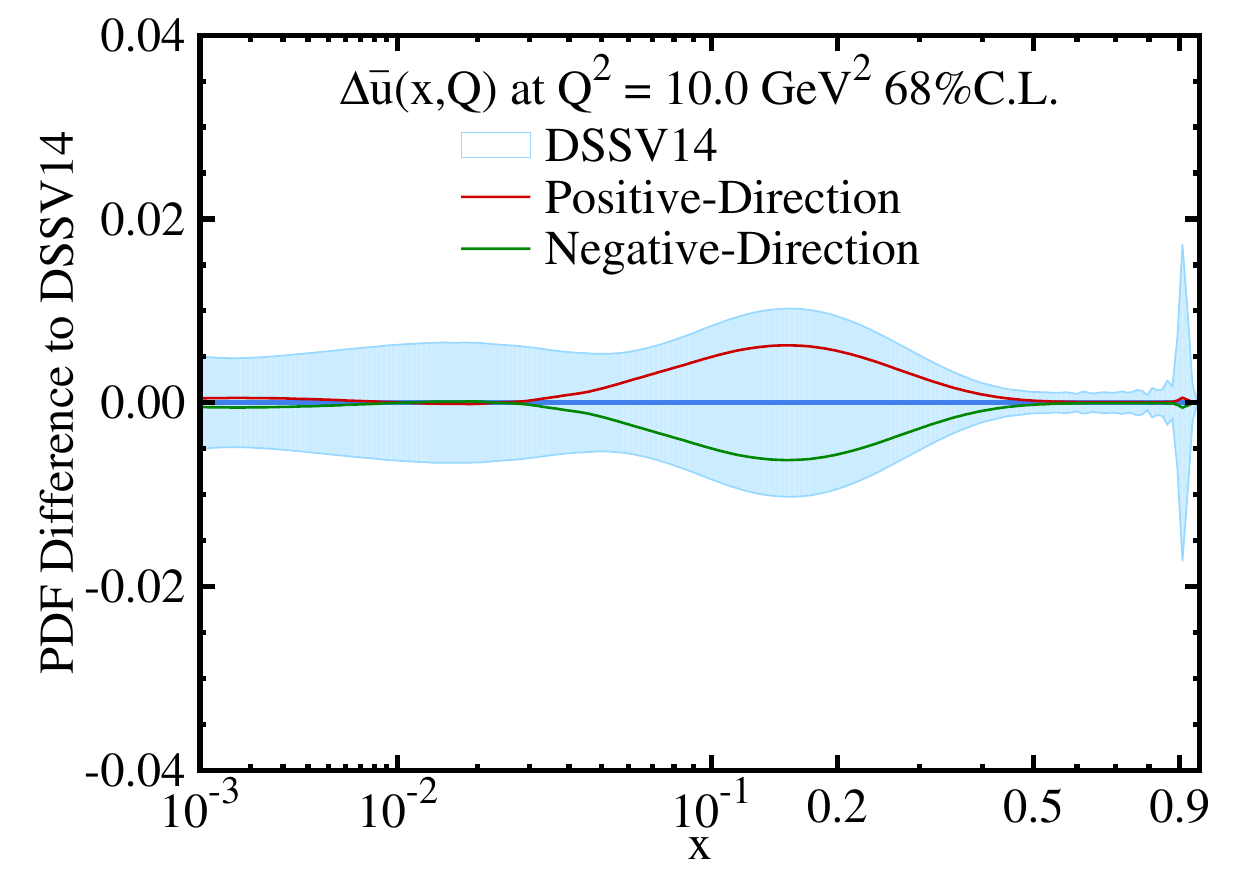}}
\subfigure[ The EV15 to $\Delta \bar{u}$]{ \label{fig:opt_ev_2:f}
\includegraphics[width=0.475\columnwidth]{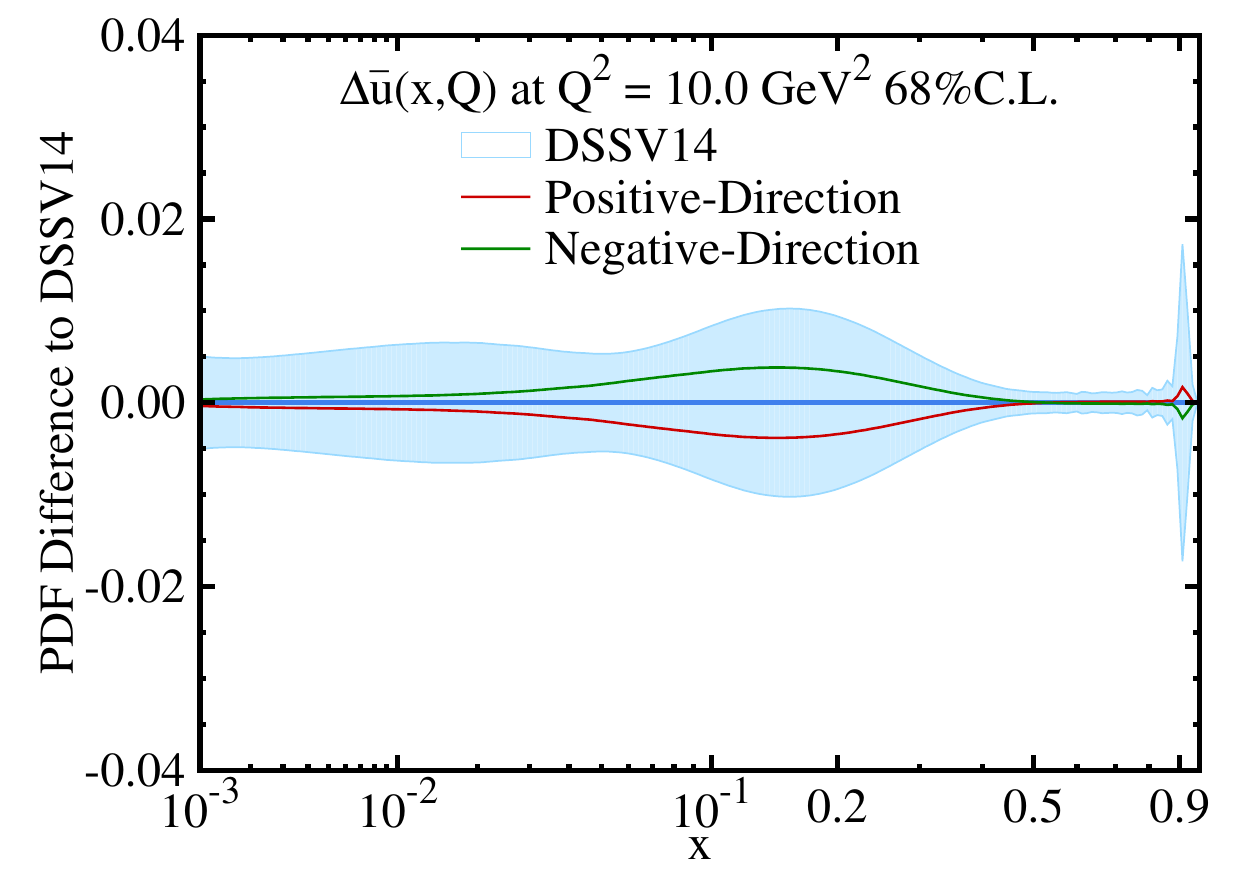}}
\caption{\label{fig:opt_ev_2}
The same as Fig. \ref{fig:opt_ev_1}, but of the sixth, tenth, thirteenth and fifteenth pairs of the optimized eigenvector PDFs.
}
\end{center}
\end{figure}

\begin{figure}[p]
\centering \subfigure[ The EV1 to SIDIS]{ \label{fig:FracCntrb1:a} 
\includegraphics[width=0.475\columnwidth]{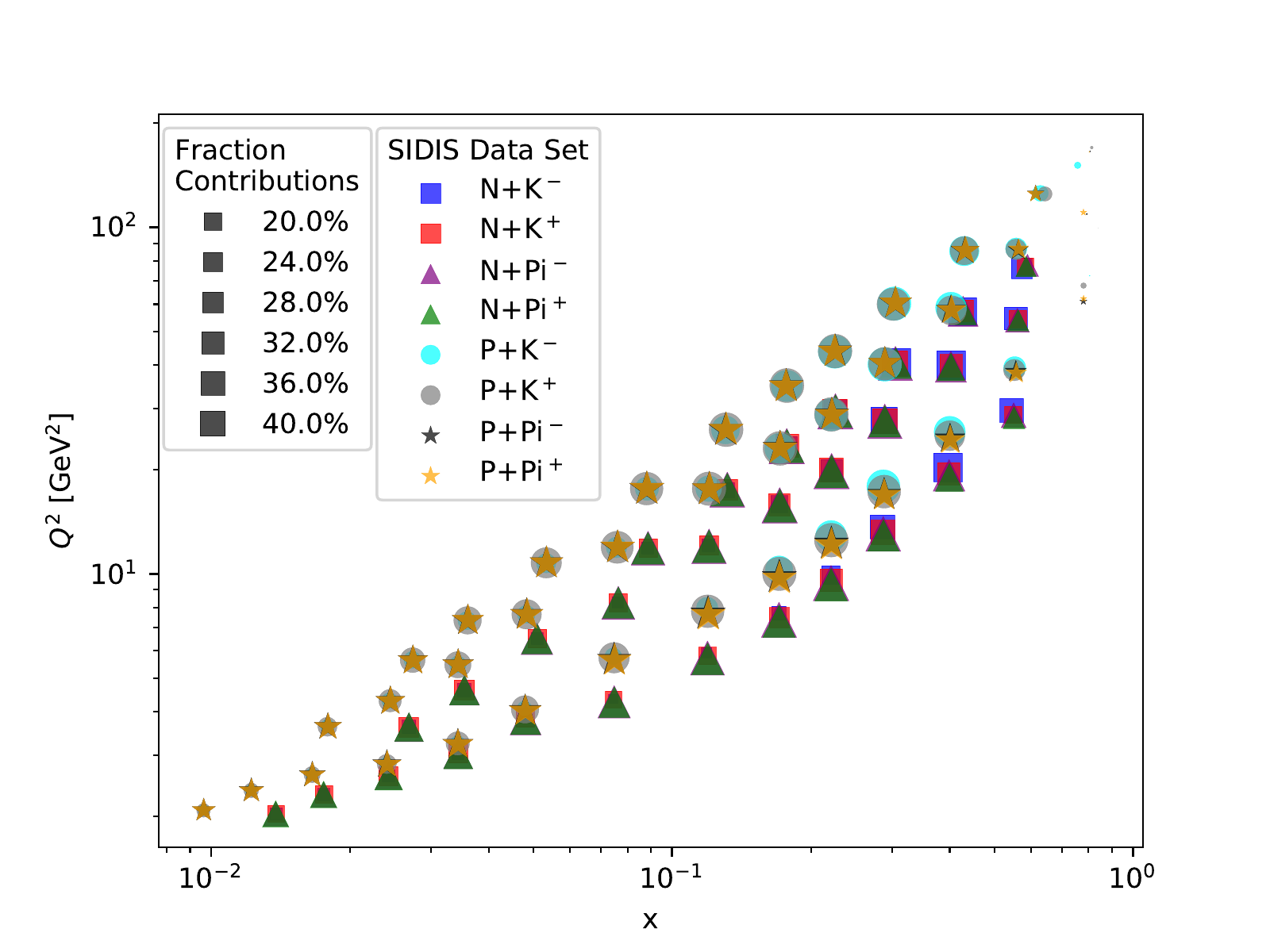}} 
\centering \subfigure[ The EV2 to SIDIS]{ \label{fig:FracCntrb1:b} 
\includegraphics[width=0.475\columnwidth]{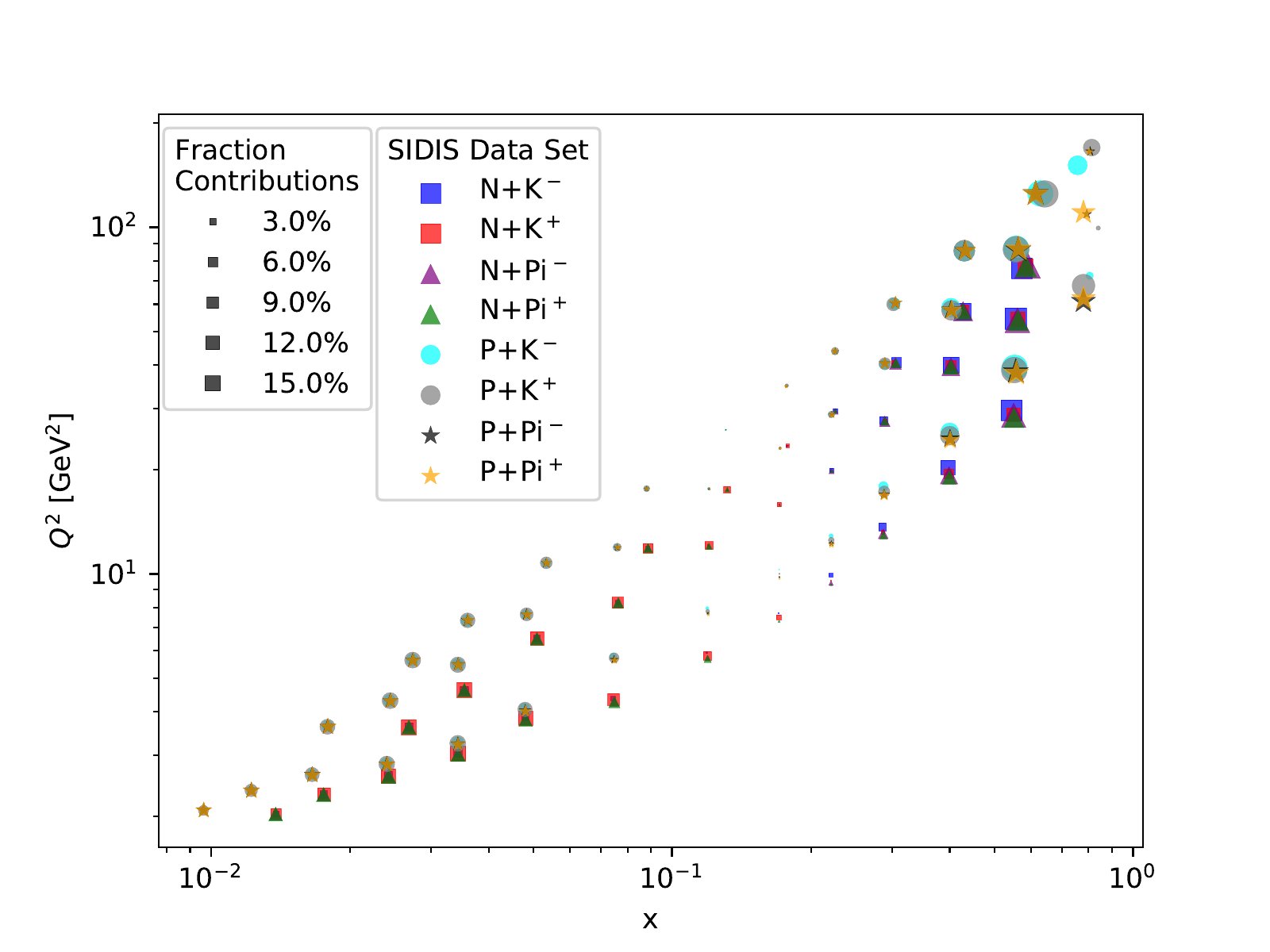}} 
\centering \subfigure[ The EV3 to SIDIS]{ \label{fig:FracCntrb1:c} 
\includegraphics[width=0.475\columnwidth]{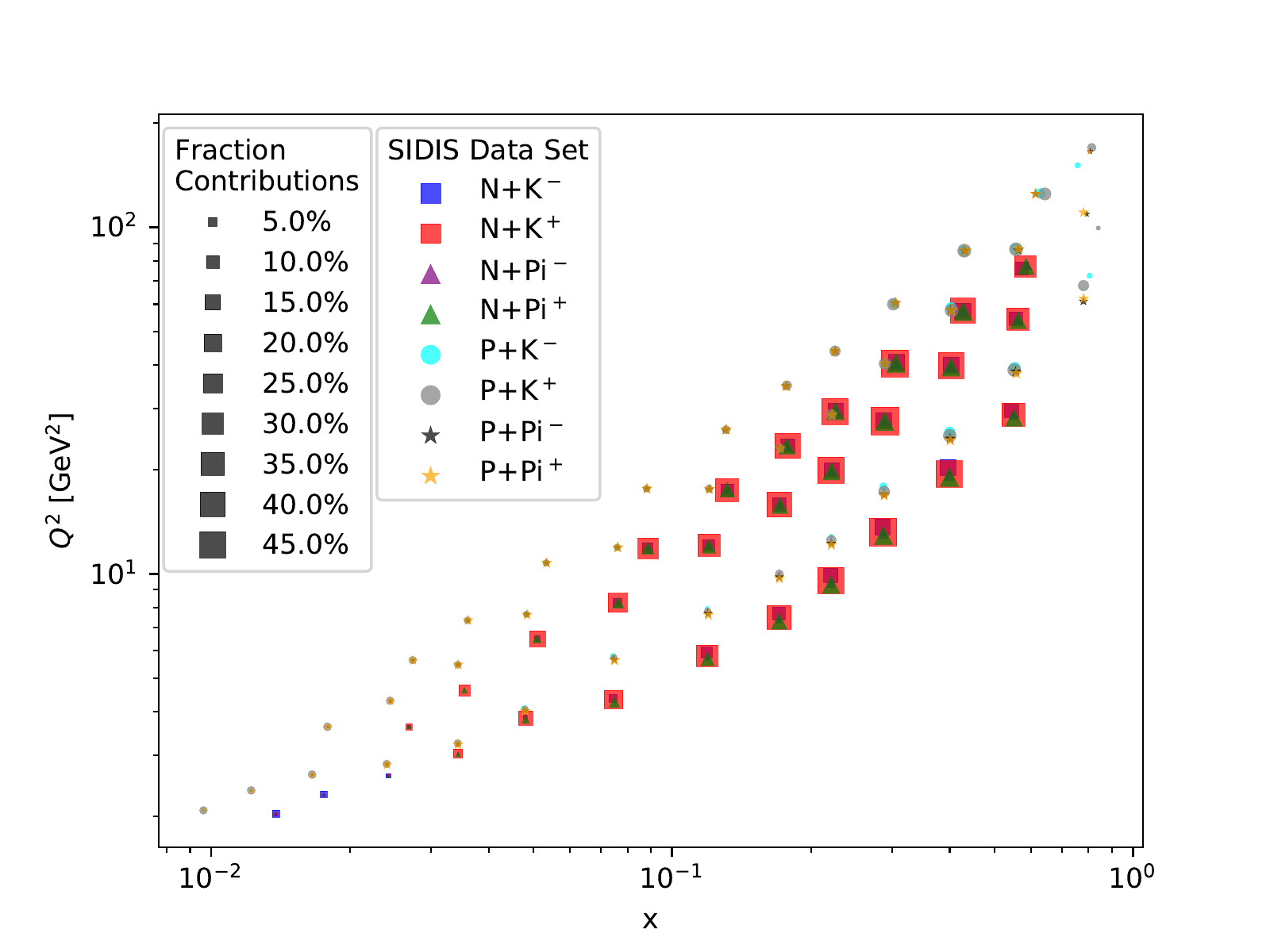}} 
\centering \subfigure[ The EV4 to SIDIS]{ \label{fig:FracCntrb1:d}
\includegraphics[width=0.475\columnwidth]{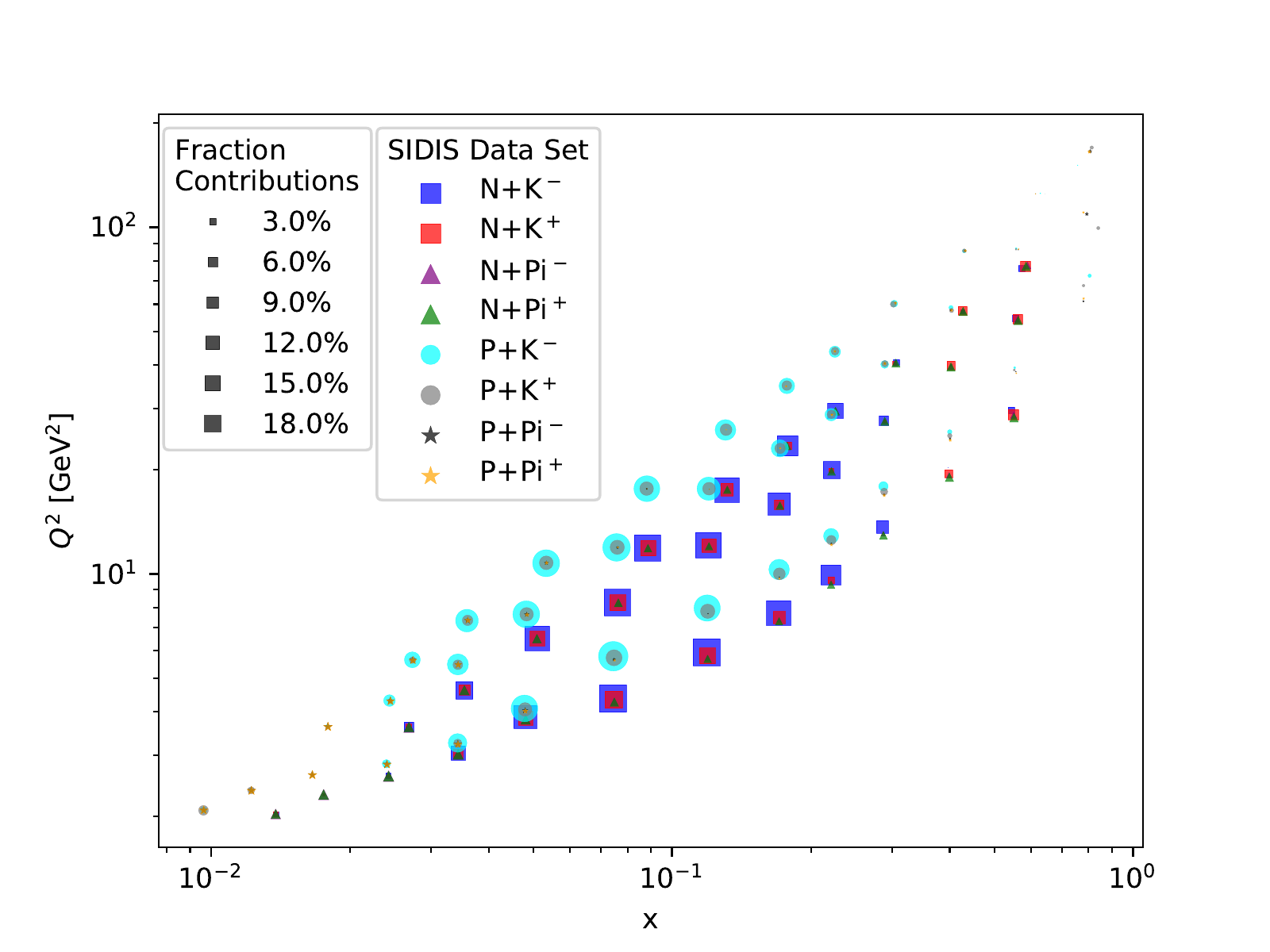}} 
\centering \subfigure[ The EV6 to SIDIS]{ \label{fig:FracCntrb1:e}
\includegraphics[width=0.475\columnwidth]{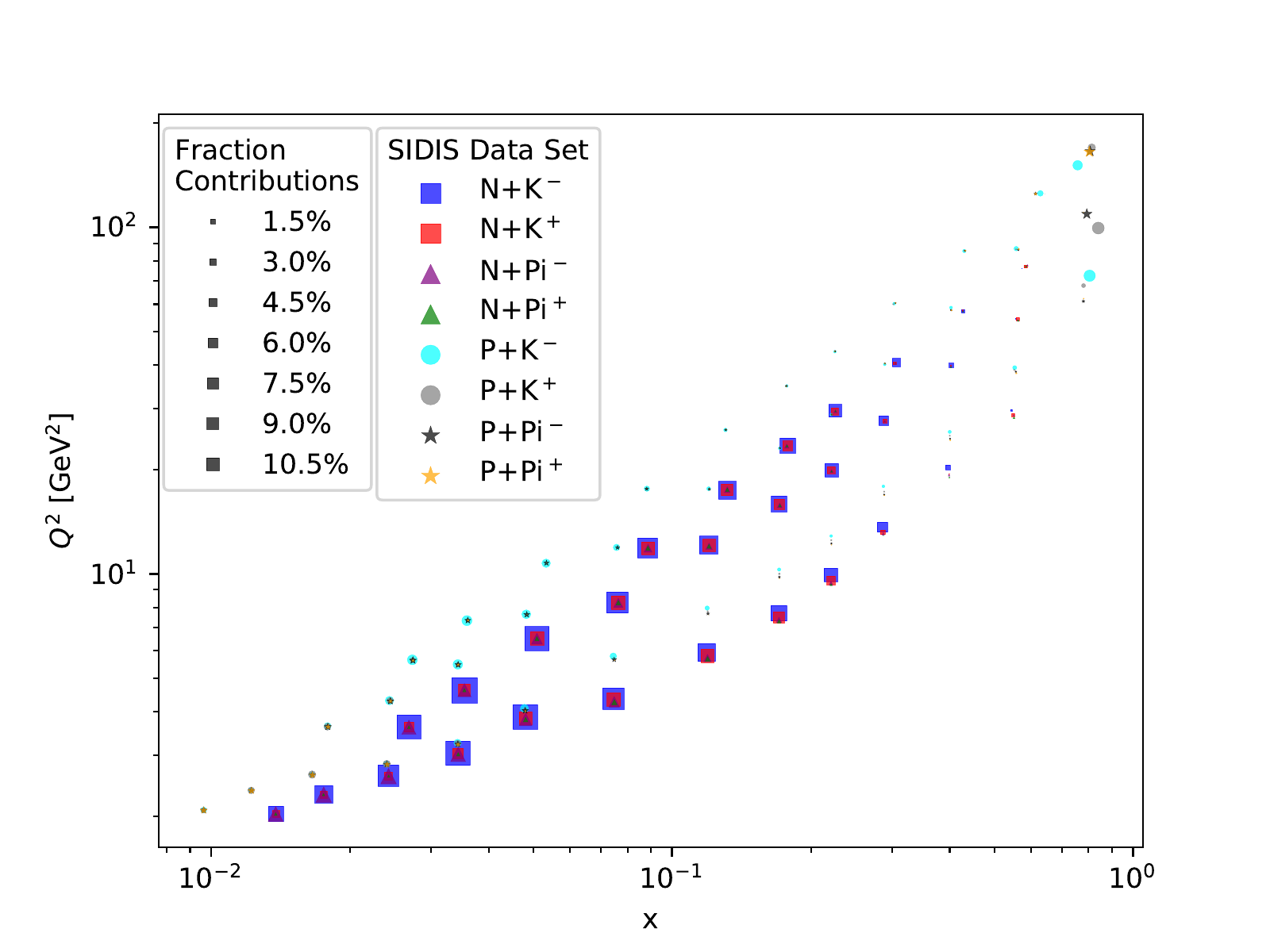}} 
\caption{\label{fig:FracCntrb1}
Fractional contributions of the first four and the sixth optimized eigenvector pairs to the PDF uncertainties of various SIDIS observables. The sizes of dots correspond to the relative contribution of optimized eigenvector pairs to these observables. In the legends of ``SIDIS Data Set'', the notation, ``N+K$^+$'' for example, stands for the pseudo-data set of the experimental observable $A_1$ for the neutron measurement, while the final hadron state is the K$^+$. The same rule applies to other data sets in the legends.
}
\end{figure}

\begin{figure}[t]
\centering \subfigure[ The EV10 to SIDIS]{ \label{fig:FracCntrb2:a} 
\includegraphics[width=0.475\columnwidth]{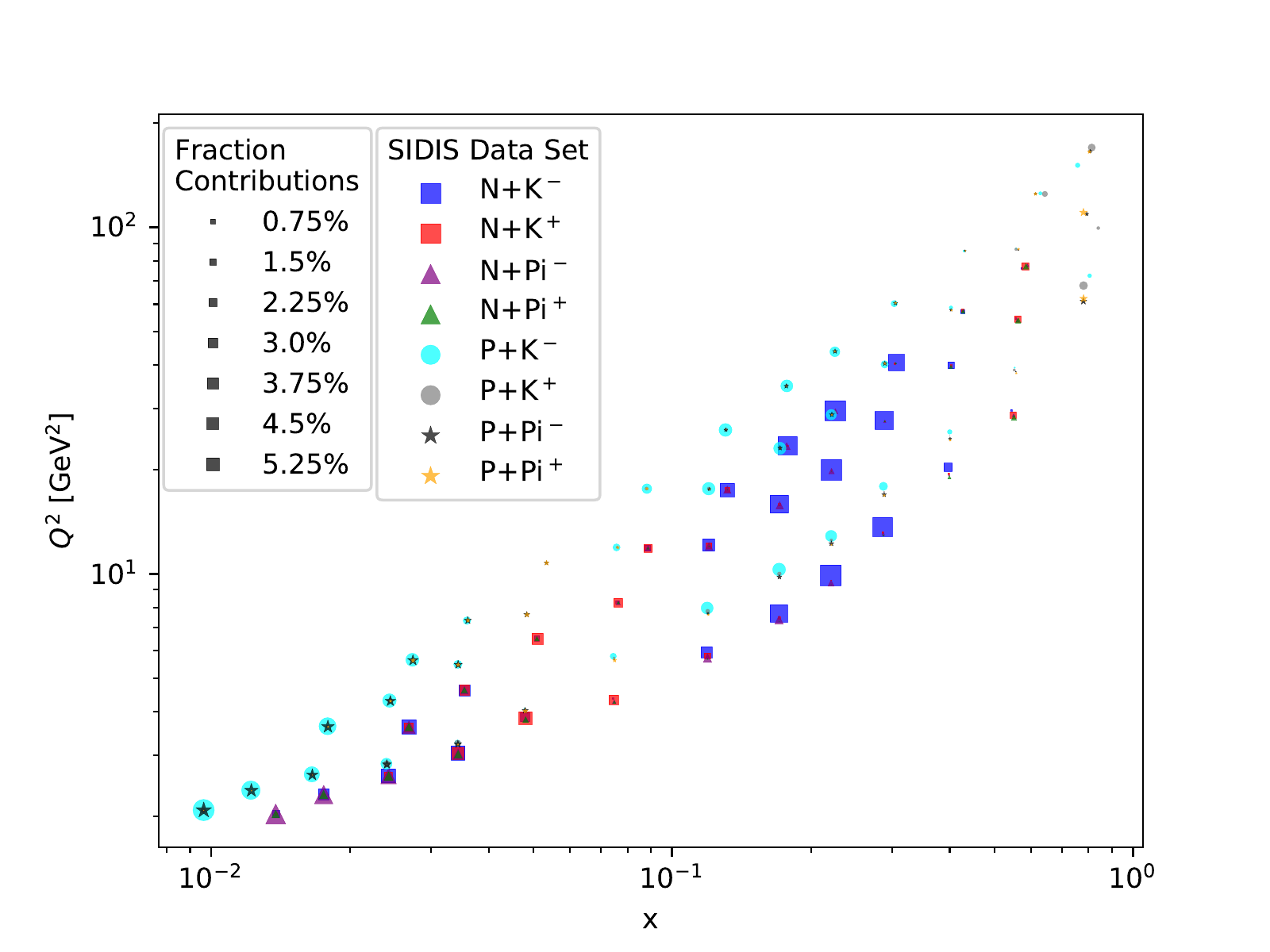}} 
\centering \subfigure[ The EV13 to SIDIS]{ \label{fig:FracCntrb2:b} 
\includegraphics[width=0.475\columnwidth]{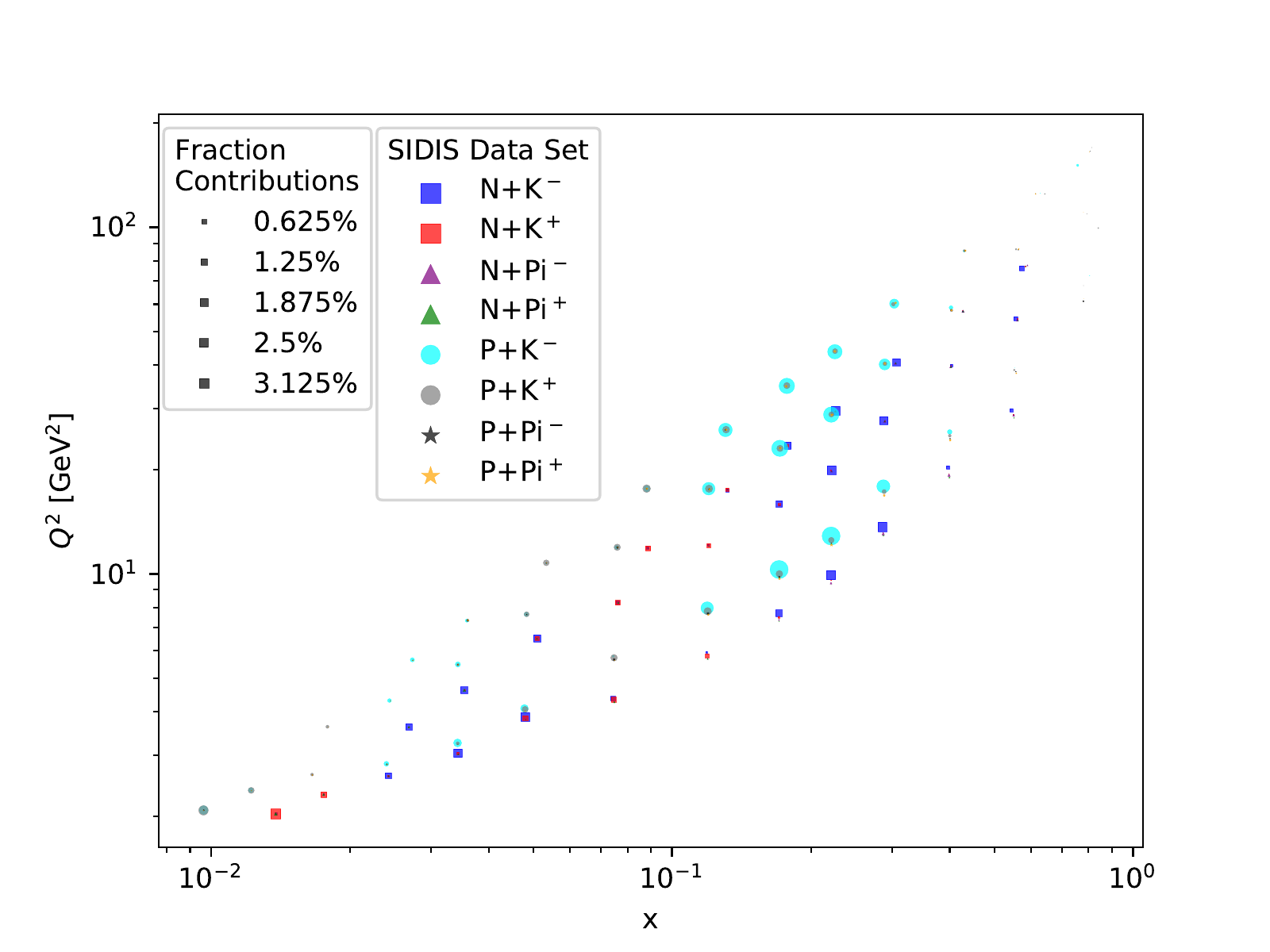}} 
\centering \subfigure[ The EV15 to SIDIS]{ \label{fig:FracCntrb2:c} 
\includegraphics[width=0.475\columnwidth]{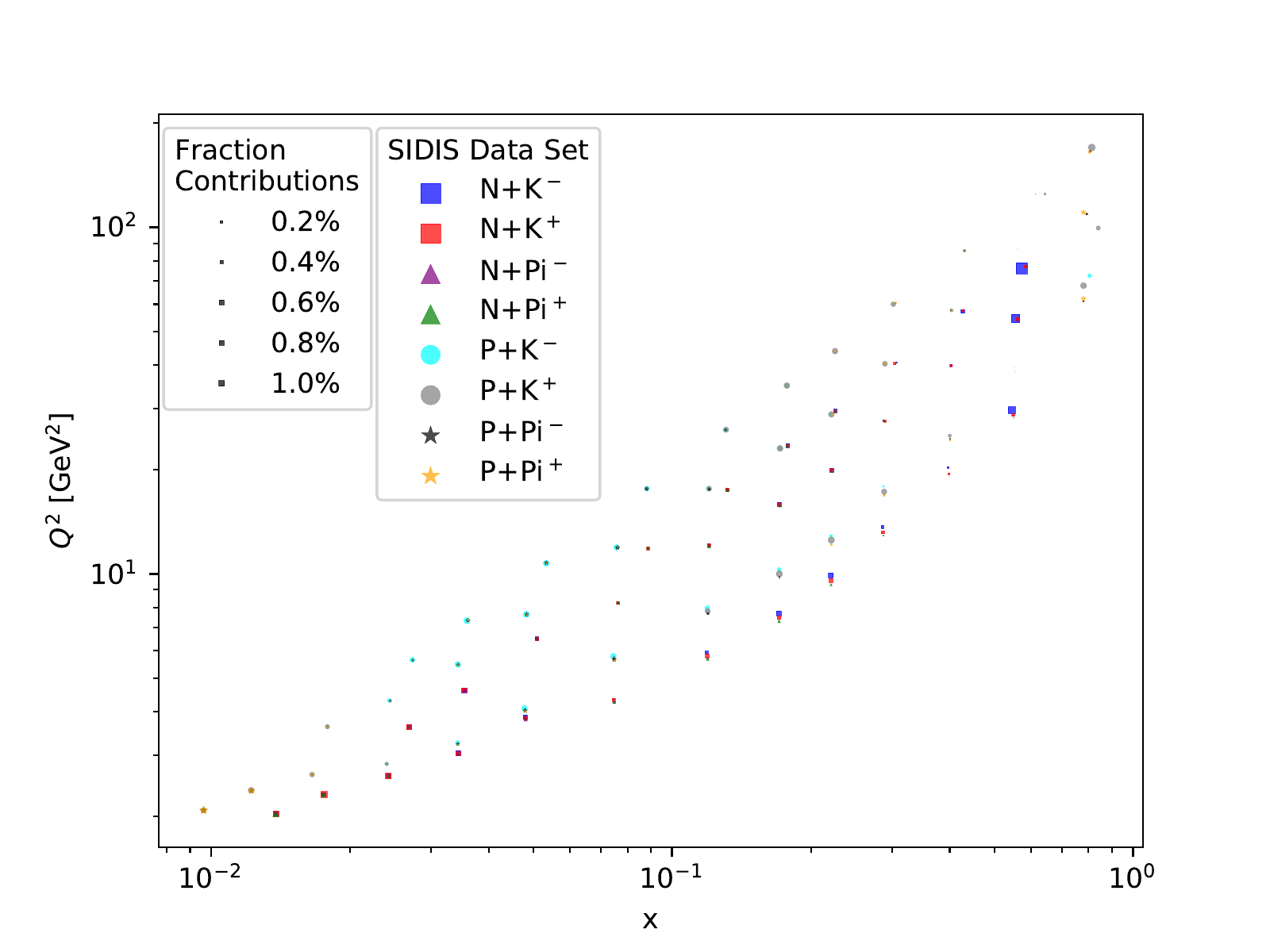}} 
\caption{\label{fig:FracCntrb2}
The same as the Fig. \ref{fig:FracCntrb1}, but of the tenth, thirteenth and fifteenth optimized eigenvectors pairs. The sizes of dots are scaled differently from the Fig. \ref{fig:FracCntrb1} for better visualization.
}
\end{figure}

To quantitatively analyze the sensitivities of individual pseudo-data set to constraining various parton flavour PDFs at certain $x$ ranges, and to demonstrate how these eight different data sets play complementary roles in reducing the PDF uncertainty in the PDF-updating procedure, we shall deploy the ePump-optimization (or PDF-rediagonization) method of the \eP code. As explained in Sec.~\ref{sec:hmo}, this application of \eP is 
based on ideas similar to that used in the data set diagonalization
method developed by Pumplin~\cite{Pumplin:2009nm}. It takes a set of Hessian error PDFs and constructs an equivalent set
of error PDFs that exactly reproduces the Hessian symmetric PDF uncertainties, but in addition, each new eigenvector pair has an eigenvalue that quantitatively describes its contribution to the PDF uncertainty of a given data set or sets. The new optimized error PDF pairs are ordered by their eigenvalues in a way that the first optimized error PDF pair possesses the largest eigenvalue while the successive error PDF pairs have smaller eigenvalues. This ordering of optimized error PDF pairs makes it easy to choose a reduced set that covers the PDF uncertainty for the data set to any desired accuracy~\cite{Schmidt:2018hvu,Hou:2019gfw}.
Higher accuracy corresponds to a choice of eigenvector pairs for which the value of the sum over their eigenvalues is closer to the total number of new data points. The contributions of new eigenvector pairs to the uncertainties of PDFs, for a given parton flavor, provide useful information of how the new data points are sensitive to PDFs, at various $x$ ranges.

By applying the \texttt{ePump} optimization method to DSSV14H PDFs with the combination of both the EicC DIS and SIDIS pseudo-data sets, which contains 332 data points totally, we found that the first six eigenvector pairs (out of the total 52 eigenvector pairs) play the most important roles in constructing the total PDF error bands. Their eigenvalues are 167.8, 38.7, 28.3,  21.5, 16.2, and 11.3 respectively. Totally these six eigenvector pairs provide 85.5\% of the total PDF error bands, while the first fifteen eigenvector pairs cover 99.1\% of the total error bands. The successive eigenvector pairs have even smaller eigenvalues. Hence, one could ignore the successive eigenvector pairs and reproduce the major part of DSSV14 PDFs uncertainties with only the first fifteen eigenvector pairs, instead of all of the $N_{eig} = 52$ eigenvector pairs or all of the $N_{rep} = 1000$ replicas.

\begin{table}[]
    \centering
    \begin{tabular}{c|cc|c}
    flavour & EV sets & $x$ range & SIDIS obs. \\ \hline\hline
    $\Delta u$ & EV1 & all & all \\
     & EV2 & $x > 0.4$ & all \\ \hline
    $\Delta d$ & EV1 & all & all \\
     & EV3 & $x > 0.1$ & N+K$^+$ \\
     & EV6 & $x < 0.1$ & N+K$^+$ \\ \hline
    $\Delta s$ & EV4 & $0.03 \sim 0.3$ & P/N+K$^-$ \\ \hline
    $\Delta \bar{u}$ & EV10 & $10^{-3} \sim 0.04$ \& $0.06 \sim 0.3$ & P+K$^-$ \\
     & & & P+$\pi^{-}$ \\
     & EV13 & $x > 0.03$ & P+K$^-$ \\
     & EV15 & $x > 0.04$ & N+K$^-$ \\ \hline
    $\Delta \bar{d}$ & EV6 & all & N+K$^-$ \\
     & EV10 & $10^{-3} \sim 0.02$ \& $0.05 \sim 0.3$ & N+K$^-$ \\
     & & & N+$\pi^{-}$
    \end{tabular}
    \caption{The leading 
    eigenvector pairs (EV sets), after the ePump-optimization, contributing to the PDF error band of each flavour,  and the SIDIS EicC pseudo-data sets (SIDIS obs.) which provide leading constraints on the specific eigenvector pair PDFs. Note that the EV6 and EV10 are sensitive to multiple flavours, therefore, when presenting the constraints of pseudo-data sets onto the EV6 and EV10, we arrange pseudo-data sets according to their flavour contents. The meaning of the notations, such as ``N+K$^+$", in the last column is the same as those in the Fig. \ref{fig:FracCntrb1}.
}
    \label{tab:flavour_ev_obs}
\end{table}

These fifteen optimized eigenvector pairs also reveal how the pseudo-data points are sensitive to different parton flavour PDFs at certain $x$ ranges. Such information can be obtained in a two-fold way. Firstly, the optimized eigenvector pairs dominating the contributions to the uncertainties of each flavour are identified. The biggest contributions of optimized eigenvector pairs to DSSV14 PDFs are shown in Figs. \ref{fig:opt_ev_1} and \ref{fig:opt_ev_2}. Secondly, the sensitivity of new data points to the optimized eigenvector pairs is assessed by how much the optimized eigenvector pairs would contribute to data points. The thickness of the dots in the $x-Q^2$ plots of Figs.~\ref{fig:FracCntrb1} and~\ref{fig:FracCntrb2} shows the fractional contributions of optimized eigenvector pairs, which are dominating DSSV14 uncertainties, to the PDF uncertainties of various EicC SIDIS pseudo-data points. The summary of this analysis is given in Table \ref{tab:flavour_ev_obs}, while our major physical findings in this section are as follows: 

\begin{itemize}
    \item The majority of EicC SIDIS pseudo-data points are essential to improve the DSSV14 $\Delta u$ distribution. Fig. \ref{fig:opt_ev_1:a} shows that the first optimized eigenvector pair (EV1) dominates the $\Delta u$ error band, while Fig. \ref{fig:FracCntrb1:a} clearly shows that EV1 contributes largely to most of SIDIS pseudo-data points. Therefore it becomes apparent that EicC SIDIS pseudo-data is pretty sensitive to the $\Delta u$ distribution.
    
    \item The EV1 also contributes to the $\Delta d$ error band. But as shown in  Fig. \ref{fig:FracCntrb1:b} it is not as dominant as it is for the $\Delta u$. The fact that the absolute value of the down quark charge is half of that of the up quark results in this difference. The majority of SIDIS pseudo-data points have the power of constraining the $\Delta u$ and the $\Delta d$ simultaneously.
    
    \item We expect that the $\Delta d$ distribution will be particularly constrained by the future EicC Neutron+K$^+$ data. As shown in Fig. \ref{fig:opt_ev_1:d}, the third optimized eigenvector pair (EV3) largely covers the $\Delta d$ error band for $x > 0.1$. We also notice in Fig. \ref{fig:FracCntrb1:c} that the Neutron+K$^+$ pseudo-data points receive large fractional contributions from EV3 for the same $x > 0.1$ region, hence indicating that the Neutron+K$^+$ pseudo-data is sensitive to the $\Delta d$. Given the flavour content of the K$^+$ meson, this is to be expected as the Neutron+K$^+$ data should be able to probe the $\Delta u$ distribution inside the neutron, which, due to the isospin symmetry, corresponds to the $\Delta d$ distribution inside the proton. 
    \item Both EicC Proton and Neutron kaon SIDIS data will be important for constraining the $\Delta s$. In Fig. \ref{fig:opt_ev_1:e}, we observe that the fourth optimized eigenvector pair (EV4) dominates the $\Delta s$ error band. Fig. \ref{fig:FracCntrb1:d} shows that EV4 is particularly sensitive to both Proton and Neutron+K$^-$ pseudo-data sets. This is consistent with the quark model picture, where the K$^-$ meson is considered to be composed by $s$ and $\bar{u}$ quarks.
    
    \item 
In the naive parton model picture, as discussed in Appendix A, one could easily conclude that the kaon data must play a decisive role in determining $\Delta s$. To check on this, we show in 
Fig. \ref{fig:opt_ev_1:f} the result of 
another \texttt{ePump} optimization study in which 
only SIDIS Kaon pseudo-data are considered. 
Unexpectedly, 
there is no single eigenvector pair dominating the PDF error band of $\Delta s$ when only the SIDIS kaon pseudo-data sets are included in ePump-optimization. By the nature of the Hessian profiling method, the eigenvectors are orthogonal to each other. 
This implies that those SIDIS kaon pseudo-data sets are providing information about  $\Delta s$ at different $x$ values. On the contrary, Fig. \ref{fig:opt_ev_1:e} shows that 
the eigenvector pair EV4 dominates the constraint on the error band of $\Delta s$ when all the DIS and SIDIS  (pion and kaon) pseudo-data are included.
Hence, there must be some other pseudo-data sets that provide an additional constraint on $\Delta s$ via some underlying correlation present in the original DSSV14 PDFs.
Since the theoretical predictions used in this study are generated with DSSV14 PDFs, it is possible that the underlying correlation comes from the original setting of DSSV14 PDFs. The identity of Eq.~(\ref{eq:su2_su3_1}) implies a correlation between $\Delta s$ and  
$\Delta {u}$, $\Delta \bar{u}$, 
$\Delta {d}$ and $\Delta \bar{d}$ imposed in the construction of DSSV14 PDFs, such that the pseudo-data sets sensitive to  $\Delta {u}$, $\Delta \bar{u}$,
$\Delta {d}$ and $\Delta \bar{d}$ are also providing information on constraining $\Delta s$. This explains why adding those non-kaon data can further constrain $\Delta s$ when using the DSSV14 PDFs.

   \item 
Fig. \ref{fig:opt_ev_2}(d)-(f) 
indicate that the 8 SIDIS pseudo-data will constrain $\Delta \bar{u}$ in different ranges of $x$, and it takes mainly three eigenvector sets (EV10, EV13 and EV15) to represent the error band of $\Delta \bar{u}$ PDF in DSSV14.
Furthermore, Fig.~\ref{fig:FracCntrb2} shows that the leading data sets that contribute to the eigenvector sets EV10, EV13 and EV15 are the kaon data.   
This is the kind of information that can not be read out from Fig. \ref{fig:updated_dist}, directly. 
Although one could perform ePump-updating by adding only one pseudo-data set at a time to study the impact from each individual data set, one could use ePump-optimization to quickly gain information about the {\it complimentary role} that each data set plays in constraining a certain flavour PDF at a given $x$ region after ePump-updating. 
The SIDIS EicC pseudo-data sets which provide leading constraints on the specific eigenvector pair PDFs can be read out from Table \ref{tab:flavour_ev_obs}.

    \item 
    For $\Delta \bar{d}$, the EicC SIDIS Neutron {K$^-$ or $\pi^-$} data will be important. In Fig. \ref{fig:opt_ev_2}, the uncertainty band of $\Delta \bar{d}$, as a function of $x$, exhibits sensitivity to EV6 and EV10. At the same time,  Fig. \ref{fig:FracCntrb1:e} shows that Neutron+K$^-$ pseudo-data provide the leading constraint on EV6 for $x < 0.3$, and Fig. \ref{fig:FracCntrb2:a} shows that both Neutron+K$^-$ and Neutron+$\pi^-$ pseudo-data also constrain EV10. Hence the $\Delta \bar{d}$ distribution is mostly constrained by the Neutron+K$^-$ data, while the Neutron+$\pi^-$ data also provide information on $\Delta \bar{d}$ with $x < 0.03$.

    \item As for the $\Delta g$, none of these fifteen optimized eigenvector pairs provide a large proportion to the error band. This is expected as the EicC SIDIS is a machine better suited to investigate the ``sea-quark'' sector rather than exploring the $\Delta g$ distribution,  which dominates the small-$x$ region and can be effectively probed at the EIC~\cite{EIC}.

\end{itemize}

In short summary, by employing the ePump-optimization procedure, we have explored the  
complementary role played by different
data sets in reducing the PDF uncertainty in the PDF-updating procedure.
In Figs.~\ref{fig:opt_ev_1} and \ref{fig:opt_ev_2}, we show the contributions provided by the leading pairs of eigenvector PDFs to the PDF error bands of various flavours, after performing ePump-optimization with the inclusion of the pseudo-data sets considered in this study.
Through this, we have identified which EV sets
dominantly constrain the PDF error bands of a given flavour. 
Furthermore, in Figs.~\ref{fig:FracCntrb1} and \ref{fig:FracCntrb2}, we have depicted the fractional contributions of the leading optimized eigenvector pairs to the PDF uncertainties of various SIDIS data included in this study. 
This tells us whether the included pseudo-data sets provide similar or independent information on reducing the PDF uncertainties in the ePump-updating procedure.  
Lastly, we note that, with the first 15 optimized EV sets, the DSSV14 error bands (calculated from a total of 52 eigenvector pairs given by the Mc2Hessian package) can be recovered as much as 99.1\%, when 
applying the \texttt{ePump} optimization procedure to DSSV14H PDFs with the combination of both the EicC DIS and SIDIS pseudo-data sets (with a total of 332 data points).

\section{Summary}
\label{sec_summary}

In this work, we have presented a study that assesses the impact of future EicC data on the uncertainty bands of the DSSV14 helicity distribution functions and their moments. With EicC pseudo-data including DIS and SIDIS processes from doubly polarized electron-proton (3.5 GeV $\times$ 20 GeV) and electron-${}^3{\rm He}$ (3.5 GeV $\times$ 40 GeV) collisions, the DSSV14 PDF sets were updated by using a hessian updating procedure via the {\tt ePump} tool.
The resulting updated hessian set of the DSSV14 PDFs, named DSSV14H PDFs, was also used to evaluate the effects of specific initial and final state combinations of DIS and SIDIS processes at the future EicC on the uncertainties of the distributions and their moments.
Moreover, the DSSV14H PDF set was rotated into an equivalent Hessian set via the \texttt{ePump}-optimization  procedure, which is employed to explore the  
complementary role played by different
data sets in reducing the PDF uncertainty in the PDF-updating procedure.	
By identifying the dominant optimized eigenvector sets to the error band of each flavour and their contributions to the pseudo-data points, the sensitivities of EicC pseudo-data points to reducing the PDF error bands over various $x$ ranges were then assessed. 

As expected from the intent embedded in the design features of the EicC, we have observed a great reduction of the uncertainties of quark helicities in the sea-quark region, especially when considering SIDIS processes. It is important to remark that both electron-proton and electron-helium data are needed to obtain a consistent reduction of uncertainties over all quark flavours. Also essential to pin down the strange quark distribution are the kaon SIDIS data. In this regard, one of the limiting factors that will inevitably hinder the accuracy of the strange distribution in a global analysis with real EicC data will be the somewhat still large uncertainty of kaon fragmentation functions. With the advent of the Electron-Ion Colliders, extracting high precision fragmentation functions will become a more and more essential task.

The reduction of uncertainty of the gluon helicity distribution, although not as impressive as the one reported for the future US EIC in Ref.~\cite{Aschenauer:2012ve}, is still very significant for the low-$x$ region. The ability of the EicC to constrain the gluon distribution at energies lower than the ones reachable by the US EIC plays, nonetheless, a fundamental role in extending the coverage of meaningful data over a larger span of the phase space that will contribute to future extraction in a global analysis of the gluon helicity distribution. 

In our discussion, we have stressed the complementary role of the two machines, the US EIC and the EicC, especially when it comes to determining the relation between the proton spin and its flavour content.

As it is well known, the current electron-proton (neutron) fixed target experiments are limited by either their low center-of-mass energy or their low luminosity which let us precisely explore only the low $Q^2$ and high $x$ region. Answering the fundamental question of the origin of the proton spin is one of the main objectives of the future electron-ion machines and the extension of the accessible $Q^2-x$ coverage to higher $Q^2$ and lower $x$ values is a key component towards this goal. In our study, we have shown how the EicC data will greatly constrain the value of the spin contributions coming from quarks and gluons with momentum fraction $x\gtrsim 10^{-3}$. On the other hand, the US EIC is better suited to constrain them for even lower values of $x$. Together, the two machines will reduce the room left to speculations and enhance our understanding on the relation between the proton spin and the quark/gluon spin as well as their orbital and angular momentum down to an extremely small momentum fraction region. 

Finally, We note that in this study the tolerance value for the {\tt ePump} updating has been set to be $\Delta\chi^2=10$, which is of the same order of magnitude as the tolerance used in the DSSV14 analysis when studying the uncertainties via means of the Lagrange multiplier's method.
Using the typical choice of $\Delta\chi^2=1$ for this update would be 
inconsistent with the DSSV14 error PDFs used for generating the DSSV14H Hessian PDFs.
As explained in Refs.~\cite{Schmidt:2018hvu,Hou:2019gfw}, using the small value of $\Delta\chi^2=1$ to updating the given error PDFs is equivalent to overweighting these pseudo-data by about a factor of 10 in this study. 
This would result in much smaller PDF error bands than what we have concluded in this paper, so that using $\Delta\chi^2=1$ would greatly overestimate the effect of these pseudo-data on reducing the PDF errors.

\vspace{3mm}
\begin{acknowledgments}
 We are grateful to W. Vogelsang for helpful discussions and for providing us with the DSSV14 replica set. This work is partially supported by the Strategic Priority Research Program of Chinese Academy of Sciences under grant number XDB34030301, the Guangdong Major Project of Basic and Applied Basic Research No. 2020B0301030008, the National Natural Science Foundation of China (NSFC) under Grants No. 12022512 and No. 12035007, the Science and Technology Program of Guangzhou No. 2019050001, the Guangdong Provincial Key Laboratory of Nuclear Science with No.2019B121203010 as well as the U.S.~National Science Foundation
under Grant No.~PHY-2013791. D.P.~Anderle is also supported by the China Postdoctoral Science Foundation under Grant No. 2020M672668. C.-P.~Yuan is also grateful for the support from the Wu-Ki Tung endowed chair in particle physics.
\end{acknowledgments}

\appendix
\section{DIS and SIDIS at Leading Order}
\label{app:A1}
In this appendix, the better flavour separation power of the SIDIS process in respect to the DIS one is made explicit by presenting explicit LO expressions for $g_1$, $g_1^h$ and finally $A_1^h$.
 
Let's start by rewriting Eq.~(\ref{eq:g1LO}) for three flavours as
\begin{equation}
g_1(x,Q^2)=\left(\pm\frac{1}{12} \Delta v^+_3+\frac{1}{36}\Delta v^+_8+ \frac{1}{9}\Delta\Sigma\right)
\end{equation}
where we have included the values of the fractional charges $e^2_u=4/9$ and $e^2_d=e^2_s=1/9$ and
\begin{eqnarray}
\label{eq:DISNS}
\Delta v^+_3&=&(\Delta u+\Delta \bar u)-(\Delta d+\Delta \bar d)\nonumber\\
\Delta v^+_8&=&(\Delta u+\Delta \bar u)+(\Delta d+\Delta \bar d)-2(\Delta s+\Delta \bar s)\nonumber\\
\Delta\Sigma&=&(\Delta u+\Delta \bar u)+(\Delta d+\Delta \bar d)+(\Delta s+\Delta \bar s).
\end{eqnarray}

The $\pm$ sign in front of $\Delta v^+_3$ is set according to the scattered nucleon, proton or neutron accordingly.

From Eq.~(\ref{eq:DISNS}), it can be inferred that only $\Delta v^+_3$ may be determined directly from measurements of proton and neutron $g_1$ structure functions. On the other hand, $\Delta v^+_8$ and $\Delta\Sigma$ may be only determined thanks to their scaling properties. Moreover, using only DIS data it is impossible to disentangle quark from anti-quark distributions as there is no weighting factor to discriminate between them. 

Combinations of the type $\Delta q - \Delta \bar q$ can be probed if flavour changing interactions or processes involving hadronization are taken into account.
In the case of SIDIS, the observation of identified hadrons in the final state, allows to distinguish between $\Delta q$ and $\Delta \bar q$ thanks to the knowledge of the observed hadron flavour content described by the fragmentation functions $D^{ q \to h}$.
The SIDIS equivalent expression to Eq.~(\ref{eq:DISNS}) reads 
\begin{equation}
\label{eq:SIDISMatrix}
g_1^h(x,z,Q^2)=
\begin{pmatrix} \Delta v^+_8 & \pm\Delta  v^+_3 & \Delta \Sigma  \end{pmatrix}
\begin{pmatrix} 
\frac{1}{72} & \frac{1}{72} & \frac{1}{108} \\[2mm]
\frac{1}{72} & \frac{1}{72} & \frac{1}{36} \\[2mm]
\frac{1}{108} & \frac{1}{36} & \frac{1}{27}
\end{pmatrix}
\begin{pmatrix}  
D_{v^+_8} \\[2mm]  
D_{v^+_8} \\[2mm]
D_{\Sigma}  
\end{pmatrix} 
+ 
\begin{pmatrix} \Delta q^v_u & \Delta  q^v_d & \Delta q^v_s  \end{pmatrix}
\begin{pmatrix} 
\frac{2}{9} & 0 & 0 \\[2mm]
0 & \frac{1}{18} & 0\\[2mm]
0 & 0 & \frac{1}{18}
\end{pmatrix}
\begin{pmatrix}  
D^v_u \\[2mm]  
D^v_d \\[2mm]
D^v_s  
\end{pmatrix},
\end{equation}
where 
\begin{eqnarray}
D_{v^+_3}&=&(D^{ u \to h}+D^{ \bar u \to h})-(D^{ d \to h}+D^{ \bar d \to h})\nonumber\\
D_{v^+_8}&=&(D^{ u \to h}+D^{ \bar u \to h})+(D^{ d \to h}+D^{ \bar d \to h})-2(D^{ s \to h}+D^{ \bar s \to h})\nonumber\\
D_{\Sigma}&=&(D^{ u \to h}+D^{ \bar u \to h})+(D^{ d \to h}+D^{ \bar d \to h})+(D^{ s \to h}+D^{ \bar s \to h})\nonumber\\
D^v_{q}&=&D^{ q \to h} - D^{\bar q \to h}\;\;\text{with  } q=u,d,s\nonumber\\
\Delta q^v_u&=&(\Delta u - \Delta \bar u)\;\;\text{for proton target}\nonumber\\
&=&(\Delta d - \Delta \bar d)\;\;\text{for neutron target}\nonumber\\
\Delta q^v_d&=&(\Delta d - \Delta \bar d)\;\;\text{for proton target}\nonumber\\
&=&(\Delta u - \Delta \bar u)\;\;\text{for neutron target}\nonumber\\
\Delta q^v_s&=&(\Delta s - \Delta \bar s).
\end{eqnarray}

Eq.~(\ref{eq:SIDISMatrix}) makes the flavour structure and the relative weights to each parton distribution functions manifest to the reader. 

To conclude, we report the full expanded expression for the asymmetries $A_1^h$.
For proton measurement, the experimental observables $A_1^h$ in the SIDIS process for identified $\pi^\pm$ and $K^\pm$  can be written as:
\begin{equation}
\small
A_{1p}^{\pi^{+}} = \frac{0.5e_u^2 D^{ u \to \pi^{+} } }{F_{1p}^{\pi^+}} \Delta u 
                  + \frac{0.5e_{\bar{u}}^2 D^{ \bar{u} \to \pi^{+} } }{F_{1p}^{\pi^+}} \Delta \bar{u} 
                  + \frac{0.5e_d^2 D^{ d \to \pi^{+} } }{F_{1p}^{\pi^+}} \Delta d
                  + \frac{0.5e_{\bar{d}}^2 D^{ \bar{d} \to \pi^{+} } }{F_{1p}^{\pi^+}} \Delta \bar{d}
                  + \frac{0.5e_s^2 ( D^{ s \to \pi^{+} }+ D^{ \bar{s} \to \pi^{+}} ) }{F_{1p}^{\pi^+}} \Delta s,
\end{equation}

\begin{equation}
\small
A_{1p}^{\pi^{-}} = \frac{0.5e_u^2 D^{ u \to \pi^{-} } }{F_{1p}^{\pi^-}} \Delta u 
                  + \frac{0.5e_{\bar{u}}^2 D^{ \bar{u} \to \pi^{-} } }{F_{1p}^{\pi^-}} \Delta \bar{u} 
                  + \frac{0.5e_d^2 D^{ d \to \pi^{-} } }{F_{1p}^{\pi^-}} \Delta d
                  + \frac{0.5e_{\bar{d}}^2 D^{ \bar{d} \to \pi^{-} } }{F_{1p}^{\pi^-}} \Delta \bar{d}
                  + \frac{0.5e_s^2 ( D^{ s \to \pi^{-} }+ D^{ \bar{s} \to \pi^{-}} ) }{F_{1p}^{\pi^-}} \Delta s,
\end{equation}

\begin{equation}
\small
A_{1p}^{K^{+}} = \frac{0.5e_u^2 D^{ u \to K^{+} } }{F_{1p}^{K^+}} \Delta u 
                  + \frac{0.5e_{\bar{u}}^2 D^{ \bar{u} \to K^{+} } }{F_{1p}^{K^+}} \Delta \bar{u} 
                  + \frac{0.5e_d^2 D^{ d \to K^{+} } }{F_{1p}^{K^+}} \Delta d
                  + \frac{0.5e_{\bar{d}}^2 D^{ \bar{d} \to K^{+} } }{F_{1p}^{K^+}} \Delta \bar{d}
                  + \frac{0.5e_s^2 ( D^{ s \to K^{+} }+ D^{ \bar{s} \to K^{+}} ) }{F_{1p}^{K^+}} \Delta s,
\end{equation}

\begin{equation}
\small
A_{1p}^{K^{-}} = \frac{0.5e_u^2 D^{ u \to K^{-} } }{F_{1p}^{K^-}} \Delta u 
                  + \frac{0.5e_{\bar{u}}^2 D^{ \bar{u} \to K^{-} } }{F_{1p}^{K^-}} \Delta \bar{u} 
                  + \frac{0.5e_d^2 D^{ d \to K^{-} } }{F_{1p}^{K^-}} \Delta d
                  + \frac{0.5e_{\bar{d}}^2 D^{ \bar{d} \to K^{-} } }{F_{1p}^{K^-}} \Delta \bar{d}
                  + \frac{0.5e_s^2 ( D^{ s \to K^{-} }+ D^{ \bar{s} \to K^{-}} ) }{F_{1p}^{K^-}} \Delta s,
\end{equation}

where 
\begin{equation}
\small
F_{1p}^{\pi^+}=0.5e_u^2 u D^{ u \to \pi^{+} } + 0.5e_{\bar{u}}^2 \bar{u} D^{ \bar{u} \to \pi^{+} } + 0.5e_d^2 d D^{ d \to \pi^{+} }+ 0.5e_{\bar{d}}^2 \bar{d} D^{ \bar{d} \to \pi^{+} } + 0.5e_s^2 s D^{ s \to \pi^{+} } +0.5e_{\bar{s}}^2 \bar{s} D^{ \bar{s} \to \pi^{+}}.
\end{equation}

For neutron measurements, if one uses iso-spin symmetry:
$\Delta u^n = \Delta d^p, \Delta \bar{u}^n = \Delta \bar{d}^p$,
$\Delta d^n = \Delta u^p, \Delta \bar{d}^n= \Delta \bar{u}^p$,
$\Delta s^n = \Delta s^p = \Delta \bar{s}^n = \Delta \bar{s}^p$,
the asymmetries can be expressed (in terms of helicity distributions in proton) as:
\begin{equation}
\small
A_{1N}^{\pi^{+}} = \frac{0.5e_d^2 D^{ d \to \pi^{+} } }{F_{1N}^{\pi^+}} \Delta u
                    +\frac{0.5e_{\bar{d}}^2 D^{ \bar{d} \to \pi^{+} } }{F_{1N}^{\pi^+}} \Delta \bar{u}
                    +\frac{0.5e_u^2 D^{ u \to \pi^{+} } }{F_{1N}^{\pi^+}} \Delta d 
                    + \frac{0.5e_{\bar{u}}^2 D^{ \bar{u} \to \pi^{+} } }{F_{1N}^{\pi^+}} \Delta \bar{d}
                    + \frac{0.5e_s^2 ( D^{ s \to \pi^{+} }+ D^{ \bar{s} \to \pi^{+}} ) }{F_{1N}^{\pi^+}} \Delta s,
\end{equation}

\begin{equation}
\small
A_{1N}^{\pi^{-}} = \frac{0.5e_d^2 D^{ d \to \pi^{-} } }{F_{1N}^{\pi^-}} \Delta u
                    +\frac{0.5e_{\bar{d}}^2 D^{ \bar{d} \to \pi^{-} } }{F_{1N}^{\pi^-}} \Delta \bar{u}
                    +\frac{0.5e_u^2 D^{ u \to \pi^{-} } }{F_{1N}^{\pi^-}} \Delta d 
                    + \frac{0.5e_{\bar{u}}^2 D^{ \bar{u} \to \pi^{-} } }{F_{1N}^{\pi^-}} \Delta \bar{d}
                    + \frac{0.5e_s^2 ( D^{ s \to \pi^{-} }+ D^{ \bar{s} \to \pi^{-}} ) }{F_{1N}^{\pi^-}} \Delta s,
\end{equation}

\begin{equation}
\small
A_{1N}^{K^{+}} = \frac{0.5e_d^2 D^{ d \to K^{+} } }{F_{1N}^{K^+}} \Delta u
                    +\frac{0.5e_{\bar{d}}^2 D^{ \bar{d} \to K^{+} } }{F_{1N}^{K^+}} \Delta \bar{u}
                    +\frac{0.5e_u^2 D^{ u \to K^{+} } }{F_{1N}^{K^+}} \Delta d 
                    + \frac{0.5e_{\bar{u}}^2 D^{ \bar{u} \to K^{+} } }{F_{1N}^{K^+}} \Delta \bar{d}
                    + \frac{0.5e_s^2 ( D^{ s \to K^{+} }+ D^{ \bar{s} \to K^{+}} ) }{F_{1N}^{K^+}} \Delta s,
\end{equation}

\begin{equation}
\small
A_{1N}^{K^{-}} = \frac{0.5e_d^2 D^{ d \to K^{-} } }{F_{1N}^{K^-}} \Delta u
                    +\frac{0.5e_{\bar{d}}^2 D^{ \bar{d} \to K^{-} } }{F_{1N}^{K^-}} \Delta \bar{u}
                    +\frac{0.5e_u^2 D^{ u \to K^{-} } }{F_{1N}^{K^-}} \Delta d 
                    + \frac{0.5e_{\bar{u}}^2 D^{ \bar{u} \to K^{-} } }{F_{1N}^{K^-}} \Delta \bar{d}
                    + \frac{0.5e_s^2 ( D^{ s \to K^{-} }+ D^{ \bar{s} \to K^{-}} ) }{F_{1N}^{K^-}} \Delta s,
\end{equation}

where (in the format of PDFs in proton)
\begin{equation}
F_{1N}^{\pi^+}=0.5e_d^2 u D^{ d \to \pi^{+} } + 0.5e_{\bar{d}}^2 \bar{u} D^{ \bar{d} \to \pi^{+} } + 0.5e_u^2 d D^{ u \to \pi^{+} }+ 0.5e_{\bar{u}}^2 \bar{d} D^{ \bar{u} \to \pi^{+} } + 0.5e_s^2 s D^{ s \to \pi^{+} } +0.5e_{\bar{s}}^2 \bar{s} D^{ \bar{s} \to \pi^{+}}. 
\end{equation}

\bibliographystyle{ieeetr}
\bibliography{main}
 
\end{document}